\documentclass[aps,prb,twocolumn,notitlepage,nofootinbib,longbibliography]{revtex4-2}
\usepackage{amsmath}
\usepackage{amssymb}
\usepackage{amsfonts}
\usepackage{graphicx}
\usepackage{hyperref}

\begin{document}
\title{Spin waves and high-frequency response in layered superconductors
with helical magnetic structure}
\author{A. E. Koshelev}
\affiliation{Materials Science Division, Argonne National Laboratory, Argonne,
Illinois 60439}
\date{\today }
\begin{abstract}
We evaluate the spin-wave spectrum and dynamic susceptibility in
a layered superconductor with helical interlayer magnetic structure.
We especially focus on the structure in which the moments rotate 90$^{\circ}$
from layer to layer realized in the iron pnictide RbEuFe$_{4}$As$_{4}$.
While in nonmagnetic superconductors low-frequency magnetic field decays on the
distance of the order of the London penetration depth,  spin waves mediate
its propagation to much larger distances limited by external dissipation mechanisms.
The spin-wave spectrum in superconductors is strongly renormalized due to the
long-range electromagnetic interactions between the oscillating magnetic moments.
This leads to strong enhancement of the frequency of the mode coupled with
uniform field and this enhancement exists only within a narrow range of the $c$-axis wave vectors of the order of the inverse London penetration depth. 
The key feature of materials like RbEuFe$_{4}$As$_{4}$ is that this uniform mode corresponds to the maximum frequency of the spin-wave spectrum with respect to
the $c$-axis wave vector.  
As a consequence, the high-frequency surface resistance acquires
a very distinct asymmetric feature spreading between the bare and
renormalized frequencies. We also consider excitation of spin waves
with the Josephson effect in a tunneling contact between helical-magnetic
and conventional superconductors and study the interplay between the spin-wave
features and geometrical cavity resonances in the current-voltage
characteristics. 
\end{abstract}
\maketitle

\section{Introduction}

Experimental realization, characterization, and understanding of quantum
materials have emerged as central topics in modern physics research.
Quantum materials have the potential to offer new functionalities
enabling novel applications and therefore provide a fundamental basis
for future technological advances. Superconductors supporting long-range
magnetic order represent a rare class of quantum materials with unique
properties caused by the interplay between magnetic and superconducting
subsystems\cite{BulaevskiiAdvPhys85,KulicBuzdinSupercondBook,WolowiecPhysC15,MapleFischerBook1982}.
As singlet superconductivity and ferromagnetism are strongly incompatible
states, the ground-state configurations are always characterized by
nonuniform structures of either magnetic moments or superconducting
gap parameter. The nature of nonuniform configurations ultimately
determines transport and thermodynamic properties of these materials.
In the case of strong superconductivity and soft magnetism, it was
theoretically predicted that the exchange interaction between two
subsystems favors a nonuniform magnetic state either in the form of
small-size domains \cite{AndersonPhysRev.116.898} or helical structure
\cite{BulaevskiiJLTP1980} for strong and weak magnetic anisotropy,
respectively.

Several classes of magnetic singlet superconductors have been discovered
and thoroughly characterized. The magnetism in these materials is
hosted in the rare-earth-element sublattice spatially separated from
the conduction-electron sublattice. In spite of high density of rare-earth
($\mathit{R\!E}$) local moments, the superconductivity survives because
the exchange interaction between two sublattices is relatively weak.
Various nonuniform magnetic structures have been revealed in the coexistence
regions. 

The first two groups of magnetic superconductors discovered half a century
ago are ternary molybdenum chalcogenides (Chevrel phases), such as
HoMo$_{6}$S$_{8}$, with superconducting transition at $T_{c}\!\approx\!1.2$ K\cite{IshikawaSSC77},
and ternary rhodium borides, such as ErRh$_{4}$B$_{4}$, with $T_{c}\!\approx\!8.5$ K\cite{FertigPhysRevLett.38.987},
see detailed reviews \cite{BulaevskiiAdvPhys85,WolowiecPhysC15,MapleFischerBook1982}.
In these materials the exchange interaction between magnetic and superconducting
subsystems is actually not very weak: the emerging ferromagnetism
at sub-Kelvin temperatures destroys superconductivity and causes the
reentrance of the normal state. Nevertheless, a narrow coexistence
region does exist near the reentrance where an intermediate oscillatory
magnetic state is formed \cite{LynnPhysRevLett.46.368,LynnPhysRevB.24.3817,MonctonPhysRevLett.45.2060},
in qualitative agreement with theoretical expectations.

Another important class of magnetic superconductors is the rare-earth
nickel borocarbides $\mathit{R\!E}$Ni$_{2}$B$_{2}$C, see reviews\cite{MullerRoPP2001,GuptaAdvPhys06,MazumdarPhysC2015}.
In contrast to the nearly cubic ternary compounds, these are layered
materials composed of magnetic $\mathit{R\!E}$C layers and conducting
Ni layers. The superconductivity coexists with different kinds of
magnetic order in four compounds with $\mathit{R\!E}$$\rightarrow$Tm,
Er, Ho, and Dy. The magnetic moments typically order ferromagnetically
within $\mathit{R\!E}$C layers and alternate from layer to layer
(A-type antiferromagnets). This basic configuration, however, is perturbed
in some compounds. Particularly interesting are Er and Ho compounds
where the magnetic transition takes place inside the superconducting
state at temperatures comparable with the superconducting transition
temperature (10.5 K and 8 K, for Er and Ho, respectively). Magnetic
structure in the ErNi$_{2}$B$_{2}$C is characterized by additional
in-plane modulation, which is probably caused by interaction with
the superconducting sublattice. In addition, a peculiar weak ferromagnetic
state appears below 2.3 K, and, contrary to HoMo$_{6}$S$_{8}$ and
ErRh$_{4}$B$_{4}$, it coexists with superconductivity at lower temperatures.
In the Ho compound the magnetic phase diagram is also very rich: the
transition to the low-temperature A-type antiferromagnetic state occurs
via two intermediate incommensurate spiral configurations, one with
helix direction along the $c$ axis and another with additional in-plane
modulation. 

Contrary to singlet Cooper pairing, a rare triplet superconducting
state may coexist with uniform ferromagnetism. Such triplet state
is realized in uranium-based compounds UGe$_{2}$, URhGe, and UCoGe,
which become superconducting at sub-Kelvin temperature range, inside
the ferromagnetic state, see reviews \cite{AokiJPSJ12,AokiJPSJ19,HuxleyPhysC2015}.
In spite of low transition temperatures, due to triplet pairing, the
superconducting state survives up to remarkably high magnetic field,
10-25 teslas. The triplet state is also likely realized in the recently
discovered compound UTe$_{2}$, even though this material is not magnetic
\cite{RanScience1019}.

Interest in the physics of magnetic singlet superconductors has been
recently reinvigorated by the discovery of the magnetically ordered
iron pnictides, in particular, europium-based 122 compounds, see review
\cite{Zapf2017}. The layered structure of these materials is similar
to borocarbides: they are composed of the magnetic Eu and conducting
FeAs layers. The parent material EuFe$_{2}$As$_{2}$ is a nonsuperconducting
compensated multiple-band metal which has the spin-density-wave transition
in the FeAs layers at 189 K and the A-type antiferromagnetic transition
in the Eu$^{2+}$ layers at 19 K with the magnetic moments aligned
along the layers \cite{Ren2008,Jeevan2008a,Xiao2010}. The superconducting
state emerges under pressure with the maximum transition temperature
reaching 30 K at 2.6 GPa exceeding the magnetic transition temperature
in the Eu sublattice \cite{MicleaPhysRevB.79.212509,*TerashimaJPSJ2009}.
Superconducting compounds with Eu magnetic order also have been obtained
by numerous chemical substitutions on different atomic sites of the
parent compound, including isovalent substitutions of P on the As site \cite{RenPhysRevLett.102.137002,Jeevan2011,Cao2011,TokiwaPhysRevB.86.220505,ZapfPhysRevLett.110.237002,NandiPhysRevB.89.014512}
and of Ru on the Fe site \cite{JiaoEPL2011}, electron doping via substitutions
of Co \cite{JiangPhysRevB.80.184514,*HeJPhys2010,*GuguchiaPhysRevB.84.144506,*JinPhysRevB.88.214516}
or Ir \cite{Paramanik2014} on the Fe site, and hole-doping via substitutions
of K \cite{Jeevan2008} or Na \cite{QiNJP2008} on the Eu site. 
The maximum superconducting transition temperature for different substitution
series ranges from 22 to 35 K exceeding the magnetic-transition temperature
in Eu layers. Therefore the key feature of these materials is that
they have magnetic transition in the Eu$^{2+}$ sublattice at the
temperature scale, comparable with the superconducting transition
in FeAs sublattice. The most studied substituted superconductor in
this family is $\mathrm{Eu}\mathrm{Fe}(\mathrm{As}_{1-x}\mathrm{P}_{x})_{2}$.
The superconducting transition temperature reaches a maximum of 26 K
for $x\approx0.3$ followed by the ferromagnetic transition at 19 K.
Contrary to the parent compound, the Eu moments align ferromagnetically
along the $c$ axis at 19 K\cite{NandiPhysRevB.89.014512}. At lower
temperatures, the coexistence of ferromagnetism with superconductivity
leads to the formation of the composite domain and vortex-antivortex
structure visualized by the decorations \cite{VeshchunovJETPLett2017}
and magnetic-force microscopy \cite{StolyarovSciAdv18}. This structure
has been explained assuming purely electromagnetic coupling between
the magnetic moments and superconducting order parameter \cite{DevizorovaPhysRevLett.122.117002}. 

A recent addition to the family of Eu-based iron pnictides is the
stoichiometric 1144 compounds \emph{A}EuFe$_{4}$As$_{4}$ with \emph{A}=Rb
\cite{Liu2016,KawashimaJPSJ2016,Bao2018,Smylie2018,Stolyarov2018}
and Cs \cite{Liu2016a,KawashimaJPSJ2016} in which every second layer
of Eu in the parent material is replaced with the layer of nonmagnetic
Rb or Cs. These materials have the superconducting transition temperature
of 36.5 K, higher than the doped 122 Eu compounds. Such high transition
temperature is achieved because of close-to-optimal hole concentration
and the absence of disorder caused by random substitutions. On the
other hand, the magnetic transition temperature 15 K is 4 K lower
than in the parent 122 compound, probably due to the weaker interaction
between the magnetic layers. These materials are characterized by
highly anisotropic easy-axis Eu magnetism \cite{Smylie2018,WillaPhysRevB.99.180502,hemmida2020topological}.
With increasing pressure the superconducting transition temperature
decreases and the magnetic transition temperature increases so that
at pressures larger that $~\sim7$ GPa the superconducting transition
already takes place in the magnetically ordered state \cite{JacksonPhysRevB.98.014518,XiangPhysRevB.99.144509}.
Recent resonant X-ray scattering and neutron diffraction measurements
demonstrated that the magnetic structure is helical: the Eu moments
align ferromagnetically inside the layers and rotate 90$^{\circ}$
from layer to layer \cite{IidaPhysRevB.100.014506,IslamPreprint2019}.

New materials frequently host new physical phenomena. In this paper
we investigate spin waves and related properties for layered superconductors
with helical magnetic structure with the modulation perpendicular
to the layer direction. Spin waves is the most important dynamic characteristic
of magnetic materials \cite{VanKranendonkRevModPhys.30.1,AkhiezerBook68,prabhakar2009spin}
and their properties are essential for the emerging spintronics\cite{WolfSci2001,ZuticRevModPhys.76.323}
and magnonics\cite{NeusserAdvMat09,KruglyakJPhys2010,ChumakNatPhys2015}
applications. As the ground-state configuration, the spin-wave spectrum
is determined by the exchange and electromagnetic interactions between
the moments and by magnetic anisotropy. A key feature of superconducting
materials is that the long-wave part of the spin-wave spectrum is
renormalized in a nontrivial way by long-range electromagnetic interactions
between the oscillating magnetic moments. In the case of a ferromagnetic
triplet superconductor with uniform magnetization, the spectrum of spin
waves, their excitation by the external electromagnetic waves, and
related features in the surface impedance have been considered in
Refs.\ \cite{BraudePhysRevLett.93.117001,BraudePhysRevB.74.054515}.
The spectrum of spin waves in antiferromagnetic singlet superconductors has been evaluated in Ref.\ \cite{BuzdinJETPLett84}.
Here we extend these considerations to superconductors with helical
magnetic structure. While some of our results are valid for a general
modulation period, we mostly focus on the case relevant for RbEuFe$_{4}$As$_{4}$,
namely, the structure in which the moments rotate 90$^{\circ}$ from layer
to layer and the easy-plane anisotropy exceeding the interlayer exchange
interaction. We evaluate the spin-wave spectrum as a function of the
$c$-axis wave vector and find that the mode having a $c$-axis uniform component
of the oscillating spins corresponds to the spectrum maximum. This
mode is strongly renormalized by the long-range electromagnetic interactions,
its frequency increases by the factor of the square root of the magnetic
permeability with respect to the bare value determined only by local
interactions. This enhancement rapidly drops when the $c$-axis wave vector
shift exceeds the inverse London penetration depth. This behavior
is qualitatively different from the case of ferromagnetic alignment \cite{BraudePhysRevLett.93.117001,BraudePhysRevB.74.054515},
where the frequency of the uniform mode is the smallest frequency
of the spectrum. We evaluate the high-frequency surface resistance
and find that it acquires a very asymmetric feature with a sharp maximum
at the bare uniform-mode frequency and a tail extending up to the
renormalized frequency.

We also investigate excitation of spin waves with AC Josephson effect
in a tunneling contact between helical-magnetic and conventional superconductors
and study the interplay between spin-wave features and geometrical
Fiske resonances in the current-voltage characteristics. This consideration
is somewhat related to the excitation of the spin waves by the Josephson
effect in the ferromagnetic interlayer in SFS junctions \cite{VolkovPhysRevLett.103.037003,MaiPhysRevB.84.144519}.
In our case, however, the spin-wave feature in current-voltage characteristic
has a very distinct shape due to the unusual spectrum in helical-magnetic
superconductor, similar to the feature in the frequency dependence
of the surface resistance. Namely, the current is sharply enhanced
when the Josephson frequency matches the bare uniform-mode frequency
and at higher frequencies this excess current has a long tail extending
up to the renormalized frequency. 

The paper is organized as follows. In Sec.~\ref{sec:Model}, we introduce
the model and write general relations determining the spin-wave spectrum
via the dynamic magnetic susceptibility. In Sec.~\ref{sec:MagnGroundState},
we consider the helical magnetic ground state. The bare spin-wave
spectrum due to the short-range interactions is derived in Sec.~\ref{sec:BareSpectrum}.
In Sec.~\ref{sec:Response}, we investigate the response to nonuniform
magnetic field and derive the nonlocal dynamic susceptibility.
Electromagnetic renormalization of the spectrum is considered in Sec.~\ref{sec:EM-renor}.
In Sec.~\ref{sec:DynEqSmooth}, we consider the dynamic equation for
smooth magnetization, derive the magnetic boundary condition, and evaluate
the frequency dependence of the surface impedance. In Sec.~\ref{sec:ExcitJos},
we investigate the excitation of spin waves by the Josephson effect.

\section{Model and general equations\label{sec:Model}}

We consider a layered magnetic superconductor described by the energy functional
\begin{align}
\mathcal{E} & =\mathcal{E}_{m}\!+\!\mathcal{E}_{s}\nonumber \\
 & \!+\!\int\!d^{3}\boldsymbol{r}\left(\!\frac{\boldsymbol{B}^{2}}{8\pi}-\boldsymbol{B}\boldsymbol{M}+2\pi\boldsymbol{M}^{2}\!-\frac{\boldsymbol{H}_{e}\boldsymbol{B}}{4\pi}\right),\label{eq:Energy}
\end{align}
where the term
\begin{equation}
\mathcal{E}_{s}=\int\!d^{3}\boldsymbol{r}\sum_{i=x,y,z}\frac{1}{8\pi\lambda_{i}^{2}}\left(A_{i}\!-\frac{\Phi_{0}}{2\pi}\nabla_{i}\varphi\right)^{2}\!\label{eq:Es}
\end{equation}
is the kinetic energy of the superconducting subsystem in the London approximation
determined by the components of the penetration depth $\lambda_{i}$
and $\boldsymbol{A}$ is the vector potential determining the local magnetic
induction, $\boldsymbol{B}=\boldsymbol{\nabla}\!\times\!\boldsymbol{A}$.
In the following, we consider the Meissner state and drop the phase of
the superconducting order parameter $\varphi$. We assume that the
magnetic subsystem is described by the classical quasi-two-dimensional easy-plane
Heisenberg model
\begin{align}
\mathcal{E}_{m} & =-\mathcal{J}\sum_{\langle\boldsymbol{i},\boldsymbol{j}\rangle,n}\boldsymbol{S}_{\boldsymbol{i},n}\boldsymbol{S}_{\boldsymbol{j},n}\nonumber \\
+ & \mathcal{K}\sum_{\boldsymbol{i},n}(2S_{z,\boldsymbol{i},n}^{2}\!-1)-\!\sum_{\boldsymbol{i},n,\ell>0}\mathcal{J}_{z,\ell}\boldsymbol{S}_{\boldsymbol{i},n}\boldsymbol{S}_{\boldsymbol{i},n+\ell},\label{eq:EnerLay}
\end{align}
where $\boldsymbol{S}_{\boldsymbol{i},n}$ is the spin at the site
$\boldsymbol{i}$ and in the layer $n$ with the absolute value equal to
$S$, $\mathcal{J}$ is the in-plane exchange constant, $\mathcal{K}$
is the easy-plane anisotropy, and $\mathcal{J}_{z,\ell}$ are the interlayer
exchange constants. %
The exchange constants likely have a substantial RKKY contribution. The behavior of $\mathcal{J}_{z,\ell}$ for $\ell>\xi_c/d$ is strongly affected by superconductivity\cite{BulaevskiiJLTP1980,KoshelevPhysRevB19}, where $d$ is the separation between the magnetic layers and $\xi_c$ is the $c$-axis coherence length. 
Local spins determine local magnetic moments $\boldsymbol{m}_{\boldsymbol{i},n}\!=\!g\mu_{B}\boldsymbol{S}_{\boldsymbol{i},n}$
where $\mu_{B}$ is the Bohr magneton. Therefore, the bulk magnetization
$\boldsymbol{M}(\boldsymbol{r})$ in Eq.~\eqref{eq:Energy} is
related to the coarse-grained spin distribution as $\boldsymbol{M}(\boldsymbol{r})=n_{M}g\mu_{B}\boldsymbol{S}(\boldsymbol{r})$,
where $n_{M}$ is the bulk density of spins. $\boldsymbol{S}(\boldsymbol{r})$
in this relation is obtained by averaging of $\boldsymbol{S}_{\boldsymbol{i},n}$
over distances much larger than neighboring spin separations.

Slowly varying in space oscillating magnetization generates macroscopic
magnetic fields which couple with this magnetization. This effect
is especially important in superconductors where it leads to significant renormalization of the
spin-wave spectrum \cite{BraudePhysRevLett.93.117001,BraudePhysRevB.74.054515}.
We will assume that the supercurrent response to the slowly oscillating
magnetization can be treated quasistatically. The corresponding equation
is obtained by variation of the energy with respect to the vector
potential $\boldsymbol{A}$,
\begin{equation}
\left(\hat{\lambda}^{-2}-\triangle\right)\boldsymbol{A}-4\pi\boldsymbol{\nabla\times}\boldsymbol{M} =0.\label{eq:VectPot}
\end{equation}
We can transform this equation into the equation connecting the local
magnetic field strength $\boldsymbol{H}=\boldsymbol{B}-4\pi\boldsymbol{M}$
and magnetization
\begin{equation}
\boldsymbol{H}+\boldsymbol{\nabla}\times\hat{\lambda}^{2}\boldsymbol{\nabla}\times\boldsymbol{H}=-4\pi\boldsymbol{M}.\label{eq:HM}
\end{equation}
For time-dependent fields, this equation is modified by quasiparticle
currents. We neglect this contribution assuming that the time variations are slow. On the other hand, the oscillating
magnetic field generates oscillating magnetization due to dynamic
magnetic response and the relation between their Fourier components is determined by the dynamic magnetic susceptibility $\hat{\chi}\left( \boldsymbol{k},\omega\right)$,
\begin{equation}
\boldsymbol{M}( \boldsymbol{k},\omega)=\!\hat{\chi}\left( \boldsymbol{k},\omega\right)\boldsymbol{H}( \boldsymbol{k},\omega).\label{eq:DynMagnResp}
\end{equation}
Note that the poles of $\hat{\chi}\left( \boldsymbol{k},\omega\right)$
give the bare spin-wave spectrum due to local interactions unrenormalized
by long-range fields. From Eqs.~\eqref{eq:HM} and \eqref{eq:DynMagnResp},
we obtain the general linear equation for $\boldsymbol{H}(q,\omega)$
which determines the spectrum of spin waves
\begin{equation}
\boldsymbol{H}- \boldsymbol{k}\times\hat{\lambda}^{2} \boldsymbol{k}\times\boldsymbol{H}  =-4\pi\hat{\chi}\left( \boldsymbol{k},\omega\right)\boldsymbol{H}.
\label{eq:SpWaveRenormGen}
\end{equation}
In the following, we consider a simple geometry of the wave vector
oriented along the $z$ direction and isotropic in-plane case, $\lambda_{x}=\lambda_{y}\equiv\lambda$. 
In this case, Eq.~\eqref{eq:SpWaveRenormGen} becomes
\[
\left(1+\lambda^{2}k_{z}^{2}\right)H_{\alpha}(k_{z},\omega)=-4\pi\chi_{\alpha\beta}\left(k_{z},\omega\right)H_{\beta}(k_{z},\omega).
\]
with $\alpha=x,y$. Note that the off-diagonal susceptibility $\chi_{xy}\left(k_{z},\omega\right)$
is finite in the helical magnetic state. Since $\chi_{yx}\left(k_{z},\omega\right)=-\chi_{xy}\left(k_{z},\omega\right)$,
we obtain the following equation for the spin-wave spectrum renormalized
by long-range electromagnetic interactions
\begin{equation}
	1\!+\frac{4\pi\left[\chi_{xx}\left(k_{z},\omega\right)\pm i\chi_{xy}\left(k_{z},\omega\right)\right]}{1+\lambda^{2}k_{z}^{2}}=\!0.\label{eq:SpWaveRenorm}
\end{equation}
The dynamic susceptibility $\chi_{\alpha\beta}\left(\boldsymbol{k},\omega\right)$
can be evaluated by solving the Landau-Lifshitz equation 
\begin{equation}
\frac{d\boldsymbol{M}}{dt}=-\gamma\left[\boldsymbol{M}\times\frac{\delta\mathcal{E}_{m}}{\delta\boldsymbol{M}}\right]+\gamma\left[\boldsymbol{M}\times\boldsymbol{H}\right]\label{eq:LL}
\end{equation}
in the linear order with respect to small deviations of the magnetization from the equilibrium configuration. Here $\gamma=g\mu_{B}/\hbar$ is the gyromagnetic factor.  We neglected the damping terms.

\section{Magnetic ground-state configuration\label{sec:MagnGroundState}}

\subsection{Arbitrary modulation wave vector}

We start with consideration of the helical interlayer magnetic ground
state determined by the energy in Eq.\ \eqref{eq:EnerLay}. In the classical
description, it is convenient to introduce the unit vectors $\boldsymbol{s}_{\boldsymbol{i},n}$$=(\cos\phi_{\boldsymbol{i},n}\cos\theta_{\boldsymbol{i},n},$ $\sin\phi_{\boldsymbol{i},n}\cos\theta_{\boldsymbol{i},n},$ $\sin\theta_{\boldsymbol{i},n})$
along the direction of $\boldsymbol{S}_{\boldsymbol{i},n}$, $\boldsymbol{S}_{\boldsymbol{i},n}\!=\!S\boldsymbol{s}_{\boldsymbol{i},n}$.
Then we can rewrite the energy in Eq.~\eqref{eq:EnerLay} as
\begin{align}
\mathcal{E}_{m} & =-J\sum_{\langle\boldsymbol{i},\boldsymbol{j}\rangle,n}\boldsymbol{s}_{\boldsymbol{i},n}\boldsymbol{s}_{\boldsymbol{j},n}+K\sum_{\boldsymbol{i},n}(2s_{z,\boldsymbol{i},n}^{2}-1)\nonumber \\
 & -\sum_{\boldsymbol{i},n,\ell>0}J_{z,\ell}\boldsymbol{s}_{\boldsymbol{i},n}\boldsymbol{s}_{\boldsymbol{i},n+\ell}.\label{eq:EnerLayRed}
\end{align}
with new parameters $J=\mathcal{J}S^{2}$, $K=\mathcal{K}S^{2}$, and $J_{z,\ell}=\mathcal{J}_{z,\ell}S^{2}$. The advantage of the constants $K$, $J$, and $J_{z,\ell}$ is that they immediately represent the energy scales of the corresponding interactions. 
Frustrated interlayer interactions may lead to the helical ground
state corresponding to $\phi_{\boldsymbol{i},n}^{(0)}=qn$ and $\theta_{\boldsymbol{i},n}^{(0)}=0$ \cite{Nagamiya1968,JohnstonPhysRevB.91.064427}.
The energy per spin for such a state is given by
\begin{equation}
E_{0}(q)=-J_{z}(q)/2\label{eq:HelEn-Q}
\end{equation}
where 
\begin{equation}
J_{z}(q)=2\sum_{\ell=1}^{\infty}J_{z,\ell}\cos(q\ell)
\label{eq:Jzq}
\end{equation}
is the discrete Fourier transform of the interlayer interactions.

The total energy also has an electromagnetic (dipole) contribution, which
is substantially affected by superconductivity. As follows from Eq.~\eqref{eq:HM},
the magnetic field generated by uniformly polarized layers with arbitrary
in-plane orientation of magnetization, $\boldsymbol{M}(z)=\sum_{n}\mathcal{\boldsymbol{\mathcal{M}}}_{n}\delta(z-z_{n})$,
is given by
\begin{equation}
	\boldsymbol{H}(z)=-\sum_{n}\left(2\pi\mathcal{\boldsymbol{\mathcal{M}}}_{n}/\lambda\right)\exp\left(-|z-z_{n}|/\lambda\right),\label{eq:H-layer}
\end{equation}
where $\mathcal{\boldsymbol{\mathcal{M}}}_{n}=dn_{M}\boldsymbol{m}_{n}$ is
the two-dimensional moment densities, $d$ is the separation between
the magnetic layers, and $z_{n}=nd$. Note that the superconducting environment
leads to the finite magnetic field outside a uniformly polarized layer,
contrary to the normal-state case, in which such field is absent. The
corresponding magnetic induction and vector-potential are 
\begin{align*}
	\boldsymbol{B}(z)\! & =\!4\pi\!\sum_{n}\mathcal{\boldsymbol{\mathcal{M}}}_{n}\left[\delta(z-z_{n})\!-\!\left(1/2\lambda\right)\exp\left(-\!|z\!-\!z_{n}|/\lambda\right)\right],\\
	\boldsymbol{A}(z) \!& =\!-2\pi\!\sum_{n}\boldsymbol{n}_{z}\!\times\!\boldsymbol{\mathcal{M}}_{n}\,\mathrm{sign}(z\!-\!z_{n})\exp\left(\!-\!|z\!-\!z_{n}|/\lambda\right).
\end{align*}
Substituting these distributions into superconducting and magnetic
energy terms in Eq.~\eqref{eq:Energy}, we derive the bulk electromagnetic
energy density
\begin{equation}
	\mathcal{F}_{\mathrm{em}}=\frac{\pi}{\lambda L_{z}}\sum_{n,m}\boldsymbol{\mathcal{M}}_{n}\boldsymbol{\mathcal{M}}_{m}\exp\left(-|z_{n}-z_{m}|/\lambda\right).\label{eq:EMenergy}
\end{equation}
Therefore, for the helical order, $\mathcal{M}_{x,n}\!=\!\mathcal{M}_{0}\cos\left(Qn\right)$, $\mathcal{M}_{y,n}\!=\!\mathcal{M}_{0}\sin\left(Qn\right)$,
the bulk electromagnetic-energy cost is 
\begin{align}
	\mathcal{F}_{\mathrm{em}}(Q) & =\frac{\pi}{\lambda d}
	\mathcal{M}_{0}^{2}\frac{\sinh\left(d/\lambda\right)}{\cosh\left(d/\lambda\right)-\cos Q}\nonumber \\
	\approx & \frac{2\pi M_{0}^{2}}{1+2\lambda^{2}\left(1-\cos Q\right)/d^{2}},\label{eq:EMEnQ}
\end{align}
where $M_{0}\!=\!\mathcal{M}_{0}/d\!=\!n_{M}m_0$ is the bulk saturation magnetization.
The corresponding energy per spin $E_{\mathrm{em}}(Q)=\mathcal{F}_{\mathrm{em}}(Q)/n_{M}$
has to be compared with the exchange energy in Eq.~\eqref{eq:HelEn-Q}.
In the range $2\lambda^{2}(1\!-\!\cos Q)/d^{2}\!\gg\!1$, this
amounts to comparison of the typical dipole energy scale $E_{d0}\!=\!\pi d^{2}n_{M}m_{0}^{2}/\lambda^{2}$
with the interlayer exchange constants $J_{z,\ell}$. Typically the
dipole interactions are much weaker than exchange ones. For example,
for parameters of RbEuFe$_{4}$As$_{4}$, $d\!=\!1.33$nm, $n_{M}\!=\!5\cdot10^{21}$cm$^{-3}$, $m_{0}\!=\!7\mu_{B}$,
and $\lambda\!=\!100$nm, we estimate $E_{d0}\!\approx \!10^{-4}$ K, while
the typical magnitude of $J_{z,\ell}$ is $0.1$--$0.2$ K. In the
following, we neglect the electromagnetic energy contribution.

We proceed with evaluation of the ground-state modulation wave vector $Q$. Equation \eqref{eq:HelEn-Q} determines the minimum-energy condition at $q\!=\!Q$,
\begin{equation}
\sum_{\ell}\ell J_{z,\ell}\sin(Q\ell)=0.\label{eq:HelMinEnCond}
\end{equation}
If we keep only three nearest neighbors, this equation becomes 
\begin{equation}
J_{z,1}+4J_{z,2}\cos Q+3J_{z,3}\left(4\cos^{2}Q-1\right)  =0.\label{eq:HelMinCond3}
\end{equation}
The energy has a minimum at $q\!=\!Q$ if
\begin{equation}
E_{0}^{\prime\prime}(Q)  =\sum_{\ell=1}^{\infty}\ell^{2}J_{z,\ell}\cos(Q\ell)>0.\label{eq:HelMinCond}
\end{equation}
The last two equations determine the optimal modulation vector in the case of frustrating interlayer exchange interactions.

\subsection{Case $Q=\pi/2$}
\label{Sec-GrStQhpi}

In the following, we will pay special attention to the case of commensurate
modulation with $Q\!=\!\pi/2$ realized in RbEuFe$_{4}$As$_{4}$.
In this case, assuming $J_{z,\ell}=0$ for $\ell>3$, the relation
in Eq.~\eqref{eq:HelMinCond3} gives $J_{z,1}=3J_{z,3}$ and 
\begin{equation}
J_{z}(q)=2J_{z,1}\left[\cos(q)+\tfrac{1}{3}\cos(3q)\right]+2J_{z,2}\cos(2q).\label{eq:Jzq-3neib}
\end{equation}
The condition for the minimum, Eq.~\eqref{eq:HelMinCond}, simply gives
$J_{z,2}<0$, i.e., the antiferromagnetic next-neighbor interaction. The
case $Q\!=\!\pi/2$, however, is special, because within the simplest
exchange model, the energy is degenerate with respect to the relative
rotation of the two sublattices composed of odd and even layers. Adding
interactions with more remote layers does not resolve this issue.
The continuous degeneracy is eliminated by the 4-fold crystal anisotropy
term, $-K_{4}(s_{x,\boldsymbol{i},n}^{4}+s_{y,\boldsymbol{i},n}^{4})$.
In addition, such anisotropy locks the $Q\!=\!\pi/2$ state within a
finite range of the interlayer exchange constants around the relation
$J_{z,1}=3J_{z,3}$. Such 4-fold anisotropy, however, does not completely
eliminate the ground-state degeneracy, because the helical state,
$\phi_{n}^{(0)}=\pi n/2$, still has the same energy as the double-periodic
state with $\phi_{n}^{(0)}=(0,0,\pi,\pi,0,0,\ldots)$. The simplest
term eliminating the latter degeneracy is the nearest-neighbor biquadratic
term $J_{z,b}\left(\boldsymbol{s}_{\boldsymbol{i},n}\boldsymbol{s}_{\boldsymbol{i},n+1}\right)^{2}$
with $J_{z,b}>0$. Without the 4-fold anisotropy term, this yields
the modified energy
\begin{align*}
E_{0}(q) & =-\sum_{\ell=1}^{\infty}J_{z,\ell}\cos(q\ell)+J_{z,b}\cos^{2}(q)
\end{align*}
and the modified ground-state condition 
\[
\sum_{\ell=1}^{\infty}\ell J_{z,\ell}\sin(Q\ell)-2J_{z,b}\cos(Q)\sin(Q)=0.
\]
For three nearest neighbors this gives 
\begin{align*}
J_{z,1}\!+\!4J_{z,2}\cos Q\!+\!3J_{z,3}\left(4\cos^{2}Q\!-\!1\right)\!-\!2J_{z,b}\cos Q & =0.
\end{align*}
For $Q\!=\!\pi/2$ the condition $J_{z,1}\!=\!3J_{z,3}$ remains unchanged,
while the condition for minimum becomes $2J_{z,2}\!-J_{z,b}\!<0$.
In the following analysis, we will assume the hierarchy of the energy
constants $J_{z,b},K_{4}\!\ll J_{z,\ell}\!<K\ll J$. In this case,
the degeneracy-breaking terms $\propto J_{z,b},K_{4}$ select the
$Q\!=\!\pi/2$ state but have only a minor impact on the properties
discussed in this paper.

\section{Bare spin-wave spectrum\label{sec:BareSpectrum}}

\begin{figure}
\includegraphics[width=2.8in]{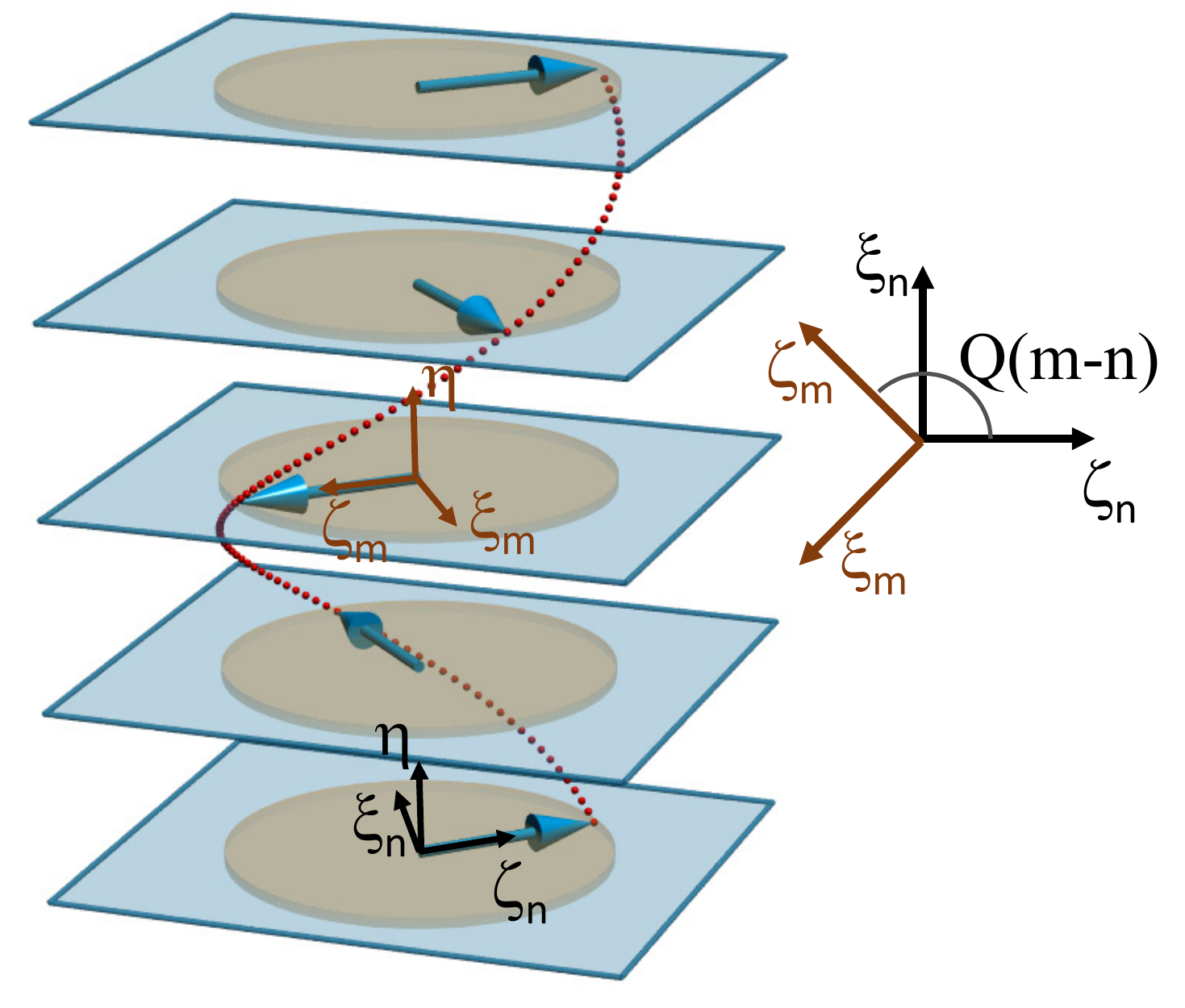}
\caption{\label{Fig:helical}Local coordinate system $(\varsigma$, $\xi$,
$\eta$) used for computation of the spin-wave spectrum. }
\end{figure}

\subsection{Arbitrary modulation wave vector}

In this section, we investigate a bare spectrum of spin waves due to
the local exchange interactions neglecting coupling to macroscopic
fields. We consider spin waves propagating along the direction of
helical modulation ($z$ axis) assuming that spin oscillations are
uniform in the layer direction. In the following derivations, we drop
the in-plane index $\boldsymbol{i}$, $\boldsymbol{S}_{\boldsymbol{i},n}\rightarrow\boldsymbol{S}_{n}$.
A useful trick allowing for analytical solution is to introduce a
local coordinate system $\varsigma$, $\xi$, $\eta$ following local
equilibrium spin orientation \cite{Nagamiya1968}. We assume that
the $\varsigma$ axis coincides with the equilibrium spin direction
at each layer, the $\xi$ axis is perpendicular to this direction
in the layer $xy$ plane, and the $\eta$ axis is parallel to the
$z$ axis, as illustrated in Fig.~\ref{Fig:helical}. Then the $\varsigma$,
$\xi$ axes at the layer $m$ are rotated with respect to those
at the layer $n$ by an angle of $Q(m-n)$ corresponding to the coordinate
transformation
\begin{align*}
\zeta_{m} & =\zeta_{n}\cos\left[Q(m-n)\right]+\xi_{n}\sin\left[Q(m-n)\right],\\
\xi_{m} & =-\zeta_{n}\sin\left[Q(m-n)\right]+\xi_{n}\cos\left[Q(m-n)\right].
\end{align*}
To fix the global coordinate system, we assume $(x,y)=(\zeta_{0},\xi_{0})$
meaning that 
\begin{align*}
\zeta_{n} & =x\cos\left(Qn\right)+y\sin\left(Qn\right),\\
\xi_{n} & =-x\sin\left(Qn\right)+y\cos\left(Qn\right).
\end{align*}
Correspondingly, the spin components in the rotated and global coordinates
are related as
\begin{subequations}
\begin{align}
S_{\zeta n} & =S_{xn}\cos\left(Qn\right)+S_{yn}\sin\left(Qn\right),\label{eq:SpinTrans-zeta}\\
S_{\xi n} & =-S_{xn}\sin\left(Qn\right)+S_{yn}\cos\left(Qn\right).\label{eq:SpinTrans-xi}
\end{align}
\end{subequations}
This and inverse transformations can also be presented
in the complex form 
\begin{subequations}
\begin{align}
S_{\zeta n}+i\delta S_{\xi n} & =\left(S_{xn}+i\delta S_{xn}\right)\exp\left(-i\delta Qn\right),\label{eq:SpinCompCompl}\\
S_{xn}+i\delta S_{yn} & =\left(S_{\zeta n}+i\delta S_{\xi n}\right)\exp\left(i\delta Qn\right)\label{eq:SpinxyComplex}
\end{align}
\end{subequations}
with $\delta=\pm1$.

The Landau-Lifshitz equation for spin dynamics in the rotated basis
can be written as
\begin{subequations}
\begin{align}
\dot{S}_{\xi n} & =S_{\eta n}h_{\zeta n}-S_{\zeta n}h_{\eta n},\label{eq:SpinDyn-xi}\\
\dot{S}_{\eta n} & =S_{\zeta n}h_{\xi n}-S_{\xi n}h_{\zeta n},\label{eq:SpinDyn-eta}
\end{align}
\end{subequations}
where $\boldsymbol{h}_{n}=-\partial\mathcal{E}_{m}/\partial\boldsymbol{S}_{n}$
is the local reduced magnetic field acting on spins in the layer $n$,
which, according to Eq.~\eqref{eq:EnerLay}, has the components
\begin{subequations}
\begin{align}
h_{\zeta n} = &\sum_{m} \mathcal{J}_{z,n-m}  \big\{  S_{\zeta m}\cos\left[Q(n\!-\!m)\right]\nonumber\\
& - S_{\xi m}\sin\left[Q(n\!-\!m)\right]\big\} ,\\
h_{\xi n}   = &\sum_{m} \mathcal{J}_{z,n-m}
  \big\{ S_{\zeta m}\sin\left[Q(n\!-\!m)\right]\nonumber\\
& + S_{\xi m}\cos\left[Q(n\!-\!m)\right]\big\} ,\\
h_{\eta n}  = &\sum_{m}\mathcal{J}_{z,n-m}S_{\eta m}-4\mathcal{K}S_{\eta n}.
\end{align}
\end{subequations}
For small spin oscillations, the local $\zeta$
component of each spin can be taken as a constant, $S_{\zeta m}\rightarrow S$.
Substituting $h_{jn}$ into Eqs.~\eqref{eq:SpinDyn-xi} and \eqref{eq:SpinDyn-eta},
we obtain equations for linear oscillations, $S_{un}(t)=S_{un}\exp(i\omega t)$
with $u=\xi,\eta$, 
\begin{subequations}
\begin{align}
i\omega S_{\xi n}\! & =\!S \sum_{m}\mathcal{J}_{z,n-m} 
\left\{ \cos\left[Q(n-m)\right]S_{\eta n}\! -\! S_{\eta m}\right\}\nonumber \\
 & +4S\mathcal{K}S_{\eta n},\label{eq:LinSp-xi}\\
i\omega S_{\eta n}\! & =\!S\sum_{m}\mathcal{J}_{z,n-m}\cos\left[Q(n\!-\!m)\right]
\left(S_{\xi m}\!-\!S_{\xi n}\right).
\label{eq:LinSp-eta}
\end{align}
\end{subequations}
We can see that, in spite of the helical structure,
in the rotating coordinates this system is uniform. Fourier transformation
$S_{u\mathfrak{q}}=\sum_{n}S_{un}\exp\left(-i\mathfrak{q}n\right)$
yields the $2\times2$ linear system 
\begin{subequations}\label{eqs:SpinDynFourier}
\begin{align}
i\omega S_{\xi\mathfrak{q}} & =S\left[\mathcal{J}_{z}(Q)-\mathcal{J}_{z}(\mathfrak{q})+4\mathcal{K}\right]S_{\eta\mathfrak{q}},\label{eq:SpinDynFourier-xi}\\
i\omega S_{\eta\mathfrak{q}} & =S\left[\frac{\mathcal{J}_{z}(Q\!+\!\mathfrak{q})\!+\!\mathcal{J}_{z}(Q\!-\!\mathfrak{q})}{2}-\!\mathcal{J}_{z}(Q)\right]S_{\xi\mathfrak{q}},\label{eq:SpinDynFourier-eta}
\end{align}
\end{subequations}
from which we obtain the spectrum
\begin{widetext}
\begin{equation}
\omega_{s}\left(\mathfrak{q}\right)\!=\!S\sqrt{\left[4\mathcal{K}\!+\!\mathcal{J}_{z}(Q)\!-\!\mathcal{J}_{z}(\mathfrak{q})\right]
	\left[\!\mathcal{J}_{z}(Q)\!-\!\frac{\mathcal{J}_{z}(Q\!+\!\mathfrak{q})\!+\!\mathcal{J}_{z}(Q\!-\!\mathfrak{q})}{2}\right]}
\label{eq:Spectrum}
\end{equation}
\end{widetext}
in terms of the reduced wave vector $\mathfrak{q}$. Since $Q$ is the ground-state modulation wave vector, $\mathcal{J}_{z}(q)$ has maximum at $q\!=\!Q$, as discussed in Sec.~\ref{sec:MagnGroundState}. This property influences the
spectrum shape near $\mathfrak{q}\!=\!0$ and $Q$. Spin oscillations
in the propagating wave have both in-plane and out-of-plane components.
Substituting $\omega_{s}(\mathfrak{q})$ into Eq.~\eqref{eq:SpinDynFourier-eta},
we derive the relation between the spin components in the mode 
\begin{equation}
S_{\eta\mathfrak{q}}=i\sqrt{\frac{\mathcal{J}_{z}(Q)-\frac{\mathcal{J}_{z}(Q+\mathfrak{q})+\mathcal{J}_{z}(Q-\mathfrak{q})}{2}}{4\mathcal{K}+\mathcal{J}_{z}(Q)-\mathcal{J}_{z}(\mathfrak{q})}}S_{\xi\mathfrak{q}}.\label{eq:ModeShape}
\end{equation}
From Eq.~\eqref{eq:SpinxyComplex}, we obtain the in-plane oscillating
spin components in real space
\begin{equation}
\left(\begin{array}{c}
S_{xn}\\
S_{yn}
\end{array}\right)=S_{\xi\mathfrak{q}}\exp\left(i\mathfrak{q}n\right)
\left(\begin{array}{c}
-\sin\left(Qn\right)\\
\cos\left(Qn\right)
\end{array}\right).
\label{eqs:SpinComp-real-space}
\end{equation}
We should emphasize that, as $\mathfrak{q}$ represents the wave vector
in the rotating-coordinates basis, the real-space spin components
$S_{x,y}$ do not behave as $\exp(i\mathfrak{q}n)$. In particular,
the mode with $\mathfrak{q}\!=\!0$ corresponding to the uniform helix
rotation does not generate spin variations uniform in
real space. 

The mode with $\mathfrak{q}\!=\!Q$ will play a key role in the following consideration.
For this mode, the in-plane spin components
\begin{equation}
\left(\begin{array}{c}
	S_{xn}\\
	S_{yn}
\end{array}\right)=\frac{S_{\xi Q}}{2}\left(\begin{array}{c}
	-i+i\exp\left(2iQn\right)\\
	1+\exp\left(2iQn\right)
\end{array}\right).
\label{eq:SxynQ}
\end{equation}
are a superposition of the uniform and $2Q$ terms. The presence of
the uniform $n$-independent component in the $\mathfrak{q}\!=\!Q$ mode implies
that it can be excited by the oscillating uniform field. The frequency
of the mode for $\mathfrak{q}\!=\!Q$ is given by the geometrical average of
the easy-plane anisotropy and combination of the interlayer exchange
constants,
\begin{equation}
\omega_{s}\left(Q\right)\!=\!S\sqrt{2\mathcal{K}\left[2\mathcal{J}_{z}(Q)\!-\!\mathcal{J}_{z}(2Q)\!-\!\mathcal{J}_{z}(0)\right]}.
\label{eq:omQ}
\end{equation}
From Eq.~\eqref{eq:ModeShape}, we also obtain the z-axis component
of this mode
\begin{equation}
S_{\eta Q}=i\sqrt{\frac{2\mathcal{J}_{z}(Q)\!-\!\mathcal{J}_{z}(2Q)\!-\!\mathcal{J}_{z}(0)}{8\mathcal{K}}}S_{\xi Q}.\label{eq:ModeShapeQ}
\end{equation}
We see that it decreases with increase of the easy-plane anisotropy. 

\subsection{Case $Q=\pi/2$ }

In the case $Q\!=\!\pi/2$ and the three-neighbor model discussed in Sec.~\ref{Sec-GrStQhpi}, using Eq.~\eqref{eq:Jzq-3neib}
and the relation $J_{z}(q)=\mathcal{J}_{z}(q)S^{2}$, we obtain the
spectrum
\begin{widetext}
\begin{equation}
\omega_{s}\left(\mathfrak{q}\right)\!=\!2S\sqrt{\left\{ 2\mathcal{K}\!-\!\mathcal{J}_{z,1}\left[\cos(\mathfrak{q})\!+\!\tfrac{1}{3}\cos(3\mathfrak{q})\right]\!+\!|\mathcal{J}_{z,2}|\left[1\!+\!\cos\left(2\mathfrak{q}\right)\right]\right\} |\mathcal{J}_{z,2}|\left[1\!-\!\cos\left(2\mathfrak{q}\right)\right]}.
\label{eq:SpectrQhpi}
\end{equation}
\end{widetext}
This frequency vanishes at $\mathfrak{q}\!=\!0$ and
$\pi$. The $\mathfrak{q}\!=\!0$ mode corresponds to uniform helix
rotation. Zero frequency at $\mathfrak{q}\!=\!\pi$ is a consequence
of the degeneracy with respect to the relative rotation of two sublattices,
which is the property of the exchange model in Eq.~\eqref{eq:EnerLay}
for $Q\!=\!\pi/2$. These degeneracies are eliminated by the additional
terms considered at the end of Sec.~\ref{sec:MagnGroundState}:
the in-plane 4-fold anisotropy and the nearest-neighbor biquadratic
term. The former term generates spin-wave gaps at both $\mathfrak{q}\!=\!0$
and $\pi$, while the latter term only generates a gap at $\mathfrak{q}\!=\!\pi$.
We assume that both these terms are small. We mostly focus on the
mode with $\mathfrak{q}\!=\!\pi/2$ which couples with uniform field. This mode
is weakly influenced by the additional degeneracy-eliminating terms
and we will neglect them below.

Consider the behavior near $\mathfrak{q}\!=\!\pi/2$. Expansion
of the frequency in Eq.\ \eqref{eq:SpectrQhpi} near this wave vector yields
\begin{equation}
\omega_{s}\!\left(\frac{\pi}{2}\!+\!q\right)\!\approx\!4S\sqrt{\mathcal{K}|\mathcal{J}_{z,2}|}\left[1\!-\!\left(1\!-\!\frac{|\mathcal{J}_{z,2}|}{\mathcal{K}}\right)\frac{q^{2}}{2}\!+\!\frac{\mathcal{J}_{z,1}}{3\mathcal{K}}q^{3}\right].
\label{eq:SpectrQhpi-qhpi}
\end{equation}
We see that the frequency has a maximum at $\mathfrak{q}\!=\!\pi/2$
for $\mathcal{K}>|\mathcal{J}_{z,2}|$. Moreover, one can check that
in this case $\omega_{s}\left(\pi/2\right)$ is the largest frequency
in the spectrum. We will focus on this case because it is likely realized in RbEuFe$_{4}$As$_{4}$.

\subsubsection{Transformation to magnetic unit cell and folded Brillouin zone\label{subsec:FoldedBZ}}

For the modulation vector $\pi/2$, the magnetic unit cell contains
four layers. Correspondingly, the folded magnetic Brillouin zone is
four times smaller than the crystalline Brillouin zone. It is therefore
useful to present the spin-wave spectrum in the folded Brillouin zone, 
which better corresponds to a standard crystallographic description. 
Introducing the index $j$ numbering magnetic unit cells, we present the
layer index as $n=4j+\nu$ with $\nu=1,2,3,4$. Correspondingly, the
spins can be represented as
\[
\boldsymbol{S}_{j,\nu}=\boldsymbol{A}_{\nu}(k)\exp\left(ikj\right),
\]
where $k$ is the wave vector within the folded Brillouin zone. Using
the presentation in Eq.~\eqref{eq:SpinxyComplex} for $S_{\xi n}=S_{0}\exp\left(i\mathfrak{q}n\right)$,
we write
\begin{align*}
S_{x,j,\nu}+i\delta S_{y,j,\nu} & =i\delta S_{0}\exp\left[i\left(\mathfrak{q}+\delta\frac{\pi}{2}\right)\left(4j+\nu\right)\right],
\end{align*}
meaning that $k=4\left(\frac{\pi}{2}m+\mathfrak{q}\right)$ and
\[
A_{x,\nu}^{(m)}(k)+i\delta A_{y,\nu}^{(m)}(k)=i\delta S_{0}\exp\left[i\left(\frac{k}{4}+m\pi+\delta\frac{\pi}{2}\right)\nu\right].
\]
The integer $m$ should be selected to reduce $k$ to the range $[-\pi,\pi]$.
This means that the four modes within such folded Brillouin zone correspond
to the frequencies \begin{subequations}
\begin{align}
\omega_{1}\left(k\right) & \!=\!\omega_{s}(k/4),\label{eq:w1}\\
\omega_{2}\left(k\right)\! & =\!\omega_{s}(\pi/2+k/4),\label{eq:w2}\\
\omega_{3}\left(k\right)\! & =\!\omega_{s}(-\pi/2+k/4),\label{eq:w3}\\
\omega_{4}\left(k\right)\! & =\!\omega_{s}(\pi+k/4),\label{eq:w4}
\end{align}
\end{subequations}where $\omega_{s}\left(\mathfrak{q}\right)$ is
the spectrum for vector $\mathfrak{q}$ within the original crystalline
Brillouin zone, Eq.~\eqref{eq:SpectrQhpi}. Note that while $\omega_{1}$
and $\omega_{4}$ are symmetric with respect to $k=0$, $\omega_{2}$
and $\omega_{3}$ do not have this symmetry and are related as $\omega_{2}\left(-k\right)=\omega_{3}\left(k\right)$.
In addition, the boundary values of the frequencies are connected
as $\omega_{1}\left(\pm\pi\right)\!=\!\omega_{2}\left(-\pi\right)\!=\!\omega_{3}\left(\pi\right)$
and $\omega_{4}\left(\pm\pi\right)\!=\!\omega_{2}\left(\pi\right)\!=\!\omega_{3}\left(-\pi\right)$.

At the center of the folded Brillouin zone, $k\!=\!0$, the first
mode corresponds to the uniform spin rotations, $A_{x,\nu}^{(0)}(0)=-S_{0}\sin(\pi\nu/2),A_{y,\nu}^{(0)}(0)=S_{0}\cos(\pi\nu/2)$.
Its frequency vanishes in the absence of the four-fold anisotropy term.
The second and third modes at $k\!=\!0$ correspond to the modes at
$\mathfrak{q}\!=\!\pm \pi/2$ coupled to the uniform field, Eq.~\eqref{eq:omQ},
$\omega_{2}(0)\!=\!\omega_{3}(0)\!=\!\omega_{s}(\pi/2)$.
The corresponding mode amplitudes are $A_{x,\nu}^{(\pm1)}(0)=\pm i\frac{\left(-1\right)^{\nu}-1}{2}S_{0},A_{y,\nu}^{(\pm1)}(0)=\frac{\left(-1\right)^{\nu}+1}{2}S_{0}$.
The fourth mode at $k\!=\!0$ corresponds to mutual rotation of odd
and even sublattices, $A_{x,\nu}^{(2)}(0)=S_{0}\sin(\pi\nu/2),A_{y,\nu}^{(2)}(0)=S_{0}\cos(\pi\nu/2)$,
and its frequency also vanishes without degeneracy-breaking terms. 

\section{Response to nonuniform oscillating magnetic field and dynamic susceptibility\label{sec:Response}}

\subsection{Arbitrary modulation wave vector}

In this section, we consider the response to the external oscillating
nonuniform magnetic field $\tilde{\boldsymbol{h}}_{n}=g\mu_{B}\tilde{\boldsymbol{H}}_{n}$.
The real-space components $(\tilde{h}_{xn},\tilde{h}_{yn},\tilde{h}_{zn})$
correspond to rotating-coordinates components $(\tilde{h}_{\xi n},\tilde{h}_{\zeta n},\tilde{h}_{\eta n})$
with \begin{subequations}
\begin{align}
\tilde{h}_{\zeta n} & =\tilde{h}_{xn}\cos\left(Qn\right)+\tilde{h}_{yn}\sin\left(Qn\right),\label{eq:th-zeta}\\
\tilde{h}_{\xi n} & =-\tilde{h}_{xn}\sin\left(Qn\right)+\tilde{h}_{yn}\cos\left(Qn\right),\label{eq:th-xi}
\end{align}
\end{subequations}
and $\tilde{h}_{\eta n}=\tilde{h}_{zn}$. In the
presence of such external field, equations for the linear spin oscillations,
Eqs.~\eqref{eqs:SpinDynFourier}, become
\begin{subequations}
\begin{align}
 & i\omega S_{\xi\mathfrak{q}}-S\left[\mathcal{J}_{z}(Q)-\mathcal{J}_{z}(\mathfrak{q})+4\mathcal{K}\right]S_{\eta\mathfrak{q}}=-S\tilde{h}_{\eta\mathfrak{q}},\label{eq:SpinDynRespFourier-xi}\\
 & i\omega S_{\eta\mathfrak{q}}\!+\!S\left[\mathcal{J}_{z}(Q)-
 \frac{\mathcal{J}_{z}(Q\!+\!\mathfrak{q})\!+\!\mathcal{J}_{z}(Q\!-\!\mathfrak{q})}{2}
 \right]S_{\xi\mathfrak{q}}\!=\!S\tilde{h}_{\xi\mathfrak{q}},\label{eq:SpinDynRespFourier}
\end{align}
\end{subequations}
where $\tilde{h}_{\alpha \mathfrak{q}}$ is the Fourier transform
of $\tilde{h}_{\alpha n}$. The solution of these equations can be presented
as\begin{subequations}
\begin{align}
S_{\xi\mathfrak{q}} & =\chi_{\xi\xi}^{S}\left(\mathfrak{q},\omega\right)\tilde{h}_{\xi\mathfrak{q}}+\chi_{\xi\eta}^{S}\left(\mathfrak{q},\omega\right)\tilde{h}_{\eta\mathfrak{q}},
\label{eq:Sxi-Response}\\
S_{\eta\mathfrak{q}} & =\chi_{\eta\xi}^{S}\left(\mathfrak{q},\omega\right)\tilde{h}_{\xi\mathfrak{q}}+\chi_{\eta\eta}^{S}\left(\mathfrak{q},\omega\right)\tilde{h}_{\eta\mathfrak{q}},
\label{eq:Seta-Response}
\end{align}
\end{subequations}
where we defined the susceptibility components in the rotating-coordinates basis,
\begin{subequations}\label{eq:chi}
\begin{align}
\chi_{\xi\xi}^{S}\left(\mathfrak{q},\omega\right) & =\!-\frac{S^{2}\left[\mathcal{J}_{z}(Q)\!-\!\mathcal{J}_{z}(\mathfrak{q})\!+\!4\mathcal{K}\right]}{\omega^{2}\!-\!\omega_{s}^{2}(\mathfrak{q})},\label{eq:chi-xixi}\\
\chi_{\eta\eta}^{S}\left(\mathfrak{q},\omega\right) & =\!-\frac{S^{2}\left[\mathcal{J}_{z}(Q)\!-\frac{\mathcal{J}_{z}(Q\!+\!\mathfrak{q})+\mathcal{J}_{z}(Q\!-\!\mathfrak{q})}{2}\right]}{\omega^{2}\!-\!\omega_{s}^{2}(\mathfrak{q})},\label{eq:chi-etaeta}\\
\chi_{\xi\eta}^{S}\left(\mathfrak{q},\omega\right) & =\!\chi_{\eta\xi}^{S*}\left(\mathfrak{q},\omega\right)\!=\!\frac{i\omega S}{\omega^{2}\!-\!\omega_{s}^{2}(\mathfrak{q})},\label{eq:chi-xieta}
\end{align}
\end{subequations}and $\omega_{s}(\mathfrak{q})$ is given by Eq.~\eqref{eq:Spectrum}.

Equations \eqref{eq:chi} give the susceptibility components in the
helically-rotating coordinates. To study interactions with macroscopic
fields, however, we need the susceptibility in real space. As follows
from Eq.~\eqref{eqs:SpinComp-real-space}, the spin Fourier components
in real coordinates are given by 
\begin{align*}
S_{xq} & =-\frac{S_{\xi,q+Q}-S_{\xi,q-Q}}{2i},\\
S_{yq} & =\frac{S_{\xi,q+Q}+S_{\xi,q-Q}}{2}.
\end{align*}
We emphasize that here and below the wave vector $q$ corresponding to real space is distinguished from the 'fractur' wave vector $\mathfrak{q}$ in the rotated basis which we used above.
We use the result for $S_{\xi\mathfrak{q}}$ from Eq.~\eqref{eq:Sxi-Response},
in which we substitute the field Fourier components
\begin{align*}
\tilde{h}_{\xi q} & =\frac{1}{2}\sum_{\delta=\pm1}\left[i\delta\tilde{h}_{x,q\!+\!\delta Q}+\tilde{h}_{y,q\!+\!\delta Q}\right],
\end{align*}
following from Eq.~\eqref{eq:th-xi}. This yields the spin response
in real coordinates
\begin{subequations} 
\begin{align}
S_{xq} & =\!\chi_{xx}^{S}\left(q,\omega\right)\tilde{h}_{x,q}+\chi_{xy}^{S}\left(q,\omega\right)\tilde{h}_{yq}\nonumber \\
+ & \sum_{\delta=\pm1}\left\{ 
\frac{\chi_{\xi\xi}^{S}\left(q\!+\!\delta Q,\omega\right)}{4}
\left[\tilde{h}_{x,q\!+\!2\delta Q}\!+\!i\delta\tilde{h}_{y,q\!+\!2\delta Q}\right]\right.\nonumber \\
- & \left.\frac{\delta\chi_{\xi\eta}^{S}\left(q+\delta Q,\omega\right)}{2i}\tilde{h}_{z,q+\delta Q}\right\} ,\label{eq:Sx-resp}\\
S_{yq} & =\!\chi_{yx}^{S}\left(q,\omega\right)\tilde{h}_{x,q}+\chi_{yy}^{S}\left(q,\omega\right)\tilde{h}_{yq}\nonumber \\
+ & \sum_{\delta=\pm1}\left\{ 
\frac{\chi_{\xi\xi}^{S}\left(q\!+\!\delta Q,\omega\right)}{4}
\left[i\delta\tilde{h}_{x,q\!+\!2\delta Q}\!+\!\tilde{h}_{y,q\!+\!2\delta Q}\right]\right.\nonumber \\
+ & \left.\frac{\chi_{\xi\eta}^{S}(q+\delta Q,\omega)}{2}\tilde{h}_{z,q+\delta Q}\right\} \label{eq:Sy-resp}
\end{align}
\end{subequations}
with the real-space spin susceptibility components
\begin{subequations}
\begin{align}
\chi_{xx}^{S}\left(q,\omega\right) & \!=\chi_{yy}^{S}\left(q,\omega\right)\!=\!\frac{\chi_{\xi\xi}^{S}\left(Q\!+\!q,\omega\right)\!+\!\chi_{\xi\xi}^{S}\left(Q\!-\!q,\omega\right)}{4},\label{eq:chi-xx}\\
\chi_{xy}^{S}\left(q,\omega\right) & \!=\!-\chi_{yx}^{S}\left(q,\omega\right)\!=\!-\frac{\chi_{\xi\xi}^{S}\left(Q\!+\!q,\omega\right)\!-\!\chi_{\xi\xi}^{S}\left(Q\!-\!q,\omega\right)}{4i}.\label{eq:chi-xy}
\end{align}
\end{subequations}
As expected, in addition to the usual diagonal
response at the same wave vector, the helical magnetic structure also
generates nondiagonal susceptibility and responses at the wave vectors
shifted by the modulation wave vector $Q$. Note that the bulk magnetic
susceptibility $\chi_{\alpha\beta}\left(k_z,\omega\right)$ in Eq.~\eqref{eq:DynMagnResp}
is related to the dynamic spin susceptibility as \footnote{In our notations, $k_z$ and $q$ in Eq.\ \eqref{eq:chi-xx} are the dimensional and dimensionless $c$-axis wave vectors, respectively, with $q=dk_z$}
\begin{equation}\chi_{\alpha\beta}\left(k_z,\omega\right)=n_{M}(g\mu_{B})^{2}\chi_{\alpha\beta}^{S}\left(dk_z,\omega\right).
\label{eq:MagnSpinSusc}
\end{equation}

We will be mostly interested in the smooth spin response to smooth
field with the wave vectors much smaller than $Q$. In this case,
we can drop the short-wave length terms $\tilde{h}_{x}(q\!\pm\!mQ)$
with $m\neq0$, i.e., keep only the first lines in Eqs.~\eqref{eq:Sx-resp}
and \eqref{eq:Sy-resp}. In addition, to obtain the long-wave length
response, we use the small-$q$ expansion (recall that $\mathcal{J}_{z}^{\prime\prime}\left(Q\right)<0$),
\begin{equation}
\chi_{xx}^{S}\left(q,\omega\right)\approx\frac{1}{4}\sum_{\delta=\pm1}\frac{-S^{2}\left[\frac{\left|\mathcal{J}_{z}^{\prime\prime}\left(Q\right)\right|}{2}q^{2}+4\mathcal{K}\right]}{\omega^{2}\!-\!\omega_{s}^{2}(Q)+\delta a_{s}q\!+\!c_{s}q^{2}},\label{eq:SuscSmoothGenqQ}
\end{equation}
with $a_{s}\!=\!2S^{2}\mathcal{K}\mathcal{J}_{z}^{\prime}\left(2Q\right)$
and 
\begin{align*}
c_{s} & =S^{2}\Biggl\{\mathcal{K}\left[\mathcal{J}_{z}^{\prime\prime}\left(0\right)+\!\mathcal{J}_{z}^{\prime\prime}\left(2Q\right)\right]\!.\\
 & -\frac{\left|\mathcal{J}_{z}^{\prime\prime}\left(Q\right)\right|}{2}\left[\mathcal{J}_{z}(Q)\!-\!\frac{\mathcal{J}_{z}(0)\!+\!\mathcal{J}_{z}(2Q)}{2}\right]\Biggr\}.
\end{align*}
Note that for incommensurate-modulation wave vector $Q$, the linear
coefficient $a_{s}$ is finite meaning that the frequency $\omega_{s}(q)$
does not have an extremum at $q\!=\!Q$. The important particular
cases of this result include the response to the uniform oscillating
field
\begin{equation}
\chi_{xx}^{S}\left(0,\omega\right)=\!-\frac{2S^{2}\mathcal{K}}{\omega^{2}\!-\!2S^{2}\mathcal{K}\left[2\mathcal{J}_{z}(Q)\!-\!\mathcal{J}_{z}(0)\!-\!\mathcal{J}_{z}(2Q)\right]},
\label{eq:DynUnifomSusc}
\end{equation}
which features the antiferromagnetic resonance at the uniform mode frequency, $\omega\!=\!\omega_{s}(Q)$, Eq.\ \eqref{eq:omQ}, and the static uniform susceptibility
\begin{equation}
\chi_{xx}^{S}\left(0,0\right)  \approx\!\frac{1}{2\mathcal{J}_{z}(Q)\!-\!\mathcal{J}_{z}(0)\!-\!\mathcal{J}_{z}(2Q)},\label{eq:StatSusc}
\end{equation}
which only depends on the interlayer exchange constants. The long-range
off-diagonal component $\chi_{xy}^{S}\left(q,\omega\right)$ is given
by
\begin{equation}
\chi_{xy}^{S}\left(q,\omega\right)\!\approx\!-\frac{i}{2}\frac{S^{2}\left[\frac{\left|\mathcal{J}_{z}^{\prime\prime}\left(Q\right)\right|}{2}q^{2}+4\mathcal{K}\right]a_{s}q}{\left[\omega^{2}\!-\!\omega_{s}^{2}(Q)\!+\!c_{s}q^{2}\right]^{2}-a_{s}^{2}q^{2}}.\label{eq:SuscSmooth-xy}
\end{equation}
It vanishes for $q\rightarrow0$ proportionally to $q$,
\begin{equation}
\chi_{xy}^{S}\left(q,\omega\right)\!\approx\!\frac{-2iS^{2}\mathcal{K}a_{s}q}{\left\{ \omega^{2}\!-\!2S^{2}\mathcal{K}\left[2\mathcal{J}_{z}(Q)\!-\!\mathcal{J}_{z}(0)\!-\!\mathcal{J}_{z}(2Q)\right]\right\} ^{2}},\label{eq:Susc-q0-xy}
\end{equation}
meaning that the transverse spin response is proportional to the field
gradient $S_{x}\propto\partial\tilde{h}_y/\partial z$.

\subsection{Case $Q=\pi/2$\label{subsec:SpSuscQhPi}}

For the commensurate modulation with $Q\!=\!\pi/2$, the spin response,
Eq.~\eqref{eq:Sx-resp}, simplifies as 
\begin{align}
S_{xq} & =\!\chi_{xx}^{S}\left(q,\omega\right)\left[\tilde{h}_{x,q}\!+\!\tilde{h}_{x,\pi\!-\!q}\right]+\chi_{xy}^{S}\left(q,\omega\right)\left[\tilde{h}_{yq}\!+\!\tilde{h}_{y,\pi\!-\!q}\right]\nonumber \\
- & \sum_{\delta=\pm1}\frac{\delta}{2i}\chi_{\xi\eta}^{S}\left(q\!+\!\delta\frac{\pi}{2},\omega\right)\tilde{h}_{z,q+\delta\frac{\pi}{2}},\label{eq:SpRespxhPi}
\end{align}
where the diagonal susceptibility, Eq,~\eqref{eq:chi-xx}, explicitly
is given by 
\begin{widetext}
\begin{equation}
\chi_{xx}^{S}\left(q,\omega\right)\!=\!\frac{1}{4}\!\sum_{\delta=\pm1}\!\frac{-S^{2}\left[\mathcal{J}_{z}\left(\pi/2\right)-\mathcal{J}_{z}(\delta q\!+\!\pi/2)+4\mathcal{K}\right]}{\omega^{2}\!-\!S^{2}\left[\mathcal{J}_{z}\left(\frac{\pi}{2}\right)\!-\!\mathcal{J}_{z}\left(\delta q\!+\!\frac{\pi}{2}\right)\!+\!4\mathcal{K}\right]\left[\mathcal{J}_{z}\left(\frac{\pi}{2}\right)\!-\!\frac{\mathcal{J}_{z}(q)+\mathcal{J}_{z}(\pi-q)}{2}\right]},\label{eq:chixx-hpi}
\end{equation}
\end{widetext}
and we used the relation $\chi_{xx}^{S}\left(\pi\!-\!q\right)\!=\!\chi_{xx}^{S}\left(q\right)$.

In the small-$q$ expansion, Eq.~\eqref{eq:SuscSmoothGenqQ}, the
linear term in the denominator $\propto a_{s}$ vanishes, since $\mathcal{J}_{z}^{\prime}\left(2Q\right)\!\equiv\!\mathcal{J}_{z}^{\prime}\left(\pi\right)\!=\!0$,
and the quadratic-term coefficient becomes 
\begin{align}
c_{s} & =S^{2}\Biggl[\mathcal{K}\left(\mathcal{J}_{z}^{\prime\prime}\left(0\right)\!+\!\mathcal{J}_{z}^{\prime\prime}\left(\pi\right)\right)\!.\nonumber \\
- & \!\frac{\left|\mathcal{J}_{z}^{\prime\prime}\left(\pi/2\right)\right|}{2}\left(\mathcal{J}_{z}(\pi/2)\!-\!\frac{\mathcal{J}_{z}(0)\!+\!\mathcal{J}_{z}(\pi)}{2}\right)\Biggr].\label{eq:cs-hpiG}
\end{align}
For the three-neighbor model, Eq.~\eqref{eq:Jzq-3neib}, this coefficient
acquires a simple form,
\begin{equation}
c_{s}=16S^{2}|\mathcal{J}_{z,2}|\left(\mathcal{K}\!-\!|\mathcal{J}_{z,2}|\right).\label{eq:cs-hpi}
\end{equation}

The behavior of the off-diagonal component is very different from the
case of incommensurate modulation. It vanishes in the static case and
for finite frequency in the small-$q$ limit it behaves as 
\begin{align*}
\chi^{S}_{xy}\left(q,\omega\right) & \approx\frac{iS^{2}\mathcal{J}_{z}^{\prime\prime\prime}(\pi/2)\omega^{2}q^{3}}{12\left(\omega^{2}\!-\!\omega_{s}^{2}\left(\frac{\pi}{2}\right)\right)^{2}}\\
\approx & -\frac{4iS^{2}\mathcal{J}_{z,1}\omega^{2}q^{3}}{3\left(\omega^{2}\!-\!16S^{2}\mathcal{K}|\mathcal{J}_{z,2}|\right)^{2}},
\end{align*}
i.e., it vanishes $\propto q^{3}$ for $q\rightarrow0$. This behavior
allows us to neglect the off-diagonal component in the further phenomenological
considerations.

\section{Electromagnetic renormalization of spectrum in superconducting state\label{sec:EM-renor}}
\begin{figure*}
	\includegraphics[width=6in]{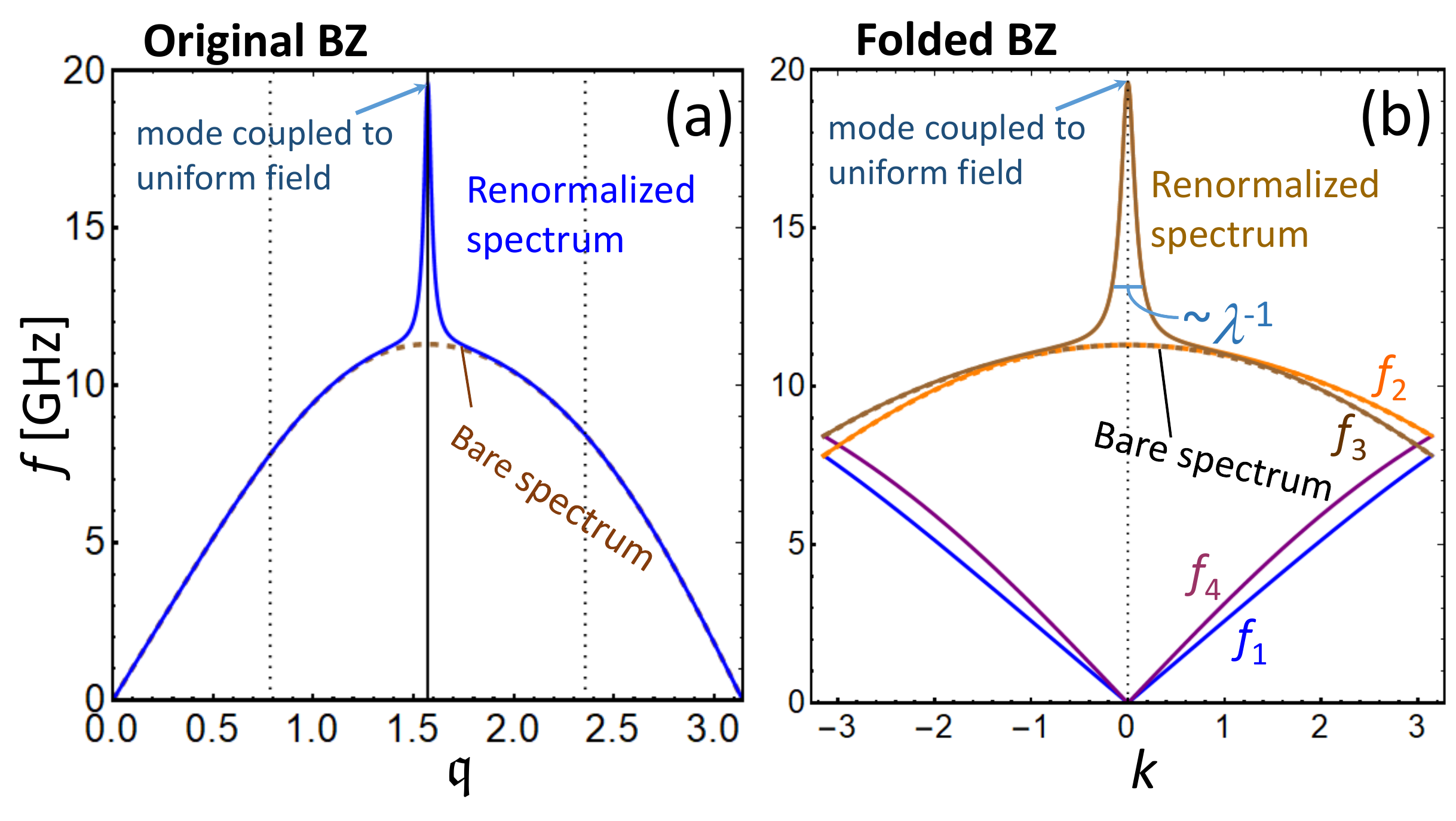}
	\caption{The representative spectrum of spin waves for the helical structure
		with $Q\!=\!\pi/2$ in (a) original and (b) folded Brillouin zones
		for the typical parameters of RbEuFe$_{4}$As$_{4}$. The dashed lines
		in both plots show the bare spectra obtained without taking into account
		the renormalization caused by the coupling to macroscopic magnetic
		field. }
	\label{Fig:Spectrum} 
\end{figure*}

\subsection{Arbitrary modulation vector}

In this section, we consider the renormalization of the spin-wave
spectrum in the superconducting state due to the long-range electromagnetic
interactions between the local moments using Eq.~\eqref{eq:SpWaveRenorm}
in terms of the reduced wave vector $q=dk_{z}$. Using Eq.~\eqref{eq:MagnSpinSusc}
connecting the spin and magnetic susceptibilities and the relation
\begin{align*}
	\chi_{xx}^{S}\left(q\right)\pm i\chi_{xy}^{S}\left(q\right) & =\frac{1}{2}\chi_{\xi\xi}^{S}\left(Q\!\mp\!q\right)\\
	=- & \frac{1}{2}\frac{S^{2}\left[\mathcal{J}_{z}(Q)\!-\!\mathcal{J}_{z}(Q\!\mp\!q)\!+\!4\mathcal{K}\right]}{\omega^{2}\!-\!\omega_{s}^{2}(Q\!\mp\!q)}
\end{align*}
following from Eqs\@.~\eqref{eq:chi-xx} and \eqref{eq:chi-xy},
we obtain the equation 
\begin{align}
	1\!+\!(\lambda/d)^{2}q^{2}\!-\!2\pi\frac{n_{M}m_{0}^{2}\left[\mathcal{J}_{z}(Q)\!-\!\mathcal{J}_{z}(Q\!\mp\!q)\!+\!4\mathcal{K}\right]}{\omega^{2}\!-\!\omega_{s}^{2}(Q\!\mp\!q)} & \!=\!0\label{eq:SpWRenormEq}
\end{align}
for the renormalized spin-wave spectrum, $\omega=\Omega_{s}(Q\!\mp\!q)$.
The solution of this equation is 
\begin{align}
	\Omega_{s}^{2}(Q\!+\!q)\! & =\!\omega_{s}^{2}(Q\!+\!q)\nonumber \\
	+ & \frac{2\pi n_{M}m_{0}^{2}\left[\mathcal{J}_{z}(Q)\!-\!\mathcal{J}_{z}(Q\!+\!q)\!+\!4\mathcal{K}\right]}{1+(\lambda/d)^{2}q^{2}}.\label{eq:FreqRenormIncomsGen}
\end{align}
The second term gives the spin-wave frequency enhancement due to the
long-range electromagnetic interactions. The maximum enhancement is
realized near $q=0$ corresponding to the uniform mode, Eqs.~\eqref{eq:SxynQ}
and \eqref{eq:omQ}. Using the presentation for the static magnetic
susceptibility $\chi_{xx}\left(0,0\right)\!=\!2n_{M}m_{0}^{2}\mathcal{K}/\omega_{s}^{2}(Q)$
following from Eqs.~\eqref{eq:MagnSpinSusc}, \eqref{eq:StatSusc},
and \eqref{eq:omQ}, we can rewrite Eq.~\eqref{eq:FreqRenormIncomsGen}
for $q=0$ as $\Omega_{s}^{2}(Q)\!=\![1\!+\!4\pi\chi_{xx}\left(0,0\right)]\omega_{s}^{2}(Q)$
meaning that the renormalized frequency of the uniform mode is
\begin{equation}
	\Omega_{s}(Q)\!=\!\sqrt{\mu_{x0}}\, \omega_{s}(Q),\label{eq:RenormUnifMode}
\end{equation}
where $\mu_{x0}\!=\!1\!+\!4\pi\chi_{xx}\left(0,0\right)$ is the static
magnetic permeability. Neglecting a weak $q$ dependence in the
nominator of the second term in Eq.~\eqref{eq:FreqRenormIncomsGen},
we can rewrite the frequency renormalization for $q\ll1$ in a somewhat
more transparent form as $\Omega_{s}^{2}(Q\!+\!q)\!\approx\!\omega_{s}^{2}(Q\!+\!q)+4\pi\chi_{xx}(0,0)\omega_{s}^{2}(Q)/[1\!+\!(\lambda/d)^{2}q^{2}]$.

\subsection{Case $Q=\pi/2$ }

The key features of the commensurate state with $Q=\pi/2$ are that
(i)the frequency $\omega_{s}\left(\mathfrak{q}\right)$ has maximum
at $\mathfrak{q}=Q$ (for $\mathcal{K}>|\mathcal{J}_{z,2}|$) and
(ii)the off-diagonal spin susceptibility vanishes as $q^{3}$ for $q\rightarrow0$
and therefore its contribution in Eq.~\eqref{eq:SpWaveRenorm} can
be neglected. Based on the results of Sec.\ \ref{subsec:SpSuscQhPi},
we can represent the dynamics magnetic susceptibility as 
\begin{equation}
	\chi_{xx}\left(k_{z},\omega\right)\approx-\frac{\chi_{xx}\left(0,0\right)\omega_{s}^{2}(Q)}{\omega^{2}\!-\!\omega_{s}^{2}(Q)\!+\!c_{s}d^{2}k_{z}^{2}}\label{eq:MagnSusc}
\end{equation}
with the static susceptibility $\chi_{xx}\left(0,0\right)\!=\!n_{M}(g\mu_{B})^{2}/(8|\mathcal{J}_{z,2}|)$
and $c_{s}$ is given by Eqs.~\eqref{eq:cs-hpiG} and \eqref{eq:cs-hpi}.
The key difference from the ferromagnetic state \cite{BraudePhysRevLett.93.117001,BraudePhysRevB.74.054515}
is the opposite sign of the quadratic coefficient, since in our case
the spin-wave frequency has a maximum at $k_{z}\!=\!0$ (corresponding
to $\mathfrak{q}=\pi/2$).

Solution of Eq.~\eqref{eq:SpWaveRenorm} gives the renormalized spectrum
in the vicinity of $\mathfrak{q}\!=\!Q$ in terms of the reduced wave
vector $q=dk_{z}$, 
\begin{equation}
	\Omega_{s}^{2}(Q\!+\!q)\!=\!\omega_{s}^{2}(Q)\left[1\!+\frac{4\pi\chi_{xx}\left(0,0\right)}{1+(\lambda/d)^{2}q^{2}}\right]\!-\!c_{s}q^{2}.\label{eq:RenormSpectrum}
\end{equation}
In particular, the renormalization of the uniform mode is again given
by Eq.~\eqref{eq:RenormUnifMode}. In the folded Brillouin zone discussed
in Sec.~\ref{subsec:FoldedBZ} this mode corresponds to second and
third modes at $k=0$, Eqs.~\eqref{eq:w2} and \eqref{eq:w3}.

Figure \ref{Fig:Spectrum} shows spectrum of spin waves in both the original
and folded Brillouin zone for the parameters typical for RbEuFe$_{4}$As$_{4}$.
Namely, we took $S\!=7/2$, $\mathcal{J}_{z,1}\!=0.05$K, $\mathcal{J}_{z,2}\!=\!-0.01$K,
$\mathcal{K}\!=0.15$K, $\lambda=70$nm, and $\mu_{x0}\!=3$. For
these parameters the bare maximum frequency is $\sim11$GHz. This frequency
is strongly enhanced in the superconducting state due to electromagnetic
renormalization. This renormalization rapidly decreases for $(\lambda/d)|q\!-\!\pi/2|,\,4(\lambda/d)k>1$.
We deliberately took a somewhat large value of $\mathcal{J}_{z,1}$
to enhance the difference between $f_{2}(k)$ and $f_{3}(k)$. For
a more realistic choice $\mathcal{J}_{z,1}\!\lesssim|\mathcal{J}_{z,2}|$
these frequencies become indistinguishable. 

\section{Dynamic equation for smooth magnetization, magnetic boundary condition,
and surface impedance\label{sec:DynEqSmooth}}

In this section we consider magnetization response to the alternating
magnetic field at the surface and derive the magnetic boundary condition.
Here and below, we limit ourselves to the commensurate case $Q\!=\!\pi/2$, for which
the frequency of the $z$-axis uniform mode is maximal. As follows
from the shape of the susceptibility, Eq.~\eqref{eq:MagnSusc}, a
phenomenological equation for the in-plane magnetization in the case
of uniform in-plane field is
\begin{equation}
\chi_{0}^{-1}\left(1+\omega_{0}^{-2}\frac{\partial^{2}}{\partial t^{2}}+\zeta_{0}^{2}\nabla_{z}^{2}\right)\boldsymbol{M}=\boldsymbol{H}\label{eq:PhenEqM-H}
\end{equation}
with $\chi_{0}=\chi_{xx}\left(0,0\right)$, $\omega_{0}^{2}=\omega_{s}^{2}(Q)$,
and $\zeta_{0}^{2}=c_{s}d^2/\omega_{s}^{2}(Q)$. This equation is only
valid for smoothly varying magnetization, i.~e.\ for $\zeta_{0}|\nabla_{z}\boldsymbol{M}|\ll1$.
On the other hand, the local magnetic field $\boldsymbol{H}$ is connected
with the magnetization as 
\begin{equation}
\left(1-\lambda^{2}\nabla_{z}^{2}\right)\boldsymbol{H}\approx-4\pi\boldsymbol{M}.\label{eq:H-M}
\end{equation}
The magnetic length scale $\zeta_{0}$ is much smaller than the London
penetration depth $\lambda$. We find the magnetization response to
the external oscillating magnetic field. This corresponds to the boundary
condition for $\boldsymbol{H}(z,t)$ at the surface, $z=0$, 
\begin{equation}
\boldsymbol{H}(0,t) =\boldsymbol{H}_{0}\exp\left(i\omega t\right),\label{eq:BoundCondH}
\end{equation}
This condition has to be supplemented by the boundary condition for
the magnetization, which is usually assumed as
\begin{equation}
\nabla_{z}\boldsymbol{M}(0,t)=0.\label{eq:BoundCondM}
\end{equation}

We look for the oscillating magnetization and field at the semispace
$z>0$ in the form
\begin{subequations} 
\begin{align}
\boldsymbol{M}(z,t) & =\sum_{\alpha}\boldsymbol{M}_{0\alpha}\exp\left(i\omega t-\kappa_{\alpha}(\omega)z\right),\label{eq:M-zt}\\
\boldsymbol{H}(z,t) & =\sum_{\alpha}\boldsymbol{H}_{0\alpha}\exp\left(i\omega t-\kappa_{\alpha}(\omega)z\right).\label{eq:H-zt}
\end{align}
\end{subequations} 
In the absence of internal dissipation mechanisms,
the parameters $\kappa_{\alpha}(\omega)$ may be either purely real
or purely imaginary. It is clear that in the former case $\kappa_{\alpha}(\omega)$
has to be positive. Care should taken to select the correct sign
for purely imaginary $\kappa_{\alpha}(\omega)$. Since for the spectrum
described by Eq.~\eqref{eq:PhenEqM-H} the group velocity has the
opposite sign with respect to the wave vector $q\!=\!\mathrm{Im}\left[\kappa_{\alpha}(\omega)\right]$,
the energy flows away from the surface for negative $\mathrm{Im}\left[\kappa_{\alpha}(\omega)\right]$.
Substituting the above distributions into Eqs.~\eqref{eq:PhenEqM-H}
and \eqref{eq:H-M}, we obtain equations connecting the vectors $\boldsymbol{M}_{0\alpha}$
and $\boldsymbol{H}_{0\alpha}$ 
\begin{subequations}\label{eqs:Decay}
\begin{align}
\left(1-\lambda^{2}\kappa_{\alpha}^{2}\right)\boldsymbol{H}_{0\alpha} & \approx-4\pi\boldsymbol{M}_{0\alpha},\label{eq:H0a}\\
\left(1-\omega^{2}/\omega_{0}^{2}+\zeta_{0}^{2}\kappa_{\alpha}^{2}\right)\boldsymbol{M}_{0\alpha} & =\chi_{0}\boldsymbol{H}_{0\alpha},\label{eq:M0a}
\end{align}
\end{subequations}
which give the quadratic equation for $\kappa_{\alpha}^{2}(\omega)$
\[
\left(1-\omega^{2}/\omega_{0}^{2}+\zeta_{0}^{2}\kappa_{\alpha}^{2}\right)\left(1-\lambda^{2}\kappa_{\alpha}^{2}\right)+\mu_{x0}-1=0.
\]
Solution of this equation is 
\begin{widetext}
\begin{align}
\kappa_{\alpha}^{2}=\frac{\lambda^{-2}\!+\!\left(\omega^{2}/\omega_{0}^{2}\!-\!1\right)\zeta_{0}^{-2}}{2}+ & \delta_{\alpha}\sqrt{\frac{\left[\lambda^{-2}\!+\!\left(\omega^{2}/\omega_{0}^{2}\!-\!1\right)\zeta_{0}^{-2}\right]^{2}}{4}\!+\!\zeta_{0}^{-2}\lambda^{-2}\left(\mu_{x0}\!-\!\omega^{2}/\omega_{0}^{2}\right)}\nonumber \\
=\frac{\lambda^{-2}\!+\!\left(\omega^{2}/\omega_{0}^{2}\!-\!1\right)\zeta_{0}^{-2}}{2}+ & \delta_{\alpha}\sqrt{\frac{\left[\lambda^{-2}\!-\!\left(\omega^{2}/\omega_{0}^{2}\!-\!1\right)\zeta_{0}^{-2}\right]^{2}}{4}\!+\!\zeta_{0}^{-2}\lambda^{-2}\left(\mu_{x0}\!-\!1\right)}.\label{eq:kapapm}
\end{align}
\end{widetext}
We select $\delta_{1}\!=\!\mathrm{sign}\left[\lambda^{-2}\!+\!\left(\omega^{2}/\omega_{0}^{2}\!-\!1\right)\zeta_{0}^{-2}\right]$
and $\delta_{2}\!=\!-\delta_{1}$. Such choice implies that $\left|\kappa_{1}(\omega)\right|>\left|\kappa_{2}(\omega)\right|$
in the whole frequency range. Note that this solution is only formally
valid in the frequency range where $\zeta_{0}\left|\kappa_{1}(\omega)\right|\!\ll1$
corresponding to the validity range of Eq.~\eqref{eq:PhenEqM-H}. In
particular, the result for $\kappa_{1}(\omega)$ is not valid for
the static case at $\omega\!=\!0$. 

Consider important special cases of Eq.~\eqref{eq:kapapm}. At the
bare uniform frequency, $\omega\!=\!\omega_{0}$, we obtain 
\begin{align}
\kappa_{\alpha}^{2}(\omega_{0})\! & =\!\frac{\lambda^{-2}}{2}\!\pm\!\frac{\lambda^{-1}}{2}\sqrt{\lambda^{-2}\!+\!4\zeta_{0}^{-2}\left(\mu_{x0}\!-\!1\right)}\nonumber \\
 & \approx\pm\zeta_{0}^{-1}\lambda^{-1}\sqrt{\mu_{x0}-1},\label{eq:kappa-a-omega0}
\end{align}
 while at renormalized frequency $\omega\!=\!\sqrt{\mu_{x0}}\omega_{0}$,
we have
\begin{align*}
\kappa_{1}^{2}(\sqrt{\mu_{x0}}\omega_{0}) & =\lambda^{-2}+\left(\mu_{x0}-1\right)\zeta_{0}^{-2},\\
\kappa_{2}^{2}(\sqrt{\mu_{x0}}\omega_{0}) & =0.
\end{align*}
However, in the latter case the value of $\kappa_{1}$ is already
beyond the applicability range of Eq.~\eqref{eq:PhenEqM-H}. Since
$\zeta_{0}\ll\lambda$, the inequality $\lambda^{-1}\ll\left|\omega^{2}/\omega_{0}^{2}\!-\!1\right|\zeta_{0}^{-1}$
is satisfied almost everywhere, except a narrow region where the frequency
is very close to $\omega_{0}$. Away from this region, we can expand
$\kappa_{\alpha}^{2}(\omega)$ with respect to $\left(\omega^{2}/\omega_{0}^{2}\!-\!1\right)^{-1}\zeta_{0}/\lambda$,
which yields
\begin{subequations}
\begin{align}
\kappa_{1}^{2} & \approx\left(\omega^{2}/\omega_{0}^{2}\!-\!1\right)\zeta_{0}^{-2}\!+\frac{\left(\mu_{x0}\!-1\right)\lambda^{-2}}{\omega^{2}/\omega_{0}^{2}-1},\label{eq:MagDecay}\\
\kappa_{2}^{2} & \approx\lambda^{-2}\frac{\omega^{2}/\omega_{0}^{2}-\mu_{x0}}{\omega^{2}/\omega_{0}^{2}-1}\label{eq:SupDecay}
\end{align}
\end{subequations}
meaning that the parameters $\kappa_{1}$ and $\kappa_{2}$
mostly describe magnetic and superconducting decay, respectively.
The approximation is valid until the second term in $\kappa_{1}^{2}$
is small with respect to the first one giving a somewhat more accurate
condition for the expansion $\left|\omega^{2}/\omega_{0}^{2}\!-\!1\right|\gg\sqrt{\mu_{x0}\!-\!1}\zeta_{0}/\lambda$.
In addition, the condition $\zeta_{0}\left|\kappa_{1}\right|\!\ll1$
implies that the result for $\kappa_{1}$ is only valid for $\left|\omega^{2}/\omega_{0}^{2}\!-\!1\right|\ll1$.
However, the result for $\kappa_{2}^{2}$ in Eq.~\eqref{eq:SupDecay}
corresponds to the approximation of local magnetic response, $\zeta_{0}\rightarrow0$,
and it remains valid even when the condition $\zeta_{0}\left|\kappa_{1}\right|\ll1$
breaks, e.g.,~in the limit $\omega\rightarrow0$. In the immediate
vicinity of the frequency $\omega_{0}$, in the range $\left|\omega^{2}/\omega_{0}^{2}\!-\!1\right|\ll\sqrt{\mu_{x0}\!-\!1}\zeta_{0}/\lambda$,
the parameters $\kappa_{\alpha}^{2}$ can be evaluated as
\begin{align*}
\kappa_{\alpha}^{2}\! & \approx\!\pm\zeta_{0}^{-1}\lambda^{-1}\sqrt{\mu_{x0}\!-\!1}\!+\frac{\lambda^{-2}\!+\!\left(\omega^{2}/\omega_{0}^{2}\!-\!1\right)\zeta_{0}^{-2}}{2}\\
 & \pm\frac{\left[\lambda^{-2}-\left(\omega^{2}/\omega_{0}^{2}-1\right)\zeta_{0}^{-2}\right]^{2}}{8\zeta_{0}^{-1}\lambda^{-1}\sqrt{\mu_{x0}-1}}.
\end{align*}
This region is characterized by a very strong mixing of spin and supercurrent
oscillations. The key observation is that, in contrast to nonmagnetic
superconductors, where low-frequency magnetic field decays on the
distance of the order of the London penetration depth, in our case
for frequency smaller than $\sqrt{\mu_{x0}}\omega_{0}$ one of the
parameters $\kappa_{\alpha}$ is complex meaning that \emph{the oscillating
magnetic field penetrates at much larger distance limited by external
dissipation mechanisms}.
\begin{figure}
	\includegraphics[width=3.2in]{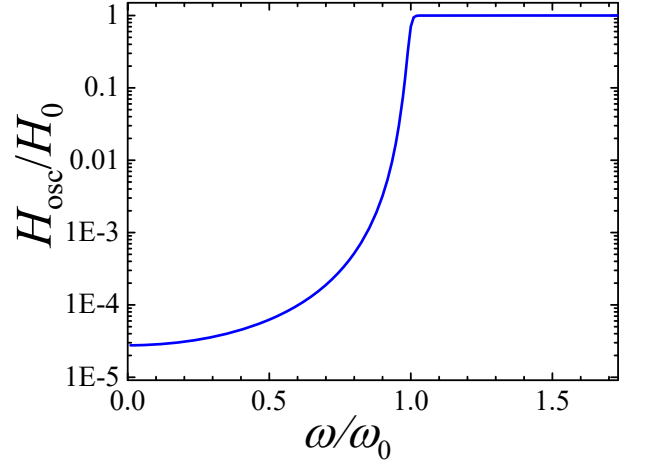}
	\caption{The frequency dependence of the amplitude of the oscillating field $H_{\mathrm{osc}}$ determining the long-range penetration of the microwave field mediated by the spin waves. We assumed $\zeta_{0}=0.02\lambda$ and $\mu_{x0}=3$. The plot terminates at $\omega/\omega_0=\sqrt{\mu_{x0}}$, where $H_{\mathrm{osc}}$ abruptly vanishes.}
	\label{Fig-Hosc-omega}
\end{figure}
\begin{figure}
	\includegraphics[width=3.2in]{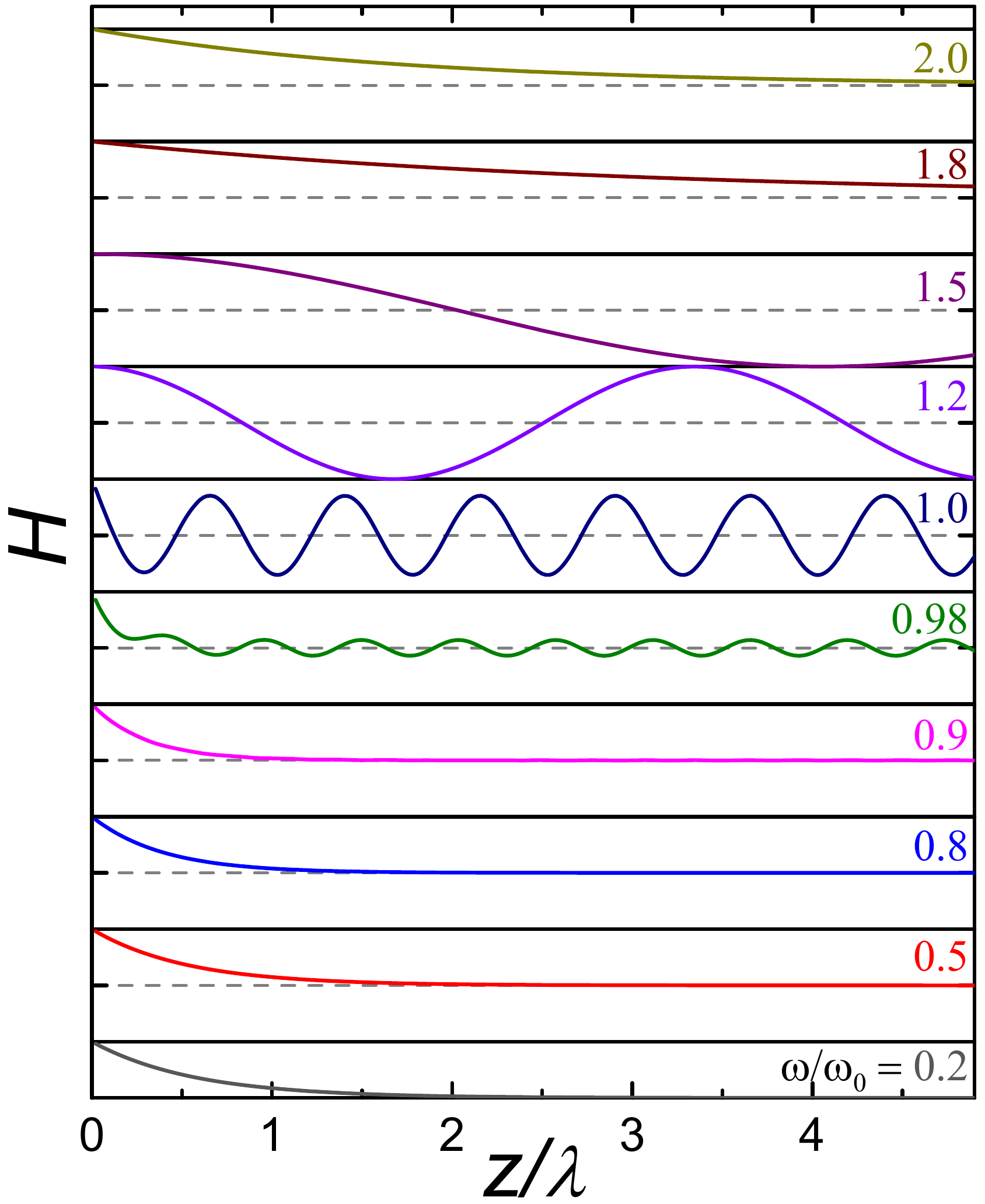}
	\caption{Series of the coordinate profiles of the microwave magnetic field inside superconductor for different frequencies. A key feature is a pronounced propagating wave in the range $1\!\lesssim \! \omega/\omega_0\! <\! \sqrt{\mu_{x0}}\!\approx 1.73\!$. 
	Such propagating wave is also present for $ \omega/\omega_0 <1$,  but, due to a very small amplitude, it is invisible for the used vertical scale. 
		}
	\label{Fig-H-z-omega}
\end{figure}

We now proceed with evaluation of the vector coefficients $\boldsymbol{M}_{0\alpha}$
and $\boldsymbol{H}_{0\alpha}$ from Eqs.~\eqref{eqs:Decay} using
the boundary conditions in Eqs.~\eqref{eq:BoundCondH} and \eqref{eq:BoundCondM}.
Substituting the relation $\boldsymbol{H}_{0\alpha}\approx-\frac{1}{1-\lambda^{2}\kappa_{\alpha}^{2}}4\pi\boldsymbol{M}_{0\alpha}$
into the boundary condition for $\boldsymbol{H},$ we obtain a $2\times2$
linear system for the magnetization coefficients
\begin{subequations}\label{eqs:M0a}
\begin{align}
\kappa_{1}\boldsymbol{M}_{01}+\kappa_{2}\boldsymbol{M}_{02} & =0,\\
-\frac{1}{1-\lambda^{2}\kappa_{1}^{2}}\boldsymbol{M}_{01}-\frac{1}{1-\lambda^{2}\kappa_{2}^{2}}\boldsymbol{M}_{02} & =\frac{\boldsymbol{H}_{0}}{4\pi},
\end{align}
\end{subequations}
which yields the solution
\begin{equation}
\left(\!\begin{array}{c}
\boldsymbol{M}_{01}\\
\boldsymbol{M}_{02}
\end{array}\!\right)\!=\!\frac{\left(-\lambda^{2}\kappa_{1}^{2}\right)\left(1-\lambda^{2}\kappa_{2}^{2}\right)\boldsymbol{H}_{0}/4\pi}
{\left(\kappa_{1}\!-\!\kappa_{2}\right)\left[1\! -\! \lambda^{2}\left(\kappa_{1}^{2}\!+\!\kappa_{1}\kappa_{2}\!+\!\kappa_{2}^{2}\right)\right]}
\left(\!\begin{array}{c}
\kappa_{2}\\
-\kappa_{1}
\end{array}\!\right).
\label{eq:M012}
\end{equation}
The corresponding field components are 
\begin{align}
\left(\begin{array}{c}
\boldsymbol{H}_{01}\\
\boldsymbol{H}_{02}
\end{array}\right) & =\frac{\boldsymbol{H}_{0}}{\left(\kappa_{1}-\kappa_{2}\right)\left[1-\lambda^{2}\left(\kappa_{1}^{2}+\kappa_{1}\kappa_{2}+\kappa_{2}^{2}\right)\right]}.\nonumber \\
\times & \left(\begin{array}{c}
-\kappa_{2}\left(1-\lambda^{2}\kappa_{2}^{2}\right)\\
\kappa_{1}\left(1-\lambda^{2}\kappa_{1}^{2}\right)
\end{array}\right).\label{eq:H012}
\end{align}
The amplitude of the field inside the superconductor with oscillating coordinate dependence, $\boldsymbol{H}_{\mathrm{osc}}$, corresponding to purely imaginary $\kappa_{\alpha}$ is given by $\boldsymbol{H}_{01}$ for $\omega<\omega_{0}$ and by $\boldsymbol{H}_{02}$  for $\omega_{0}<\omega<\sqrt{\mu_{x0}}\omega_{0}$. It determines the long-propagating microwave field mediated by the spin waves. Figure \ref{Fig-Hosc-omega} shows the frequency dependence of the ratio $H_{\mathrm{osc}}/H_0$. We can see that the oscillatory component rapidly increases when the frequency approaches $\omega_{0}$ from below and becomes very close to unity within the range $\omega_{0}\!<\!\omega\!<\!\sqrt{\mu_{x0}}\omega_{0}$. It abruptly vanishes at $\omega=\sqrt{\mu_{x0}}\omega_{0}$. Figure  \ref{Fig-H-z-omega} illustrates the coordinate profiles of the microwave magnetic field inside superconductor, $H(z)=\mathrm{Re}[H_1\exp(-\kappa_1z)+H_2\exp(-\kappa_2z)]$, for different frequencies. We see that the pronounced oscillating contribution emerges near $\omega\sim\omega_{0}$ and dominates in the range $\omega_{0}\!<\!\omega\!<\!\sqrt{\mu_{x0}}\omega_{0}\approx 1.73\omega_{0}$, while the corresponding wave length of oscillations increases as the frequency approaches $\sqrt{\mu_{x0}}\omega_{0}$. Slightly above this frequency, the microwave field  monotonically decreases but with very large decay length. 

\begin{figure}
\includegraphics[width=3.4in]{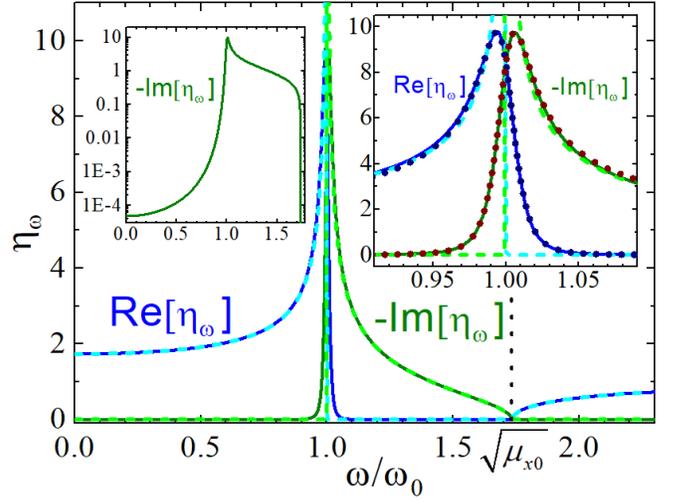}\caption{The frequency dependences of the real and imaginary part of the parameter
$\eta_{\omega}$, Eq.~\eqref{eq:eta-omega}, which determines the
dynamic magnetic boundary condition, Eq.~\eqref{eq:DynBoundCond}.
The dashed lines show the approximate result in Eq.~\eqref{eq:eta-omega-Approx}
valid away from the frequency $\omega_{0}$. The right inset
shows zoom in the region near the frequency $\omega_{0}$. The navy
and wine dotted lines in this inset show the approximate scaling result
in Eq.~\eqref{eq:eta-scaling}. The left inset shows the logarithmic
plot of $-\mathrm{Im}[\eta_{\omega}]$ to illustrate that it remains
finite down to zero frequency. The plots are made for $\zeta_{0}=0.02\lambda$
and $\mu_{x0}=3$. }
\label{Fig-eta-omega}
\end{figure}
The interaction between the magnetic superconductor and outside world
can be conveniently formulated in terms the boundary condition connecting
the gradient $\nabla_{z}\boldsymbol{H}$ with the field at the surface.
From Eq.~\eqref{eq:H012}, we obtain
\begin{align*}
\nabla_{z}\boldsymbol{H}|_{z=0} & =-\kappa_{1}\boldsymbol{H}_{01}-\kappa_{2}\boldsymbol{H}_{02}\\
= & -\frac{\kappa_{1}\kappa_{2}\left(\kappa_{1}+\kappa_{2}\right)}{\kappa_{1}^{2}+\kappa_{1}\kappa_{2}+\kappa_{2}^{2}-\lambda^{-2}}\boldsymbol{H}_{0}
\end{align*}
Using the relations $\kappa_{1}\kappa_{2}=-i\zeta_{0}^{-1}\lambda^{-1}\sqrt{\mu_{x0}\!-\!\omega^{2}/\omega_{0}^{2}}$
and $\kappa_{1}^{2}\!+\!\kappa_{2}^{2}\!-\!\lambda^{-2}\!=\left(\omega^{2}/\omega_{0}^{2}\!-\!1\right)\zeta_{0}^{-2},$
we can rewrite this boundary condition as 
\begin{subequations}
\begin{equation}
\nabla_{z}\boldsymbol{H}=-\eta_{\omega}\boldsymbol{H}/\lambda\label{eq:DynBoundCond}
\end{equation}
with 
\begin{equation}
\eta_{\omega}=\frac{-i\zeta_{0}\sqrt{\mu_{x0}-\omega^{2}/\omega_{0}^{2}}\left(\kappa_{1}+\kappa_{2}\right)}{\omega^{2}/\omega_{0}^{2}\!-\!1\!-i\zeta_{0}/\lambda\sqrt{\mu_{x0}\!-\!\omega^{2}/\omega_{0}^{2}}}.\label{eq:eta-omega}
\end{equation}
\end{subequations}
In the range $\left|\omega^{2}/\omega_{0}^{2}\!-\!1\right|\!\gg\!\sqrt{\mu_{x0}}\zeta_{0}/\lambda$
the parameters $\kappa_{\alpha}$ are given by Eqs.~\eqref{eq:MagDecay}
and \eqref{eq:SupDecay}. In this case $|\kappa_{1}|\gg|\kappa_{2}|,\lambda^{-1}$
and we obtain a simple approximate result
\begin{equation}
\eta_{\omega}\approx\lambda\kappa_{2}\approx-i\sqrt{\frac{\mu_{x0}-\omega^{2}/\omega_{0}^{2}}{\omega^{2}/\omega_{0}^{2}-1}}.\label{eq:eta-omega-Approx}
\end{equation}
Note that this result corresponds to the approximation of local magnetic
response and it remains valid even in the regime where $\zeta_{0}|\kappa_{1}|>1$.
In particular, it gives correctly the static-case result $\eta_{\omega=0}=\sqrt{\mu_{x0}}$.
On the other hand, at $\omega\!=\!\omega_{0}$, using Eq.~\eqref{eq:kappa-a-omega0},
we obtain 
\begin{equation}
\eta_{\omega_{0}}\approx\left(1-i\right)\left(\mu_{x0}-1\right)^{1/4}\sqrt{\lambda/\zeta_{0}}.\label{eq:eta-omega0}
\end{equation}
In the range $\omega^{2}/\omega_{0}^{2}\!-\!1\!\ll\!2\left(\mu_{x0}\!-\!1\right)$,
we derive the following approximate scaling form, 
\begin{align}
\eta_{\omega} & \approx\sqrt{\lambda/\zeta_{0}}\left(\mu_{x0}\!-\!1\right)^{1/4}v\!\left(\frac{\omega^{2}/\omega_{0}^{2}\!-\!1}{\zeta_{0}/\lambda\sqrt{\mu_{x0}-1}}\right),\label{eq:eta-scaling}\\
v(u) & =\frac{1}{1\!+\!iu}\left(\sqrt{\sqrt{\frac{u^{2}}{4}\!+\!1}\!+\frac{u}{2}}\!-i\sqrt{\sqrt{\frac{u^{2}}{4}\!+\!1}\!-\frac{u}{2}}\right).\nonumber 
\end{align}
The real and imaginary parts of the complex function $v(u)$ are connected
by the relation $\mathrm{Re}\left[v(-u)\right]=-\mathrm{Im}\left[v(u)\right]$.
The real part reaches the maximum value equal to $1.162$ at $u\approx-0.436$.
The asymptotics of $v(u)$ in the range $u\gg1$ is $v(u)\simeq-i/\sqrt{u}$
yielding $\eta_{\omega}\approx-i\sqrt{\mu_{x0}\!-1}/\sqrt{\omega^{2}/\omega_{0}^{2}\!-\!1}$.
This matches the result in Eq.~\eqref{eq:eta-omega-Approx} in the
range $\omega^{2}/\omega_{0}^{2}\!-\!1\ll1$. In the large negative
region, $u<0$, $|u|\gg1$, the imaginary part of $v(u)$ decays as
$\mathrm{Im}[v(u)]\simeq-|u|^{-7/2}$.

Figure \ref{Fig-eta-omega} shows plots of the real and imaginary
part of the parameter $\eta_{\omega}$, Eq.~\eqref{eq:eta-omega},
computed using typical parameters $\zeta_{0}=0.02\lambda$ and $\mu_{x0}=3$.
We also show in the figure the approximate result, Eq.~\eqref{eq:eta-omega-Approx},
valid for frequencies not very close to $\omega_{0}$, and, in the
upper right inset, the approximate scaling result in Eq.~\eqref{eq:eta-scaling}
describing behavior near $\omega_{0}$. The frequency dependence of
$\eta_{\omega}$ can be summarized as follows. In the range $\omega<\omega_{0}$,
the real part of $\eta_{\omega}$ is much larger than $-\mathrm{Im}(\eta_{\omega})$.
Both parts increase for $\omega\rightarrow\omega_{0}$ and become
equal in absolute value at $\omega\!=\!\omega_{0}$. The real part
reaches maximum $1.162\sqrt{\lambda/\zeta_{0}}\left(\mu_{x0}\!-\!1\right)^{1/4}$
slightly below $\omega_{0}$, at $\omega\!\approx\omega_{0}(1\!-\!0.218\zeta_{0}/\lambda\sqrt{\mu_{x0}\!-\!1})$,
while $-\mathrm{Im}(\eta_{\omega})$ reaches the same maximum slightly
above $\omega_{0}$, at $\omega\!\approx\omega_{0}(1\!+\!0.218\zeta_{0}/\lambda\sqrt{\mu_{x0}\!-\!1})$.
In the range $\omega_{0}<\omega<\sqrt{\mu_{x0}}\omega_{0}$, the real
part of $\eta_{\omega}$ is much smaller than $-\mathrm{Im}(\eta_{\omega})$.
Finally, in the region $\omega>\sqrt{\mu_{x0}}\omega_{0}$ the imaginary
part is zero, while the real part monotonically increases asymptotically
approaching unity.

\begin{figure}
\includegraphics[width=3.4in]{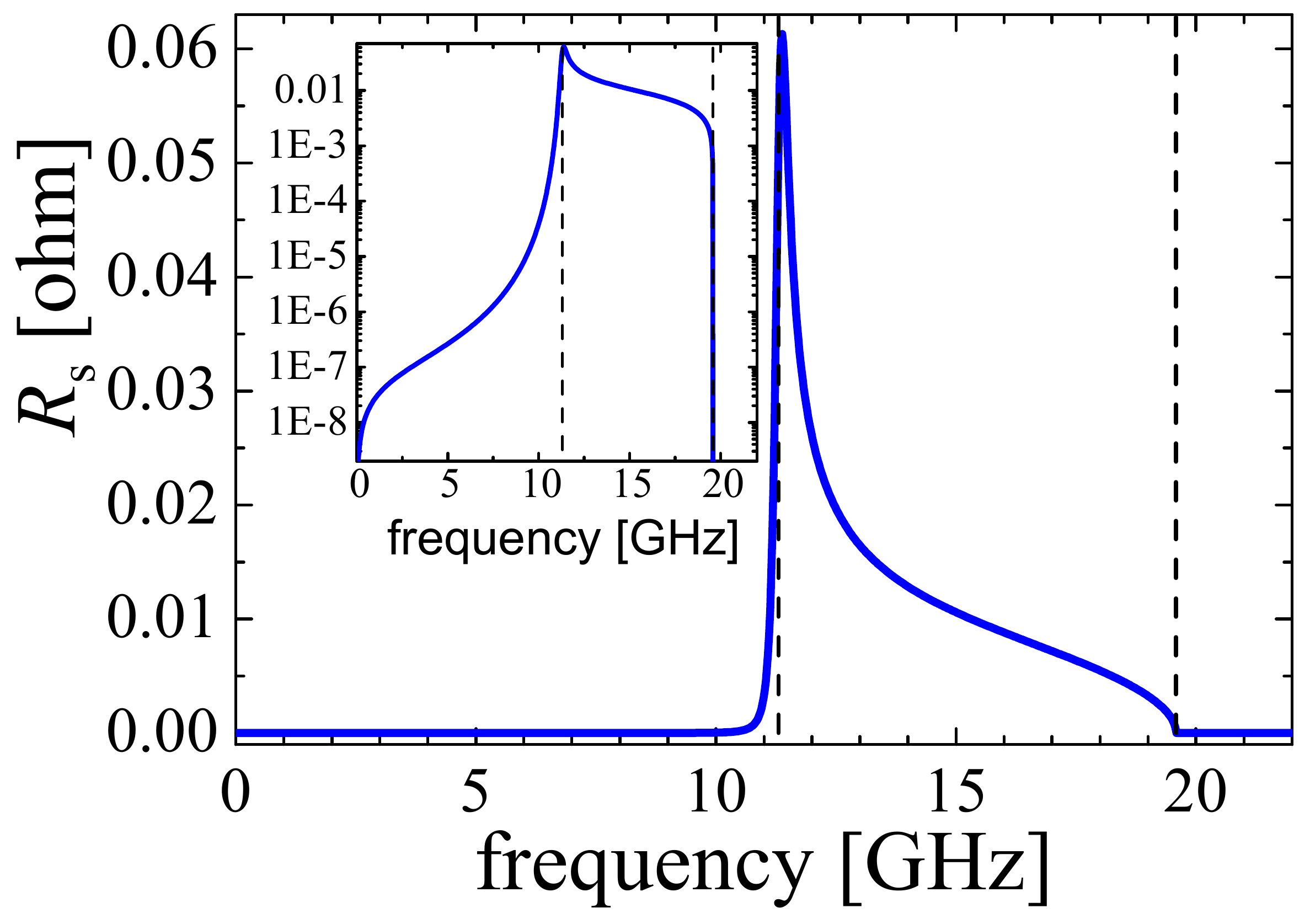}\caption{The frequency dependence of the surface resistivity using the same
parameters as in Figs.~\ref{Fig:Spectrum} and \ref{Fig-eta-omega}
corresponding to RbEuFe$_{4}$As$_{4}$. The inset shows the same
plot in logarithmic scale for the better presentation of the low-frequency
behavior. The vertical dashed lines show locations of the frequencies
$f_{0}\!=\!\omega_{0}/2\pi$ and $\sqrt{\mu_{x0}}f_{0}$.}
\label{Fig:Rs}
\end{figure}
The parameter $\eta_{\omega}$ is directly connected with the conventional
parameter characterizing the microwave response, surface impedance
\begin{equation}
Z=\frac{E_{x}}{\int_{0}^{\infty}j_{x}(z)dz}=\frac{4\pi}{\mathrm{c}}\frac{E_{x}}{H_{y}}.\label{eq:SurfImpDef}
\end{equation}
To establish this connection, we have to relate the tangential electric
field with the normal gradient of the magnetic field. At small frequencies,
we can use the London relation $\frac{4\pi}{\mathrm{c}}\frac{\partial j_{x}}{\partial t}\approx c\lambda^{-2}E_{x}$
neglecting a small contribution from the quasiparticle current and
the Maxwell equation $-\nabla_{z}H_{y}=\frac{4\pi}{\mathrm{c}}j_{x}$ omitting
the displacement current. This gives $-\nabla_{z}H_{y}=\frac{\mathrm{c}}{i\omega\lambda^{2}}E_{x}$
and from Eq.~\eqref{eq:DynBoundCond} we obtain 
\begin{equation}
Z  =4\pi i\omega\eta_{\omega}\lambda/\mathrm{c}^{2}.
\label{eq:SurfImpRelation}
\end{equation}
The real part of this equation, $R_s=\mathrm{Re}(Z)$, can be converted to the practical formula
for surface resistivity $R_{s}\left[\mathrm{ohm}\right]=-8\pi^{2}10^{-4}$
$\mathrm{Im}\left[\eta_{\omega}\right]f[\mathrm{GHz}]\lambda[\mathrm{\mu m}]$. 

Figure \ref{Fig:Rs} shows the frequency dependence of the surface
resistivity using the same parameters as in Figs.~\ref{Fig:Spectrum}
and \ref{Fig-eta-omega}. We can see that the surface resistivity
has a very distinct shape. It is very small at small frequencies $f<f_{0}\!=\!\omega_{0}/2\pi$
and starts to increase sharply when the frequency approaches $f_{0}$.
After reaching a peak value $\sim0.06$ ohm slightly above $f_{0}$,
it slowly decreases within extended frequency range $f_{0}<f<\sqrt{\mu_{x0}}f_{0}$,
and abruptly vanishes at $\sqrt{\mu_{x0}}f_{0}$.

\section{Excitation of spin waves with AC Josephson effect \label{sec:ExcitJos}}

\begin{figure}
\includegraphics[width=3in]{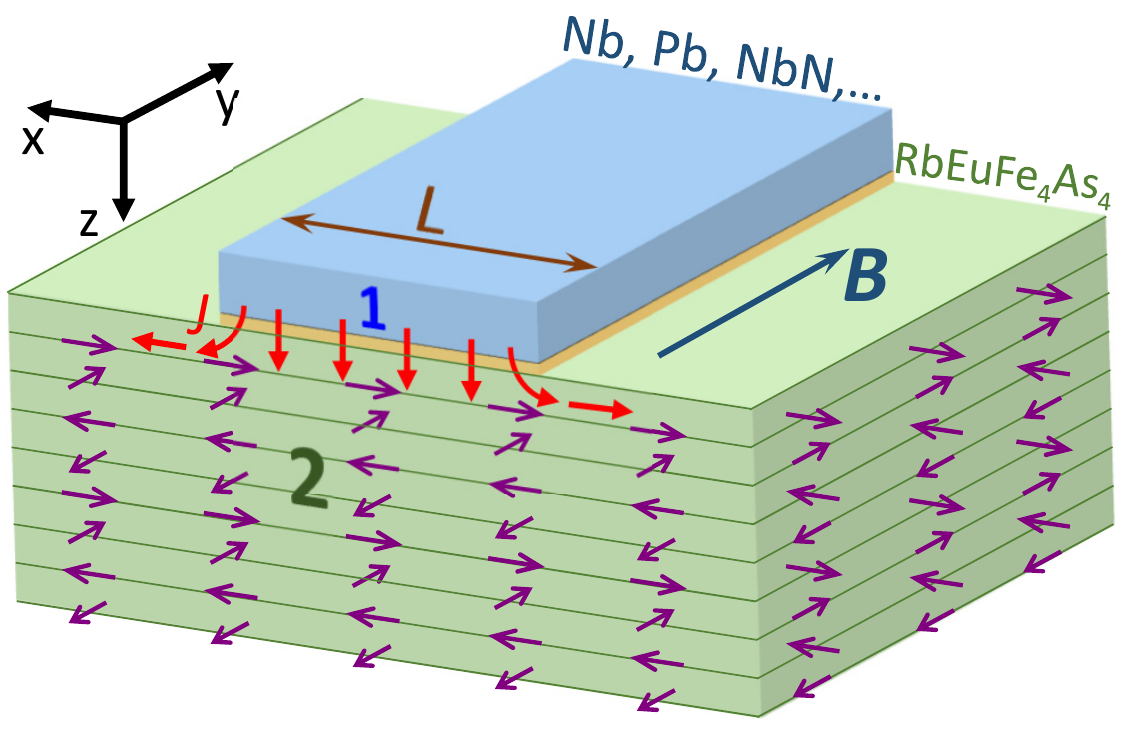}\caption{Illustration of a planar tunneling contact between a conventional superconductor
(1) and magnetic superconductor with helical magnetic structure (2).
Purple arrows illustrate orientation of the magnetic moments. \label{fig:JJ-FM-SC}}
\end{figure}
The presence of magnetic order inside superconducting material provides a unique possibility 
to generate and manipulate magnons using the AC Josephson effect.   
In this section, we consider the excitation of spin waves in a tunneling
contact between a conventional superconductor marked by the index
$1$ and a superconductor with helical magnetic structure marked by
the index $2$, as illustrated in Fig.~\ref{fig:JJ-FM-SC}.
We assume that the system is uniform along the $y$ direction and 
the interlayer with thickness $\mathfrak{t}$ is insulating and nonmagnetic.
The magnetic and conventional superconductors occupy the regions $z>0$
and $z<-\mathfrak{t}$, respectively.

\subsection{Dynamic equation for the Josephson phase }

We follow the standard derivation of the dynamic equation for the
gauge-invariant phase difference between two superconductors 
$\theta=\phi_{2}-\phi_{1}-\frac{2\pi \mathfrak{t}}{\Phi_{0}}A_{z}$
taking into account the dynamic magnetization response. The starting point
of derivation is the $z$ component of the Maxwell equation, 
\begin{equation}
\nabla_{x}H_{y}=\frac{4\pi}{\mathrm{c}}j_{z}+\frac{\varepsilon}{\mathrm{c}}\frac{\partial E_{z}}{\partial t},
\label{eq:Maxw-z}
\end{equation}
in which the total current density across the junction is composed
of the superconducting and quasiparticle contributions, $j_{z}=j_{s,z}+j_{n,z}$,
where the superconducting contribution is given by the DC Josephson
relation, 
\begin{equation}
j_{s,z} =j_{J}\sin\theta,\label{eq:Jos-dc}
\end{equation}
and the quasiparticle contribution is determined by tunneling conductivity
$\sigma$, $j_{n,z}=\sigma E_{z}$. The electric field is related
to the phase by the AC Josephson relation
\begin{equation}
E_{z}=\frac{\Phi_{0}}{2\pi cd}\frac{\partial\theta}{\partial t}.\label{eq:Jos-ac}
\end{equation}
To relate $\nabla_{x}H_{y}$ in Eq.~\eqref{eq:Maxw-z} with the phase
gradient, we use the $x$ component of the Maxwell equations $-\nabla_{z}H_{y}=\frac{4\pi}{\mathrm{c}}j_{x}$
and the London relation for supercurrents along the junction  $\frac{4\pi}{\mathrm{c}}j_{x}\approx\lambda_{i}^{-2}\left(\frac{\Phi_{0}}{2\pi}\nabla_{x}\phi-A_{x}\right)$.
Here, we neglected the displacement current assuming small frequencies
and quasiparticle current inside the superconductors. This leads to
the relation between the in-plane phase gradient and magnetic fields
\begin{align}
\nabla_{x}\theta & \!=\frac{8\pi^{2}}{\mathrm{c}\Phi_{0}}\left(\lambda_{2}^{2}j_{x,2}\!-\!\lambda_{1}^{2}j_{x,1}\right)\!+\frac{2\pi \mathfrak{t}}{\Phi_{0}}B_{y}\nonumber \\
=\!- & \frac{2\pi}{\Phi_{0}}\left(\lambda_{2}^{2}\nabla_{z}H_{y,2}\!-\!\lambda_{1}^{2}\nabla_{z}H_{y,1}\right)\!+\frac{2\pi \mathfrak{t}}{\Phi_{0}}B_{y},\label{eq:grad-theta}
\end{align}
where $j_{x,i}$ and $H_{y,i}$ are the current densities and the
magnetic fields at the surfaces of two superconductors and $B_{y}$
is the magnetic induction inside the junction. We assume a nonmagnetic
interlayer meaning that $B_{y}\!=H_{y}$. Also, for the nonmagnetic
superconductor in the Meissner state at $z<-\mathfrak{t}$, we have $\nabla_{z}H_{y,1}\!=B_{y}/\lambda_{1}$.
To obtain the close system, we need the boundary condition connecting
$\nabla_{z}H_{y,2}$ with $H_{y}\!=B_{y}$ at the surface of the magnetic
superconductor at $z=0$. Note that $H_{y}(z)$ is continuous, while
$B_{y}(z)$ has a jump at $z\!=\!0.$ Due to magnetization dynamics,
this boundary condition is frequency dependent. At fixed frequency,
such boundary condition has been derived in Sec.~\ref{sec:DynEqSmooth}
and is given by Eq.~\eqref{eq:DynBoundCond}, which in our case becomes
$\nabla_{z}H_{y,2}=-\eta_{\omega}B_{y}/\lambda_{2}$. The complex
parameter $\eta_{\omega}$ is determined by the general result in
Eq.~\eqref{eq:eta-omega}. In the approximation of local magnetic
response valid for frequencies not too close to the bare uniform-mode
frequency $\omega_{0}$, it has a much simpler approximate presentation
in Eq.~\eqref{eq:eta-omega-Approx}. Therefore, Eq.~\eqref{eq:grad-theta}
at finite frequency becomes
\begin{equation}
\nabla_{x}\theta=\!\frac{2\pi(\mathfrak{t}\!+\!\lambda_{1}\!+\!\eta_{\omega}\lambda_{2})}{\Phi_{0}}H_{y}\label{eq:grad2-theta}
\end{equation}
Applying $\nabla_{x}$ to both sides, substituting $\nabla_{x}H_{y}$ from Eq.~\eqref{eq:Maxw-z}, and
using the Josephson relations for current and electric field, Eqs.~\eqref{eq:Jos-dc}
and \eqref{eq:Jos-ac}, we obtain the dynamic phase equation at finite
frequency in the form
\begin{equation}
\frac{1}{\mathfrak{t}\!+\!\lambda_{1}\!+\!\eta_{\omega}\lambda_{2}}\nabla_{x}^{2}\theta=\frac{8\pi^{2}}{\mathrm{c}\Phi_{0}}j_{J}\left[\sin\theta\right]_{\omega}\!-\frac{\varepsilon_{\omega}\omega^{2}}{\mathfrak{t}c^{2}}\theta\label{eq:DynJos-Freq}
\end{equation}
where $\varepsilon_{\omega}\!\equiv\!\varepsilon-4\pi i\sigma/\omega$
and $\left[\sin\theta\right]_{\omega}$ notates the time Fourier transform
of $\sin\left[\theta(x,t)\right]$. The only difference from the standard
phase-dynamics Sine-Gordon equation \cite{kulik1972josephson,BaroneBook}
is the presence of the complex factor $\eta_{\omega}$ with complicated
frequency dependence, see Fig.~\ref{Fig-eta-omega}. In the static
case, the phase equation is 
\begin{equation}
\frac{1}{\mathfrak{t}\!+\!\lambda_{1}+\!\sqrt{\mu_{x0}}\lambda_{2}}\nabla_{x}^{2}\theta=\frac{8\pi^{2}}{\mathrm{c}\Phi_{0}}j_{J}\sin\theta.\label{eq:StaticPhaseEq}
\end{equation}
Therefore, the effective junction interlayer width 
$\tilde{\mathfrak{t}}=\mathfrak{t}\!+\!\lambda_{1}\!+\!\sqrt{\mu_{x0}}\lambda_{2}$
is enlarged by the magnetic response. From the last equation, we can
evaluate the static Josephson length 
\begin{equation}
\lambda_{J}\!=\!\left\{ \frac{\mathrm{c}\Phi_{0}}{\left[8\pi^{2}\left(\mathfrak{t}\!+\!\lambda_{1}\!+\!\sqrt{\mu_{x0}}\lambda_{2}\right)j_{J}\right]}\right\} ^{1/2}.\label{eq:JosLength}
\end{equation}
In the next subsection we consider the influence of magnetic response
on the spectrum and damping of the electromagnetic wave propagating through
the Josephson junction. 

\subsection{Spectrum and damping of the Josephson plasma mode}

\begin{figure}
\includegraphics[width=3.4in]{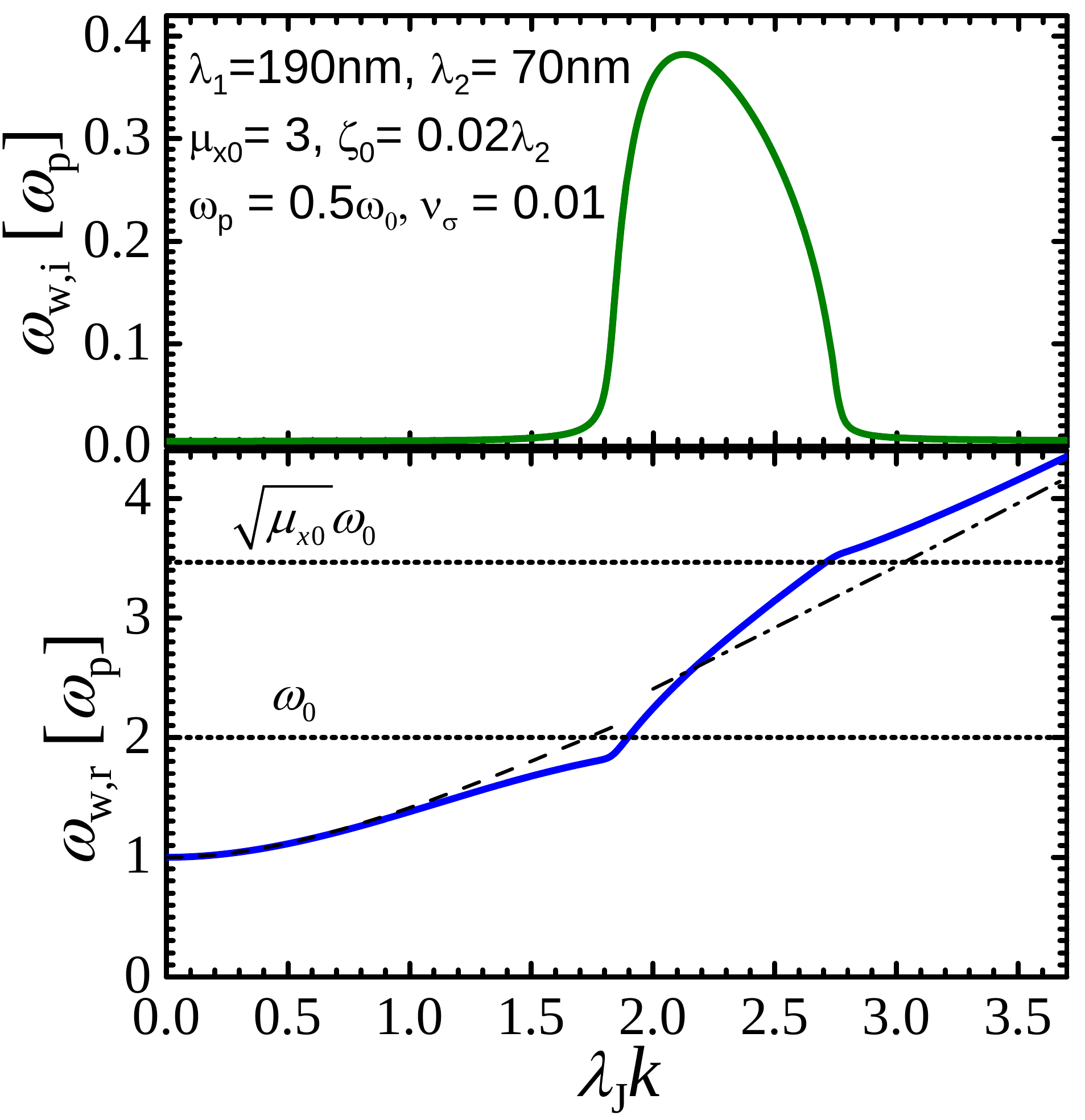}
\caption{Spectrum and damping of the electromagnetic wave inside a Josephson
junction between conventional and helical magnetic superconductors,
Eq.~\eqref{eq:SpectrumEqPl}. In the lower plot, the dashed and dash-dotted
lines show spectra corresponding to low-frequency and high-frequency
limits, respectively.}
\label{fig:JJwaverp05}
\end{figure}
The superconductor-insulator-superconductor sandwich structure with
sufficiently large width is a waveguide capable of supporting a traveling
electromagnetic wave \cite{SwihartJAP61,BaroneBook} with the phase
$\theta(x,t)\propto\exp\left[i\left(\omega_{w}t\pm kx\right)\right]$.
Such a wave can be resonantly excited by the AC Josephson effect.
For the fixed real wave vector $k$, Eq.~\eqref{eq:DynJos-Freq}
gives the following equation for the complex frequency $\omega_{w}(k)=\omega_{w,r}(k)+i\omega_{w,i}(k)$,
with the real and imaginary part giving the wave spectrum and its
damping, respectively,
\begin{equation}
\omega_{w}^{2}\!-\frac{4\pi\sigma}{\varepsilon}i\omega_{w}=\omega_{p}^{2}+
\frac{\mathfrak{t}}{\mathfrak{t}\!+\!\lambda_{1}\!+\!\eta_{\omega}\lambda_{2}}\frac{\mathrm{c}^{2}}{\varepsilon}k^{2},\label{eq:SwModePl}
\end{equation}
where
\begin{equation}
\omega_{p}=\sqrt{\frac{8\pi^{2}c\mathfrak{t}}{\varepsilon\Phi_{0}}j_{J}}\label{eq:PlasmaFreq}
\end{equation}
is the Josephson plasma frequency. Note that the magnetic response does not modify this parameter. It is convenient to rewrite Eq.~\eqref{eq:SwModePl}
in the reduced form 
\begin{equation}
\frac{\omega_{w}^{2}}{\omega_{p}^{2}}-i\nu_{\sigma}\frac{\omega_{w}}{\omega_{p}}=1+\frac{\lambda_{1}+\sqrt{\mu_{x0}}\lambda_{2}}{\lambda_{1}+\eta_{\omega}\lambda_{2}}\lambda_{J}^{2}k^{2}\label{eq:SpectrumEqPl}
\end{equation}
with the static Josephson length $\lambda_{J}$, Eq.\ \eqref{eq:JosLength},
and the dumping parameter
\begin{equation}
\nu_{\sigma}=\frac{4\pi\sigma}{\varepsilon\omega_{p}}.\label{eq:DumpParam}
\end{equation}
The parameter $\eta_{\omega}$ has the strongest feature around
$\omega=\omega_{0}$. Therefore, the spectrum of the Josephson plasmon
is substantially affected only if $\omega_{p}<\omega_{0}$.

Figure \ref{fig:JJwaverp05} shows the spectrum and damping of the
propagating wave computed from Eq.~\eqref{eq:SpectrumEqPl} for parameters
corresponding to the contact between NbN and RbEuFe$_{4}$As$_{4}$,
$\lambda_{1}\!=\!190$nm, $\lambda_{2}\!=\!70$nm, $\mu_{x0}\!=\!3$, and $\varsigma_{0}\!=\!0.02\lambda_{2}$.
We also assume $\omega_{p}\!=\!0.5\omega_{0}$ and $\nu_{\sigma}\!=\!0.01$.
One can distinguish three regions with qualitatively different behavior.
In the low-frequency region $\omega_{w,r}(k)<\omega_{0}$, the spectrum
is approximately $\omega_{w,r}(k)\simeq\sqrt{\omega_{p}^{2}+c_{s0}^{2}k^{2}}$,
where $c_{s0}$ is the low-frequency Swihart velocity
\begin{equation}
c_{s0}\!=\!\lambda_{J}\omega_{p}\!=\sqrt{\frac{\mathfrak{t}}{\lambda_{1}\!+\!\sqrt{\mu_{x0}}\lambda_{2}}}\frac{\mathrm{c}}{\sqrt{\varepsilon}}.\label{eq:SwihVel}
\end{equation}
In this region the spin waves give a small contribution to the mode
damping. The intermediate region $\omega_{0}<\omega_{w,r}(k)<\sqrt{\mu_{x0}}\omega_{0}$
is characterized by a sharp enhancement of the damping caused by excitation of the spin waves.
Finally, in the high-frequency region $\omega_{w,r}(k)>\sqrt{\mu_{x0}}\omega_{0}$
the damping caused by spin waves is absent and the spectrum approaches
the high-frequency limit $\omega_{w,r}(k)\simeq\sqrt{\omega_{p}^{2}+c_{s1}^{2}k^{2}}\simeq c_{s1}k$,
where $c_{s1}$ is the high-frequency mode velocity, 
\begin{align}
c_{s1}\! & =\!\sqrt{\frac{d}{\lambda_{1}\!+\!\lambda_{2}}}\frac{\mathrm{c}}{\sqrt{\varepsilon}}\nonumber \\
 & =\!\sqrt{\frac{\lambda_{1}\!+\!\sqrt{\mu_{x0}}\lambda_{2}}{\lambda_{1}\!+\!\lambda_{2}}}c_{s0}.\label{eq:cs-high-2}
\end{align}
In this limit the influence of magnetism is weak.

\subsection{Current-voltage characteristics and Fiske resonances in finite magnetic
field}

Transport properties of a Josephson junction in the magnetic field
directly probe its dynamic response \cite{kulik1972josephson,BaroneBook,CirilloPhysRevB.58.12377,KirtleyInBook2019}.
In particular, one can directly excite collective modes in superconducting
materials and the spectrum of these modes can be inferred from the dynamic
features in the current-voltage characteristics \cite{CarlsonJLTP1976}.
In this section, we evaluate the current-voltage characteristics for
our system using the standard approach of the expansion with respect
to the Josephson current assuming fixed voltage \cite{KulikJETPLett65}. Consider a junction in finite magnetic field $B_{y}$ and in the resistive state with
finite voltage drop across the junction, $V=\mathfrak{t}E_{z}$. In this state,
in the zeroth order with respect to the Josephson current, the phase has
the shape of a traveling wave
\begin{equation}
\theta_{0}(x,t) =k_{B}x+\omega t\label{eq:PhZeroOrder}
\end{equation}
with the wave vector 
\begin{equation}
k_{B}  =\frac{2\pi}{\Phi_{0}}\left(\mathfrak{t}\!+\!\lambda_{1}\!+\!\sqrt{\mu_{x0}}\lambda_{2}\right)B_{y},\label{eq:kB}
\end{equation}
and the Josephson frequency
\begin{equation}
\omega=\frac{2\pi \mathrm{c}}{\Phi_{0}}V.\label{eq:omegaJ}
\end{equation}
Representing $\sin\theta_{0}(x,t)\!=\!\mathrm{Re}\left[\!-i\exp\left(ik_{B}x\!+\!i\omega t\right)\right]$,
we obtain from Eq.~\eqref{eq:DynJos-Freq} the equation for the first-order
correction to the dynamic phase, $\tilde{\theta}(x,t)=\mathrm{Re}\left[\tilde{\theta}(x)\exp\left(i\omega t\right)\right]$,
which we present as 
\begin{equation}
\nabla_{x}^{2}\tilde{\theta}+p_{\omega}^{2}\tilde{\theta}=-ir_{\omega}\lambda_{J}^{-2}\exp\left(ik_{B}x\right)\label{eq:EqPhaseOmega1}
\end{equation}
with 
\begin{subequations}
\begin{align}
r_{\omega} & \!\equiv\frac{\mathfrak{t}\!+\!\lambda_{1}\!+\!\eta_{\omega}\lambda_{2}}{\mathfrak{t}\!+\!\lambda_{1}\!+\!\sqrt{\mu_{x0}}\lambda_{2}},
\label{eq:romega}\\
p_{\omega}^{2} & \equiv\frac{\varepsilon_{\omega}\omega^{2}}{\mathrm{c}^{2}}\frac{\mathfrak{t}\!+\!\lambda_{1}\!+\!\eta_{\omega}\lambda_{2}}{\mathfrak{t}}\notag\\
 & =\frac{\omega^{2}\!-\!\left(4\pi\sigma/\varepsilon\right)i\omega}{c_{s0}^{2}}r_{\omega},
\label{eq:pomega}
\end{align}
\end{subequations}
where $c_{s0}$ is the low-frequency Swihart velocity, Eq.~\eqref{eq:SwihVel}. 

We look for the solution of Eq.~\eqref{eq:EqPhaseOmega1} in the
form 
\begin{equation}
\tilde{\theta}(x)\!=\!\frac{r_{\omega}\lambda_{J}^{-2}}{k_{B}^{2}\!-\!p_{\omega}^{2}}\left[A_{c}\cos\left(p_{\omega}x\right)\!+\!A_{s}\sin\left(p_{\omega}x\right)\!+i\exp\left(ik_{B}x\right)\right].\label{eq:SolForm}
\end{equation}
Assuming the nonradiative boundary conditions, $\nabla_{x}\tilde{\theta}\!=\!0$
for $x\!=\!0,\,L$, we find the coefficients $A_{c}$ and $A_{s}$,
\begin{subequations}\label{eqs:AcAs}
\begin{align}
A_{s}\! & =k_{B}/p_{\omega},\\
A_{c}\sin\left(p_{\omega}L\right) & \!=\frac{k_{B}}{p_{\omega}}\left[\cos\left(p_{\omega}L\right)\!-\!\exp\left(ik_{B}L\right)\right]
\end{align}
\end{subequations} 
and substitute them into Eq.~\eqref{eq:SolForm}.
This yields the oscillating phase
\begin{widetext}
\begin{align}
\tilde{\theta}(x) & \!=\!\frac{r_{\omega}\lambda_{J}^{-2}}{k_{B}^{2}\!-\!p_{\omega}^{2}}\left[\frac{k_{B}}{p_{\omega}}\frac{\cos\left[p_{\omega}\left(L\!-\!x\right)\right]\!-\!\exp\left(ik_{B}L\right)\cos\left(p_{\omega}x\right)}{\sin\left(p_{\omega}L\right)}+\!i\exp\left(ik_{B}x\right)\right].\label{eq:DynPhaseResult}
\end{align}
\end{widetext}
The average Josephson current density is given by
\begin{align}
\delta j\! & =\!\frac{j_{J}}{L}\int\limits _{0}^{L}\left\langle \sin\left(k_{B}x\!+\!\omega t\!+\!\mathrm{Re}\left[\tilde{\theta}(x)\exp\left(i\omega t\right)\right]\right)\right\rangle _{t}dx\nonumber \\
 & \approx\frac{j_{J}}{2L}\int\limits _{0}^{L}\mathrm{Re}\left[\tilde{\theta}(x)\exp\left(-ik_{B}x\right)\right]dx\label{eq:AvJCurrDen}.
\end{align}
Substituting $\tilde{\theta}(x)$ from Eq.~\eqref{eq:DynPhaseResult},
we obtain
\begin{align}
\delta j & \!\approx\!\frac{j_{J}\lambda_{J}^{-2}}{2}\label{eq:deltaI}\\
\times & \mathrm{Im}\left\{ \left[1\!+\!\frac{\cos\left(p_{\omega}L\right)\!-\!\cos\left(k_{B}L\right)}{p_{\omega}L\sin\left(p_{\omega}L\right)}\frac{2k_{B}^{2}}{p_{\omega}^{2}\!-\!k_{B}^{2}}\right]\frac{r_{\omega}}{p_{\omega}^{2}\!-\!k_{B}^{2}}\right\} .\nonumber 
\end{align}
The key difference from the standard result \cite{KulikJETPLett65}
is the presence of the complex factor $\eta_{\omega}$ in the parameters
$p_{\omega}$ and $r_{\omega}$ in Eqs.\ \eqref{eq:romega} and \eqref{eq:pomega} from the magnetic boundary condition
describing the excitation of spin waves in the magnetic superconductor. The
location of the Fiske peaks corresponding to excitation of the standing
electromagnetic waves inside the junction is determined by the condition
$\mathrm{Re}[p_{\omega}]L=\pi n$. In the regions $\omega<\omega_{0}$
and $\omega>\sqrt{\mu_{x0}}\omega_{0}$ where $\mathrm{Im}(\eta_{\omega})\ll \mathrm{Re}(\eta_{\omega})$, 
this condition gives the equation for the resonance frequencies
\begin{equation}
\omega_{n}=\sqrt{\frac{\mathfrak{t}\!+\!\lambda_{1}\!+\!\sqrt{\mu_{x0}}\lambda_{2}}{\mathfrak{t}\!+\!\lambda_{1}\!+\!\mathrm{Re}(\eta_{\omega})\lambda_{2}}}\frac{\pi n\lambda_{J}}{L}\omega_{p},\label{eq:ResonFreqs}
\end{equation}
where $\omega_{p}$ is the Josephson plasma frequency, Eq.~\eqref{eq:PlasmaFreq}.

To facilitate numerical calculations, we rewrite Eq.~\eqref{eq:deltaI}
in the reduced form. We define the dimensionless size $\tilde{L}=L/\lambda_{J}$
and frequency $\tilde{\omega}=\omega/\omega_{p}$. We
also introduce the reduced wave-vector parameters $\tilde{k}_{B}=\lambda_{J}k_{B}$
and
\[
\tilde{p}_{\omega}\!=\lambda_{J}p_{\omega}=\sqrt{\left(\tilde{\omega}^{2}\!-\!i\nu_{\sigma}\tilde{\omega}\right)r_{\omega}},
\]
where $\nu_{\sigma}$ is the dimensionless damping parameter, Eq.~\eqref{eq:DumpParam}.
With these variables, we rewrite Eq\@.~\eqref{eq:deltaI} as
\begin{equation}
\frac{\delta j}{j_{J}}\!=\!\tfrac{1}{2}\mathrm{Im}\!\left\{ \!\left[1\!+\!\frac{\cos(\tilde{p}_{\omega}\tilde{L})\!-\!\cos(\tilde{k}_{B}\tilde{L})}{\tilde{p}_{\omega}\tilde{L}\sin(\tilde{p}_{\omega}\tilde{L})}\frac{2\tilde{k}_{B}^{2}}{\tilde{p}_{\omega}^{2}\!-\!\tilde{k}_{B}^{2}}\right]\frac{r_{\omega}}{\tilde{p}_{\omega}^{2}\!-\!\tilde{k}_{B}^{2}}\!\right\} .\label{eq:deltaIreduced}
\end{equation}
The product $\tilde{k}_{B}\tilde{L}$ here may be related to the
magnetic field as $\tilde{k}_{B}\tilde{L}\!=\!\pi B_{y}/B_{L}\!=\!2\pi\Phi_{y}/\Phi_{0}$, where
\begin{equation}
B_{L}  =\frac{\Phi_{0}}{2L\left(\mathfrak{t}\!+\!\lambda_{1}\!+\!\sqrt{\mu_{x0}}\lambda_{2}\right)}
\label{eq:BL}
\end{equation}
is the size-dependent scale determining periodicity of magnetic oscillations of the Fiske resonances  and $\Phi_{y}\!=\!L\left(\mathfrak{t}\!+\!\lambda_{1}\!+\!\sqrt{\mu_{x0}}\lambda_{2}\right)B_{y}$ is the magnetic flux through the junction. For frequency in Eq.\ \eqref{eq:ResonFreqs}, the strongest resonance is realized at $B=nB_L$. For other Fiske resonances, odd peaks with $n\!=\!2m+1$ are maximal for  $B_y=2j B_L$ ($\Phi_{y}/\Phi_{0}\!=\!j$) while even peaks with $n\!=\!2m$ are maximal for $B_y\!=\!(2j+1)B_L$ ($\Phi_{y}/\Phi_{0}\!=\!j\!+\!1/2$) \cite{KulikJETPLett65,kulik1972josephson,BaroneBook}.
Adding the tunnel quasiparticle
current, $j_{n}\!=\!\sigma E_{z}$, we obtain the total current in the
reduced form
\begin{equation}
\frac{j}{j_{J}}=\nu_{\sigma}\tilde{\omega}+\frac{\delta j}{j_{J}}.\label{eq:TotalCurrReduced}
\end{equation}
This equation together with Eq.~\eqref{eq:deltaIreduced} determines
the current-voltage characteristic in the reduced form in the second
order with respect to the Josephson current. 

\begin{figure}
\includegraphics[width=3.4in]{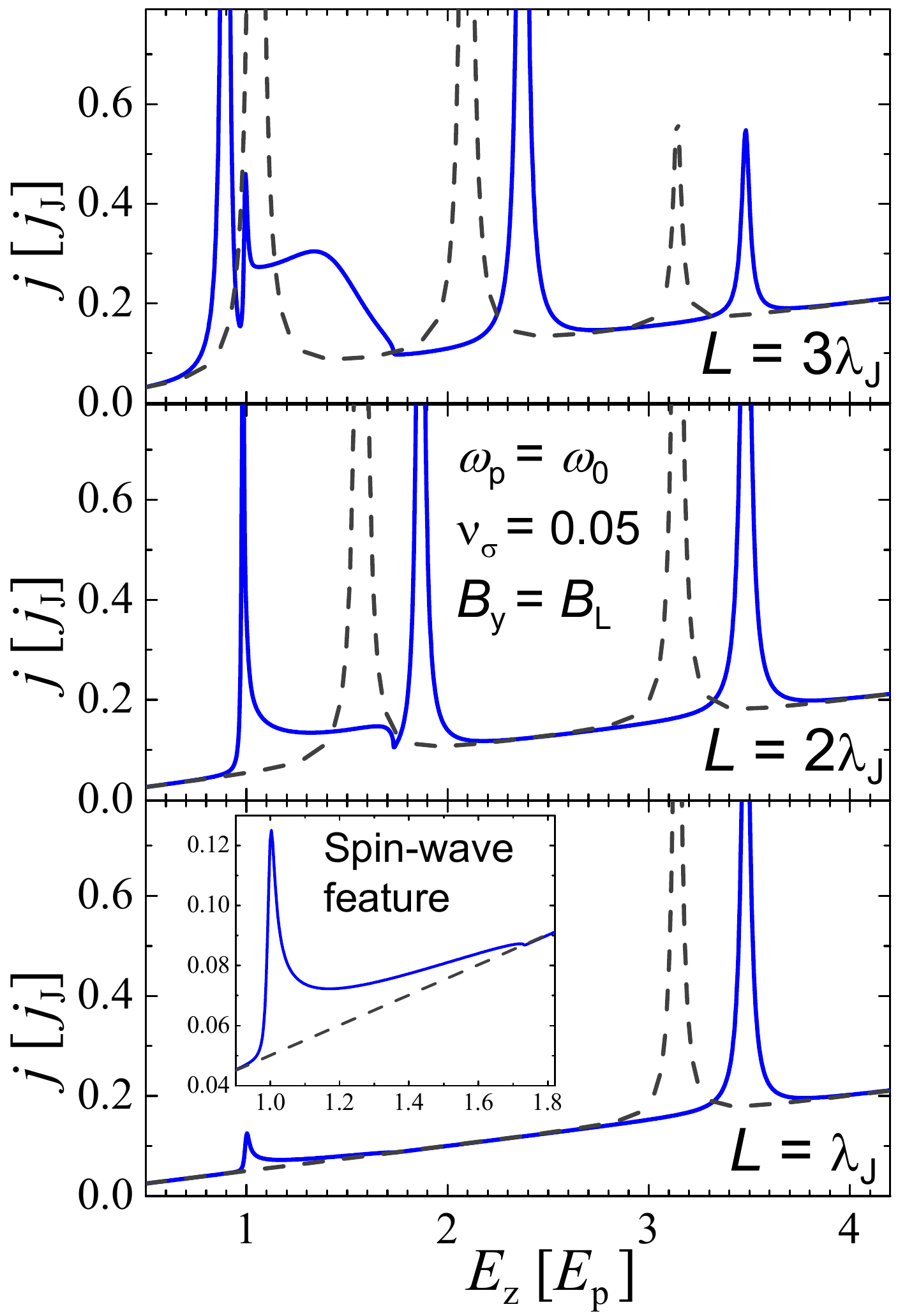}\caption{Representative current-voltage characteristics for junctions with
different lateral sizes $L$. The horizontal-axis scale $E_{p}$ is
the electric field at which the Josephson frequency equals $\omega_{p}$,
$E_{p}\!=\!\Phi_{0}\omega_{p}/(2\pi c \mathfrak{t})$. For comparison, the dashed
lines show the current-voltage characteristics without dynamic magnetic
response using static parameters $c_{so}$ and $\lambda_{J}$. They
display a usual sequence of the Fiske resonances. The inset in the
bottom plot zooms into the spin-wave feature.}
\label{Fig:IVz002nu005mu3rp1B1BL}
\end{figure}
The shape of the current-voltage characteristic mostly depends on the
relation between the Josephson plasma frequency $\omega_{p}$, the
location of the first Fiske resonance $\omega_{1}\!=\!\pi c_{s1}/L$,
and the two typical spin-wave frequencies $\omega_{0}$ and $\sqrt{\mu_{x0}}\omega_{0}$.
As the resistive state is stable until the Josephson frequency exceeds
the plasma frequency $\omega_{p}$, Eq.~\eqref{eq:PlasmaFreq},  spin waves can be excited only if $\omega_{p}$ is at least smaller than $\sqrt{\mu_{x0}}\omega_{0}$.
The clearest spin-wave features can be observed if $\omega_{p}\lesssim\omega_{0}$.
In addition, the behavior is also very sensitive to the junction size
$L$. For junctions narrower than the typical size $L_{c}\!=\!c_{s1}\pi/(\sqrt{\mu_{x0}}\omega_{0})$,
the whole spin-wave region $\omega_{0}<\omega<$$\sqrt{\mu_{x0}}\omega_{0}$
is located below the Fiske resonances allowing for its clear resolution.
For wider junctions the behavior is more complicated, because in this
case the Fiske resonances are located both above and below the spin-wave
region and some of them may fall inside this region. A very special
situation is realized for the particular junction size $L_{\mathrm{res}}$, at
which the the first Fiske resonance is very close to $\omega_{0}$.
To estimate this junction size, we substitute the maximum value of
$\mathrm{Re}\left(\eta_{\omega}\right)$ $\sim\!\sqrt{\lambda/\zeta_{0}}\left(\mu_{x0}\!-\!1\right)^{1/4}$
to Eq.~\eqref{eq:ResonFreqs} at $n=1$ yielding
\begin{equation}
L_{\mathrm{res}}=\pi\lambda_{J}\frac{\omega_{p}}{\omega_{0}}\sqrt{\frac{\mathfrak{t}\!+\!\lambda_{1}\!+\!\sqrt{\mu_{x0}}\lambda_{2}}{\mathfrak{t}\!+\!\lambda_{1}\!+\!\left(\mu_{x0}\!-\!1\right)^{1/4}\lambda_{2}^{3/2}/\sqrt{\zeta_{0}}}}.
\label{eq:Lres}
\end{equation}
For this size, at the Josephson frequency slightly below $\omega_{0}$
the excited cavity mode generates the strongest spin wave inside the
magnetic superconductor.

\begin{figure*}
\includegraphics[width=5.8in]{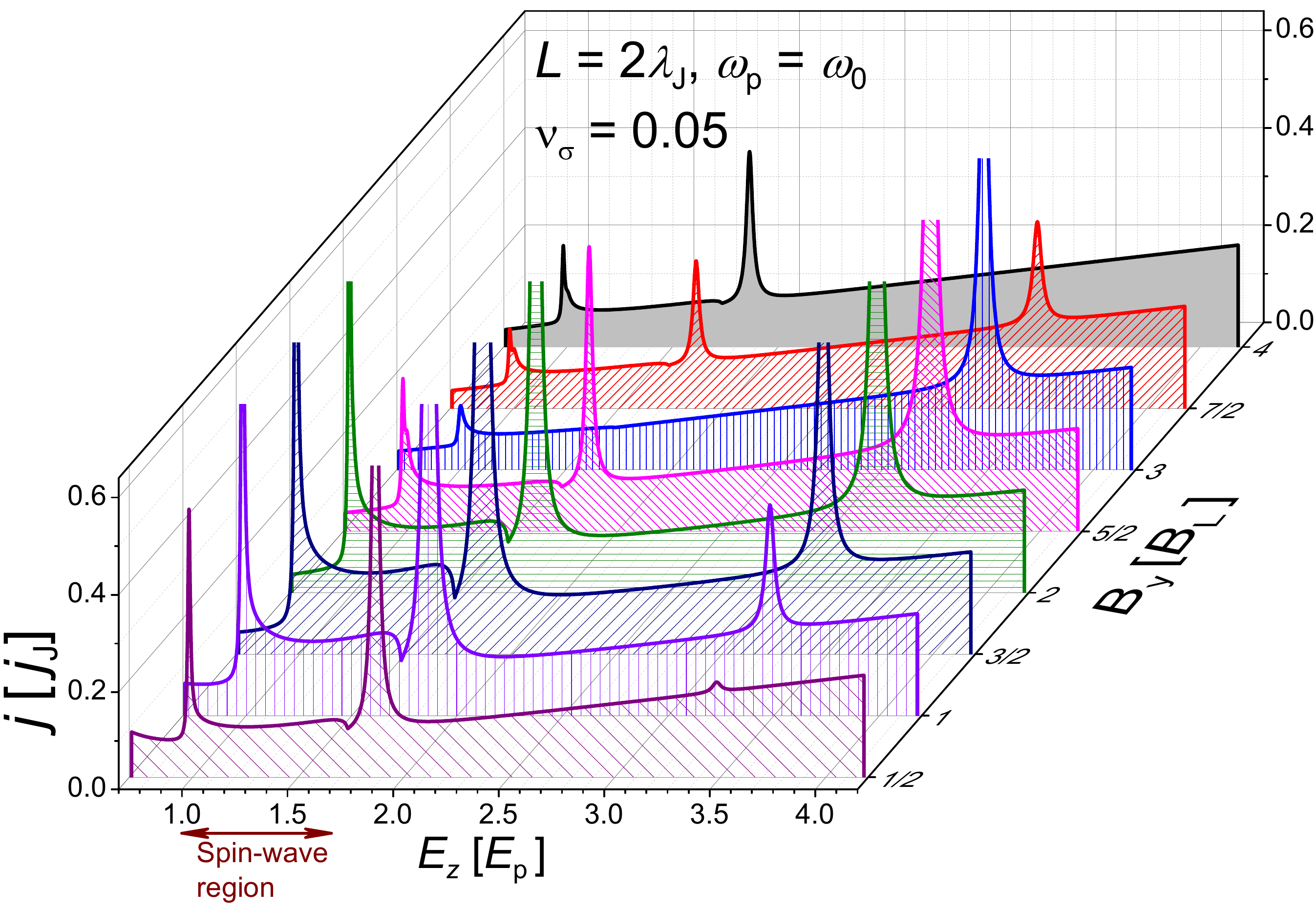}
\caption{The magnetic-field evolution of the current-voltage characteristics for
junction with the parameters shown in the plot. The spin-wave feature is located in the region $1\!\lesssim\!E_z/E_p\!\lesssim \!1.7$. As the Fiske resonances, it is strongly modulated by the magnetic field. }
\label{Fig:z002nu005mu3rb1L2Waterfall}
\end{figure*}
Figure \ref{Fig:IVz002nu005mu3rp1B1BL} shows the representative current-voltage
characteristics computed for the parameters $\omega_{p}\!=\!\omega_{0}$,
$\nu_{\sigma}\!=0.05$, three different sizes, $L/\lambda_{J}\!=$1,
$2$, and $3$, and the magnetic field $B_{y}=B_{L}$. For reference,
we also show by the dashed lines the current-voltage characteristics
computed without dynamic magnetic response using static junction parameters.
Note that (i) only the ascending left-side parts of the peaks are usually observed experimentally and (ii) the used linear approximation breaks in the middle of resonances meaning that the approximation overestimates the peak heights. 
We can see that there are substantial qualitative differences between
the three shown cases. The junction size for the smallest junction is smaller
than $L_{c}$ and therefore the spin-wave region is well below the
Fiske resonances. The spin-wave feature has the same asymmetric shape
as the surface resistivity in Fig.~\ref{Fig:Rs}, it has a sharp peak
when the Josephson frequency matches $\omega_{0}$ followed by an
extended tail up to frequency $\sqrt{\mu_{x0}}\omega_{0}$, see the
inset in the bottom plot. The junction size $2\lambda_{J}$ (middle
plot) is very close to the resonance value $L_{\mathrm{res}}$, Eq.~\eqref{eq:Lres},
meaning that the spin-wave resonance at $\omega=\omega_{0}$ coincides
with the first Fiske resonance leading to the very strong peak. A very
peculiar feature of this case is that, due to strongly nonmonotonic behavior of $\mathrm{Re}(\eta_{\omega})$
near the frequency $\omega_{0}$, the condition for the first resonance
in Eq.~\eqref{eq:ResonFreqs} is satisfied at two frequencies, slightly
below $\omega_{0}$ and slightly above $\sqrt{\mu_{x0}}\omega_{0}$.
Correspondingly, two strong peaks are realized at both frequencies.
The largest size $3\lambda_{J}$ exceeds both $L_{c}$ and $L_{\mathrm{res}}$
(top plot). The first Fiske resonance in this case is located below
$\omega_{0}$ and is slightly separated from the peak marking the
onset of the spin-wave region. Correspondingly, the spin-wave region
is located in between the first and second Fiske resonances. We also
observe larger amplitude of the spin-wave feature in the region $\omega>\omega_{0}$.
The reason is that the condition for the first resonance in Eq.~\eqref{eq:ResonFreqs}
is also formally satisfied in the range $\omega_{0}<\omega<\sqrt{\mu_{x0}}\omega_{0}$
where the absolute value of $\mathrm{Im}(\eta_{\omega})$ is large marking very strong spin-wave damping of the resonance. 
As the resonance takes place in the overdamped region, it is seen
as a shallow maximum. 

The amplitudes of the Fiske resonances have oscillating dependence on the
magnetic field \cite{KulikJETPLett65,kulik1972josephson,BaroneBook}.
Figure \ref{Fig:z002nu005mu3rb1L2Waterfall} shows the magnetic-field
evolution of the current-voltage characteristics for the junction with
$L=2\lambda_{J}$. We see the familiar modulation of the resonances
with magnetic field but with specific features. We see that the first
two peaks have a similar dependence on the magnetic field, since they
both represent the first Fiske resonance, while the third peak representing
the second Fiske resonance is shifted by a half period. Note that the
maximums of the first two peaks at $B=B_{L}$ and maximum of the third
peak at $B=2B_{L}$ are out of this general trend because they correspond
to Eck resonance, $\omega=c_{si}k_{B}$.

We demonstrated that the AC Josephson effect in a tunneling contact 
between conventional and helical-magnetic superconductor can be utilized for the excitation of spin waves.
Such excitation is most efficient when the Josephson frequency is in the range between the two typical spin-wave frequencies $\omega_0$ and $\sqrt{\mu_{x0}}\omega_0$. In this range the current-voltage characteristic has a distinct feature similar to one in the frequency dependence of the surface resistance. In addition, the spin-wave feature may strongly perturb the shape of Fiske resonances and the power of the excited spin wave may be enhanced when the Fiske resonance falls into the spin-wave region. 

\section{Summary and discussion}

In summary, in this paper we consider spin waves and related observable
effects in superconductors with helical magnetic order. Most computed specific results
correspond to the structure realized in the iron pnictide RbEuFe$_{4}$As$_{4}$,
in which the moments rotate 90$^{\circ}$ from layer to layer, Fig.~\ref{Fig:helical}.
The key feature of such materials is that the mode coupled with uniform
field corresponds to the maximum frequency of the spin-wave spectrum
with respect to the $c$-axis wave vector. The frequency of this mode is strongly enlarged
by the long-range electromagnetic interactions between the oscillating
magnetic moments and this enlargement rapidly vanishes when the $c$-axis
wave-vector mismatch exceeds the inverse London penetration depth, see Fig.~\ref{Fig:Spectrum}.
For the parameters of RbEuFe$_{4}$As$_{4}$, we estimate the bare
uniform-mode frequency $f_{0}$ as $\sim$11 GHz and the renormalized one as
$\sim$19 GHz, meaning that these frequencies are located within a convenient
microwave range. We evaluate the frequency dependence of the surface resistance
and find that it has a very distinct asymmetric spin-wave feature
spreading between the bare and renormalized frequencies, see Fig.~\ref{Fig:Rs}.

We also investigate excitation of spin waves with the AC Josephson effect in
a tunneling contact between helical-magnetic and conventional superconductors. For the most efficient excitation of spin waves, the Josephson plasma frequency has to be smaller than the bare uniform-mode frequency $\omega_0$. 
In addition, the features in the current-voltage characteristics are very sensitive to the junction size due to the interplay between the spin-wave excitation and Fiske resonances. The simplest behavior is realized in small-size junctions, when the renormalized frequency $\sqrt{\mu_{x0}}\omega_0$ is below the lowest Fiske resonance. In this case, the whole spin-wave region is separated from the Fiske resonances and has a strongly asymmetric shape resembling the feature in the surface resistivity, see the inset in Fig.\ \ref{Fig:IVz002nu005mu3rp1B1BL}(bottom). In larger junctions, the Fiske resonances may fall inside the spin-wave region leading to more complicated behavior, see Fig.\ \ref{Fig:IVz002nu005mu3rp1B1BL}(top and middle). The strongest excitation of the spin wave can be achieved in the situation when the Fiske resonance frequency is slightly below $\omega_0$ corresponding to the junction size in Eq.\ \eqref{eq:Lres}.
As the Fiske resonances, the shape and amplitude of the spin-wave feature are modulated by magnetic field, see Fig.\ \ref{Fig:z002nu005mu3rb1L2Waterfall}. We conclude that the AC Josephson effect provides a unique way to excite and manipulate spin waves in magnetic superconductors.

The author would like to acknowledge discussions with Ulrich Welp, Wai K. Kwok, Yi Li, and Valentine Novosad on possible experimental observations of the effects discussed in this paper. This work was supported by the US Department of Energy, Office of
Science, Basic Energy Sciences, Materials Sciences and Engineering Division.

\bibliography{SpinWavesSC}

\begin{thebibliography}{76}%
\makeatletter
\providecommand \@ifxundefined [1]{%
 \@ifx{#1\undefined}
}%
\providecommand \@ifnum [1]{%
 \ifnum #1\expandafter \@firstoftwo
 \else \expandafter \@secondoftwo
 \fi
}%
\providecommand \@ifx [1]{%
 \ifx #1\expandafter \@firstoftwo
 \else \expandafter \@secondoftwo
 \fi
}%
\providecommand \natexlab [1]{#1}%
\providecommand \enquote  [1]{``#1''}%
\providecommand \bibnamefont  [1]{#1}%
\providecommand \bibfnamefont [1]{#1}%
\providecommand \citenamefont [1]{#1}%
\providecommand \href@noop [0]{\@secondoftwo}%
\providecommand \href [0]{\begingroup \@sanitize@url \@href}%
\providecommand \@href[1]{\@@startlink{#1}\@@href}%
\providecommand \@@href[1]{\endgroup#1\@@endlink}%
\providecommand \@sanitize@url [0]{\catcode `\\12\catcode `\$12\catcode
  `\&12\catcode `\#12\catcode `\^12\catcode `\_12\catcode `\%12\relax}%
\providecommand \@@startlink[1]{}%
\providecommand \@@endlink[0]{}%
\providecommand \url  [0]{\begingroup\@sanitize@url \@url }%
\providecommand \@url [1]{\endgroup\@href {#1}{\urlprefix }}%
\providecommand \urlprefix  [0]{URL }%
\providecommand \Eprint [0]{\href }%
\providecommand \doibase [0]{https://doi.org/}%
\providecommand \selectlanguage [0]{\@gobble}%
\providecommand \bibinfo  [0]{\@secondoftwo}%
\providecommand \bibfield  [0]{\@secondoftwo}%
\providecommand \translation [1]{[#1]}%
\providecommand \BibitemOpen [0]{}%
\providecommand \bibitemStop [0]{}%
\providecommand \bibitemNoStop [0]{.\EOS\space}%
\providecommand \EOS [0]{\spacefactor3000\relax}%
\providecommand \BibitemShut  [1]{\csname bibitem#1\endcsname}%
\let\auto@bib@innerbib\@empty
\bibitem [{\citenamefont {Bulaevskii}\ \emph {et~al.}(1985)\citenamefont
  {Bulaevskii}, \citenamefont {Buzdin}, \citenamefont {Kuli{\'{c}}},\ and\
  \citenamefont {Panjukov}}]{BulaevskiiAdvPhys85}%
  \BibitemOpen
  \bibfield  {author} {\bibinfo {author} {\bibfnamefont {L.}~\bibnamefont
  {Bulaevskii}}, \bibinfo {author} {\bibfnamefont {A.}~\bibnamefont {Buzdin}},
  \bibinfo {author} {\bibfnamefont {M.}~\bibnamefont {Kuli{\'{c}}}},\ and\
  \bibinfo {author} {\bibfnamefont {S.}~\bibnamefont {Panjukov}},\ }\bibfield
  {title} {\bibinfo {title} {Coexistence of superconductivity and magnetism
  theoretical predictions and experimental results},\ }\href
  {https://doi.org/10.1080/00018738500101741} {\bibfield  {journal} {\bibinfo
  {journal} {Adv. Phys.}\ }\textbf {\bibinfo {volume} {34}},\ \bibinfo {pages}
  {175} (\bibinfo {year} {1985})}\BibitemShut {NoStop}%
\bibitem [{\citenamefont {Kuli\'{c}}\ and\ \citenamefont
  {Buzdin}(2008)}]{KulicBuzdinSupercondBook}%
  \BibitemOpen
  \bibfield  {author} {\bibinfo {author} {\bibfnamefont {M.}~\bibnamefont
  {Kuli\'{c}}}\ and\ \bibinfo {author} {\bibfnamefont {A.~I.}\ \bibnamefont
  {Buzdin}},\ }\href@noop {} {\emph {\bibinfo {title} {Superconductivity}}},\
  edited by\ \bibinfo {editor} {\bibfnamefont {K.~H.}\ \bibnamefont
  {Bennemann}}\ and\ \bibinfo {editor} {\bibfnamefont {J.~B.}\ \bibnamefont
  {Ketterson}}\ (\bibinfo  {publisher} {Springer},\ \bibinfo {address}
  {Berlin},\ \bibinfo {year} {2008})\ Chap.\ \bibinfo {chapter} {4. Coexistence
  of Singlet Superconductivity andMagnetic Order in Bulk Magnetic
  Superconductors and SF Heterostructures}, p.\ \bibinfo {pages}
  {163}\BibitemShut {NoStop}%
\bibitem [{\citenamefont {Wolowiec}\ \emph {et~al.}(2015)\citenamefont
  {Wolowiec}, \citenamefont {White},\ and\ \citenamefont
  {Maple}}]{WolowiecPhysC15}%
  \BibitemOpen
  \bibfield  {author} {\bibinfo {author} {\bibfnamefont {C.~T.}\ \bibnamefont
  {Wolowiec}}, \bibinfo {author} {\bibfnamefont {B.~D.}\ \bibnamefont
  {White}},\ and\ \bibinfo {author} {\bibfnamefont {M.~B.}\ \bibnamefont
  {Maple}},\ }\bibfield  {title} {\bibinfo {title} {Conventional magnetic
  superconductors},\ }\href
  {https://doi.org/https://doi.org/10.1016/j.physc.2015.02.050} {\bibfield
  {journal} {\bibinfo  {journal} {Physica C}\ }\textbf {\bibinfo {volume}
  {514}},\ \bibinfo {pages} {113 } (\bibinfo {year} {2015})}\BibitemShut
  {NoStop}%
\bibitem [{\citenamefont {Maple}\ and\ \citenamefont
  {Fischer}(1982)}]{MapleFischerBook1982}%
  \BibitemOpen
  \bibinfo {editor} {\bibfnamefont {M.~B.}\ \bibnamefont {Maple}}\ and\
  \bibinfo {editor} {\bibfnamefont {{\O}.}~\bibnamefont {Fischer}},\ eds.,\
  \href@noop {} {\emph {\bibinfo {title} {Superconductivity in Ternary
  Compounds II, Superconductivity and Magnetism}}}\ (\bibinfo  {publisher}
  {Springer},\ \bibinfo {address} {Berlin},\ \bibinfo {year}
  {1982})\BibitemShut {NoStop}%
\bibitem [{\citenamefont {Anderson}\ and\ \citenamefont
  {Suhl}(1959)}]{AndersonPhysRev.116.898}%
  \BibitemOpen
  \bibfield  {author} {\bibinfo {author} {\bibfnamefont {P.~W.}\ \bibnamefont
  {Anderson}}\ and\ \bibinfo {author} {\bibfnamefont {H.}~\bibnamefont
  {Suhl}},\ }\bibfield  {title} {\bibinfo {title} {Spin alignment in the
  superconducting state},\ }\href {https://doi.org/10.1103/PhysRev.116.898}
  {\bibfield  {journal} {\bibinfo  {journal} {Phys. Rev.}\ }\textbf {\bibinfo
  {volume} {116}},\ \bibinfo {pages} {898} (\bibinfo {year}
  {1959})}\BibitemShut {NoStop}%
\bibitem [{\citenamefont {Bulaevskii}\ \emph {et~al.}(1980)\citenamefont
  {Bulaevskii}, \citenamefont {Rusinov},\ and\ \citenamefont
  {Kuli{\'{c}}}}]{BulaevskiiJLTP1980}%
  \BibitemOpen
  \bibfield  {author} {\bibinfo {author} {\bibfnamefont {L.~N.}\ \bibnamefont
  {Bulaevskii}}, \bibinfo {author} {\bibfnamefont {A.~I.}\ \bibnamefont
  {Rusinov}},\ and\ \bibinfo {author} {\bibfnamefont {M.}~\bibnamefont
  {Kuli{\'{c}}}},\ }\bibfield  {title} {\bibinfo {title} {Helical ordering of
  spins in a superconductor},\ }\href {https://doi.org/10.1007/BF00115620}
  {\bibfield  {journal} {\bibinfo  {journal} {J. Low Temp. Phys.}\ }\textbf
  {\bibinfo {volume} {39}},\ \bibinfo {pages} {255} (\bibinfo {year}
  {1980})}\BibitemShut {NoStop}%
\bibitem [{\citenamefont {Ishikawa}\ and\ \citenamefont
  {Fischer}(1977)}]{IshikawaSSC77}%
  \BibitemOpen
  \bibfield  {author} {\bibinfo {author} {\bibfnamefont {M.}~\bibnamefont
  {Ishikawa}}\ and\ \bibinfo {author} {\bibfnamefont {{\O}.}~\bibnamefont
  {Fischer}},\ }\bibfield  {title} {\bibinfo {title} {Destruction of
  superconductivity by magnetic ordering in {Ho$_{1.2}$Mo$_6$S$_8$}},\ }\href
  {https://doi.org/https://doi.org/10.1016/0038-1098(77)90625-1} {\bibfield
  {journal} {\bibinfo  {journal} {Solid State Commun.}\ }\textbf {\bibinfo
  {volume} {23}},\ \bibinfo {pages} {37 } (\bibinfo {year} {1977})}\BibitemShut
  {NoStop}%
\bibitem [{\citenamefont {Fertig}\ \emph {et~al.}(1977)\citenamefont {Fertig},
  \citenamefont {Johnston}, \citenamefont {DeLong}, \citenamefont {McCallum},
  \citenamefont {Maple},\ and\ \citenamefont
  {Matthias}}]{FertigPhysRevLett.38.987}%
  \BibitemOpen
  \bibfield  {author} {\bibinfo {author} {\bibfnamefont {W.~A.}\ \bibnamefont
  {Fertig}}, \bibinfo {author} {\bibfnamefont {D.~C.}\ \bibnamefont
  {Johnston}}, \bibinfo {author} {\bibfnamefont {L.~E.}\ \bibnamefont
  {DeLong}}, \bibinfo {author} {\bibfnamefont {R.~W.}\ \bibnamefont
  {McCallum}}, \bibinfo {author} {\bibfnamefont {M.~B.}\ \bibnamefont
  {Maple}},\ and\ \bibinfo {author} {\bibfnamefont {B.~T.}\ \bibnamefont
  {Matthias}},\ }\bibfield  {title} {\bibinfo {title} {Destruction of
  superconductivity at the onset of long-range magnetic order in the compound
  {Er${\mathrm{Rh}}_{4}$${\mathrm{B}}_{4}$}},\ }\href
  {https://doi.org/10.1103/PhysRevLett.38.987} {\bibfield  {journal} {\bibinfo
  {journal} {Phys. Rev. Lett.}\ }\textbf {\bibinfo {volume} {38}},\ \bibinfo
  {pages} {987} (\bibinfo {year} {1977})}\BibitemShut {NoStop}%
\bibitem [{\citenamefont {Lynn}\ \emph
  {et~al.}(1981{\natexlab{a}})\citenamefont {Lynn}, \citenamefont {Shirane},
  \citenamefont {Thomlinson},\ and\ \citenamefont
  {Shelton}}]{LynnPhysRevLett.46.368}%
  \BibitemOpen
  \bibfield  {author} {\bibinfo {author} {\bibfnamefont {J.~W.}\ \bibnamefont
  {Lynn}}, \bibinfo {author} {\bibfnamefont {G.}~\bibnamefont {Shirane}},
  \bibinfo {author} {\bibfnamefont {W.}~\bibnamefont {Thomlinson}},\ and\
  \bibinfo {author} {\bibfnamefont {R.~N.}\ \bibnamefont {Shelton}},\
  }\bibfield  {title} {\bibinfo {title} {Competition between ferromagnetism and
  superconductivity in {Ho${\mathrm{Mo}}_{6}$${\mathrm{S}}_{8}$}},\ }\href
  {https://doi.org/10.1103/PhysRevLett.46.368} {\bibfield  {journal} {\bibinfo
  {journal} {Phys. Rev. Lett.}\ }\textbf {\bibinfo {volume} {46}},\ \bibinfo
  {pages} {368} (\bibinfo {year} {1981}{\natexlab{a}})}\BibitemShut {NoStop}%
\bibitem [{\citenamefont {Lynn}\ \emph
  {et~al.}(1981{\natexlab{b}})\citenamefont {Lynn}, \citenamefont {Shirane},
  \citenamefont {Thomlinson}, \citenamefont {Shelton},\ and\ \citenamefont
  {Moncton}}]{LynnPhysRevB.24.3817}%
  \BibitemOpen
  \bibfield  {author} {\bibinfo {author} {\bibfnamefont {J.~W.}\ \bibnamefont
  {Lynn}}, \bibinfo {author} {\bibfnamefont {G.}~\bibnamefont {Shirane}},
  \bibinfo {author} {\bibfnamefont {W.}~\bibnamefont {Thomlinson}}, \bibinfo
  {author} {\bibfnamefont {R.~N.}\ \bibnamefont {Shelton}},\ and\ \bibinfo
  {author} {\bibfnamefont {D.~E.}\ \bibnamefont {Moncton}},\ }\bibfield
  {title} {\bibinfo {title} {Magnetic properties of the reentrant ferromagnetic
  superconductor {Ho${\mathrm{Mo}}_{6}$${\mathrm{S}}_{8}$}},\ }\href
  {https://doi.org/10.1103/PhysRevB.24.3817} {\bibfield  {journal} {\bibinfo
  {journal} {Phys. Rev. B}\ }\textbf {\bibinfo {volume} {24}},\ \bibinfo
  {pages} {3817} (\bibinfo {year} {1981}{\natexlab{b}})}\BibitemShut {NoStop}%
\bibitem [{\citenamefont {Moncton}\ \emph {et~al.}(1980)\citenamefont
  {Moncton}, \citenamefont {McWhan}, \citenamefont {Schmidt}, \citenamefont
  {Shirane}, \citenamefont {Thomlinson}, \citenamefont {Maple}, \citenamefont
  {MacKay}, \citenamefont {Woolf}, \citenamefont {Fisk},\ and\ \citenamefont
  {Johnston}}]{MonctonPhysRevLett.45.2060}%
  \BibitemOpen
  \bibfield  {author} {\bibinfo {author} {\bibfnamefont {D.~E.}\ \bibnamefont
  {Moncton}}, \bibinfo {author} {\bibfnamefont {D.~B.}\ \bibnamefont {McWhan}},
  \bibinfo {author} {\bibfnamefont {P.~H.}\ \bibnamefont {Schmidt}}, \bibinfo
  {author} {\bibfnamefont {G.}~\bibnamefont {Shirane}}, \bibinfo {author}
  {\bibfnamefont {W.}~\bibnamefont {Thomlinson}}, \bibinfo {author}
  {\bibfnamefont {M.~B.}\ \bibnamefont {Maple}}, \bibinfo {author}
  {\bibfnamefont {H.~B.}\ \bibnamefont {MacKay}}, \bibinfo {author}
  {\bibfnamefont {L.~D.}\ \bibnamefont {Woolf}}, \bibinfo {author}
  {\bibfnamefont {Z.}~\bibnamefont {Fisk}},\ and\ \bibinfo {author}
  {\bibfnamefont {D.~C.}\ \bibnamefont {Johnston}},\ }\bibfield  {title}
  {\bibinfo {title} {Oscillatory magnetic fluctuations near the
  superconductor-to-ferromagnet transition in
  er${\mathrm{rh}}_{4}$${\mathrm{b}}_{4}$},\ }\href
  {https://doi.org/10.1103/PhysRevLett.45.2060} {\bibfield  {journal} {\bibinfo
   {journal} {Phys. Rev. Lett.}\ }\textbf {\bibinfo {volume} {45}},\ \bibinfo
  {pages} {2060} (\bibinfo {year} {1980})}\BibitemShut {NoStop}%
\bibitem [{\citenamefont {M{\"u}ller}\ and\ \citenamefont
  {Narozhnyi}(2001)}]{MullerRoPP2001}%
  \BibitemOpen
  \bibfield  {author} {\bibinfo {author} {\bibfnamefont {K.-H.}\ \bibnamefont
  {M{\"u}ller}}\ and\ \bibinfo {author} {\bibfnamefont {V.~N.}\ \bibnamefont
  {Narozhnyi}},\ }\bibfield  {title} {\bibinfo {title} {Interaction of
  superconductivity and magnetism in borocarbide superconductors},\ }\href
  {https://doi.org/10.1088/0034-4885/64/8/202} {\bibfield  {journal} {\bibinfo
  {journal} {Rep. Prog. Phys.}\ }\textbf {\bibinfo {volume} {64}},\ \bibinfo
  {pages} {943} (\bibinfo {year} {2001})}\BibitemShut {NoStop}%
\bibitem [{\citenamefont {Gupta}(2006)}]{GuptaAdvPhys06}%
  \BibitemOpen
  \bibfield  {author} {\bibinfo {author} {\bibfnamefont {L.~C.}\ \bibnamefont
  {Gupta}},\ }\bibfield  {title} {\bibinfo {title} {Superconductivity and
  magnetism and their interplay in quaternary borocarbides {RNi$_2$B$_2$C}},\
  }\href {https://doi.org/10.1080/00018730601061130} {\bibfield  {journal}
  {\bibinfo  {journal} {Adv. Phys.}\ }\textbf {\bibinfo {volume} {55}},\
  \bibinfo {pages} {691} (\bibinfo {year} {2006})}\BibitemShut {NoStop}%
\bibitem [{\citenamefont {Mazumdar}\ and\ \citenamefont
  {Nagarajan}(2015)}]{MazumdarPhysC2015}%
  \BibitemOpen
  \bibfield  {author} {\bibinfo {author} {\bibfnamefont {C.}~\bibnamefont
  {Mazumdar}}\ and\ \bibinfo {author} {\bibfnamefont {R.}~\bibnamefont
  {Nagarajan}},\ }\bibfield  {title} {\bibinfo {title} {Quaternary
  borocarbides: Relatively high {T$_c$} intermetallic superconductors and
  magnetic superconductors},\ }\href
  {http://www.sciencedirect.com/science/article/pii/S0921453415000611}
  {\bibfield  {journal} {\bibinfo  {journal} {Physica C}\ }\textbf {\bibinfo
  {volume} {514}},\ \bibinfo {pages} {173} (\bibinfo {year}
  {2015})}\BibitemShut {NoStop}%
\bibitem [{\citenamefont {Aoki}\ and\ \citenamefont
  {Flouquet}(2012)}]{AokiJPSJ12}%
  \BibitemOpen
  \bibfield  {author} {\bibinfo {author} {\bibfnamefont {D.}~\bibnamefont
  {Aoki}}\ and\ \bibinfo {author} {\bibfnamefont {J.}~\bibnamefont
  {Flouquet}},\ }\bibfield  {title} {\bibinfo {title} {Ferromagnetism and
  superconductivity in uranium compounds},\ }\href
  {https://doi.org/10.1143/JPSJ.81.011003} {\bibfield  {journal} {\bibinfo
  {journal} {J. Phys. Soc. Jpn.}\ }\textbf {\bibinfo {volume} {81}},\ \bibinfo
  {pages} {011003} (\bibinfo {year} {2012})}\BibitemShut {NoStop}%
\bibitem [{\citenamefont {Aoki}\ \emph {et~al.}(2019)\citenamefont {Aoki},
  \citenamefont {Ishida},\ and\ \citenamefont {Flouquet}}]{AokiJPSJ19}%
  \BibitemOpen
  \bibfield  {author} {\bibinfo {author} {\bibfnamefont {D.}~\bibnamefont
  {Aoki}}, \bibinfo {author} {\bibfnamefont {K.}~\bibnamefont {Ishida}},\ and\
  \bibinfo {author} {\bibfnamefont {J.}~\bibnamefont {Flouquet}},\ }\bibfield
  {title} {\bibinfo {title} {Review of {U}-based ferromagnetic superconductors:
  Comparison between {UGe$_2$}, {URhGe}, and {UCoGe}},\ }\href
  {https://doi.org/10.7566/JPSJ.88.022001} {\bibfield  {journal} {\bibinfo
  {journal} {J. Phys. Soc. Jpn.}\ }\textbf {\bibinfo {volume} {88}},\ \bibinfo
  {pages} {022001} (\bibinfo {year} {2019})}\BibitemShut {NoStop}%
\bibitem [{\citenamefont {Huxley}(2015)}]{HuxleyPhysC2015}%
  \BibitemOpen
  \bibfield  {author} {\bibinfo {author} {\bibfnamefont {A.~D.}\ \bibnamefont
  {Huxley}},\ }\bibfield  {title} {\bibinfo {title} {Ferromagnetic
  superconductors},\ }\href
  {http://www.sciencedirect.com/science/article/pii/S0921453415000532}
  {\bibfield  {journal} {\bibinfo  {journal} {Physica C}\ }\textbf {\bibinfo
  {volume} {514}},\ \bibinfo {pages} {368} (\bibinfo {year}
  {2015})}\BibitemShut {NoStop}%
\bibitem [{\citenamefont {Ran}\ \emph {et~al.}(2019)\citenamefont {Ran},
  \citenamefont {Eckberg}, \citenamefont {Ding}, \citenamefont {Furukawa},
  \citenamefont {Metz}, \citenamefont {Saha}, \citenamefont {Liu},
  \citenamefont {Zic}, \citenamefont {Kim}, \citenamefont {Paglione},\ and\
  \citenamefont {Butch}}]{RanScience1019}%
  \BibitemOpen
  \bibfield  {author} {\bibinfo {author} {\bibfnamefont {S.}~\bibnamefont
  {Ran}}, \bibinfo {author} {\bibfnamefont {C.}~\bibnamefont {Eckberg}},
  \bibinfo {author} {\bibfnamefont {Q.-P.}\ \bibnamefont {Ding}}, \bibinfo
  {author} {\bibfnamefont {Y.}~\bibnamefont {Furukawa}}, \bibinfo {author}
  {\bibfnamefont {T.}~\bibnamefont {Metz}}, \bibinfo {author} {\bibfnamefont
  {S.~R.}\ \bibnamefont {Saha}}, \bibinfo {author} {\bibfnamefont {I.-L.}\
  \bibnamefont {Liu}}, \bibinfo {author} {\bibfnamefont {M.}~\bibnamefont
  {Zic}}, \bibinfo {author} {\bibfnamefont {H.}~\bibnamefont {Kim}}, \bibinfo
  {author} {\bibfnamefont {J.}~\bibnamefont {Paglione}},\ and\ \bibinfo
  {author} {\bibfnamefont {N.~P.}\ \bibnamefont {Butch}},\ }\bibfield  {title}
  {\bibinfo {title} {Nearly ferromagnetic spin-triplet superconductivity},\
  }\href {https://doi.org/10.1126/science.aav8645} {\bibfield  {journal}
  {\bibinfo  {journal} {Science}\ }\textbf {\bibinfo {volume} {365}},\ \bibinfo
  {pages} {684} (\bibinfo {year} {2019})}\BibitemShut {NoStop}%
\bibitem [{\citenamefont {Zapf}\ and\ \citenamefont
  {Dressel}(2017)}]{Zapf2017}%
  \BibitemOpen
  \bibfield  {author} {\bibinfo {author} {\bibfnamefont {S.}~\bibnamefont
  {Zapf}}\ and\ \bibinfo {author} {\bibfnamefont {M.}~\bibnamefont {Dressel}},\
  }\bibfield  {title} {\bibinfo {title} {Europium-based iron pnictides: a
  unique laboratory for magnetism, superconductivity and structural effects},\
  }\href {http://stacks.iop.org/0034-4885/80/i=1/a=016501} {\bibfield
  {journal} {\bibinfo  {journal} {Rep. Prog. Phys.}\ }\textbf {\bibinfo
  {volume} {80}},\ \bibinfo {pages} {016501} (\bibinfo {year}
  {2017})}\BibitemShut {NoStop}%
\bibitem [{\citenamefont {Ren}\ \emph {et~al.}(2008)\citenamefont {Ren},
  \citenamefont {Zhu}, \citenamefont {Jiang}, \citenamefont {Xu}, \citenamefont
  {Tao}, \citenamefont {Wang}, \citenamefont {Feng}, \citenamefont {Cao},\ and\
  \citenamefont {Xu}}]{Ren2008}%
  \BibitemOpen
  \bibfield  {author} {\bibinfo {author} {\bibfnamefont {Z.}~\bibnamefont
  {Ren}}, \bibinfo {author} {\bibfnamefont {Z.}~\bibnamefont {Zhu}}, \bibinfo
  {author} {\bibfnamefont {S.}~\bibnamefont {Jiang}}, \bibinfo {author}
  {\bibfnamefont {X.}~\bibnamefont {Xu}}, \bibinfo {author} {\bibfnamefont
  {Q.}~\bibnamefont {Tao}}, \bibinfo {author} {\bibfnamefont {C.}~\bibnamefont
  {Wang}}, \bibinfo {author} {\bibfnamefont {C.}~\bibnamefont {Feng}}, \bibinfo
  {author} {\bibfnamefont {G.}~\bibnamefont {Cao}},\ and\ \bibinfo {author}
  {\bibfnamefont {Z.}~\bibnamefont {Xu}},\ }\bibfield  {title} {\bibinfo
  {title} {Antiferromagnetic transition in
  {${\text{EuFe}}_{2}{\text{As}}_{2}$}: A possible parent compound for
  superconductors},\ }\href {https://doi.org/10.1103/PhysRevB.78.052501}
  {\bibfield  {journal} {\bibinfo  {journal} {Phys. Rev. B}\ }\textbf {\bibinfo
  {volume} {78}},\ \bibinfo {pages} {052501} (\bibinfo {year}
  {2008})}\BibitemShut {NoStop}%
\bibitem [{\citenamefont {Jeevan}\ \emph
  {et~al.}(2008{\natexlab{a}})\citenamefont {Jeevan}, \citenamefont {Hossain},
  \citenamefont {Kasinathan}, \citenamefont {Rosner}, \citenamefont {Geibel},\
  and\ \citenamefont {Gegenwart}}]{Jeevan2008a}%
  \BibitemOpen
  \bibfield  {author} {\bibinfo {author} {\bibfnamefont {H.~S.}\ \bibnamefont
  {Jeevan}}, \bibinfo {author} {\bibfnamefont {Z.}~\bibnamefont {Hossain}},
  \bibinfo {author} {\bibfnamefont {D.}~\bibnamefont {Kasinathan}}, \bibinfo
  {author} {\bibfnamefont {H.}~\bibnamefont {Rosner}}, \bibinfo {author}
  {\bibfnamefont {C.}~\bibnamefont {Geibel}},\ and\ \bibinfo {author}
  {\bibfnamefont {P.}~\bibnamefont {Gegenwart}},\ }\bibfield  {title} {\bibinfo
  {title} {Electrical resistivity and specific heat of single-crystalline
  {${\text{EuFe}}_{2}{\text{As}}_{2}$}: A magnetic homologue of
  {${\text{SrFe}}_{2}{\text{As}}_{2}$}},\ }\href
  {https://doi.org/10.1103/PhysRevB.78.052502} {\bibfield  {journal} {\bibinfo
  {journal} {Phys. Rev. B}\ }\textbf {\bibinfo {volume} {78}},\ \bibinfo
  {pages} {052502} (\bibinfo {year} {2008}{\natexlab{a}})}\BibitemShut
  {NoStop}%
\bibitem [{\citenamefont {Xiao}\ \emph {et~al.}(2010)\citenamefont {Xiao},
  \citenamefont {Su}, \citenamefont {Schmidt}, \citenamefont {Schmalzl},
  \citenamefont {Kumar}, \citenamefont {Price}, \citenamefont {Chatterji},
  \citenamefont {Mittal}, \citenamefont {Chang}, \citenamefont {Nandi},
  \citenamefont {Kumar}, \citenamefont {Dhar}, \citenamefont {Thamizhavel},\
  and\ \citenamefont {Brueckel}}]{Xiao2010}%
  \BibitemOpen
  \bibfield  {author} {\bibinfo {author} {\bibfnamefont {Y.}~\bibnamefont
  {Xiao}}, \bibinfo {author} {\bibfnamefont {Y.}~\bibnamefont {Su}}, \bibinfo
  {author} {\bibfnamefont {W.}~\bibnamefont {Schmidt}}, \bibinfo {author}
  {\bibfnamefont {K.}~\bibnamefont {Schmalzl}}, \bibinfo {author}
  {\bibfnamefont {C.~M.~N.}\ \bibnamefont {Kumar}}, \bibinfo {author}
  {\bibfnamefont {S.}~\bibnamefont {Price}}, \bibinfo {author} {\bibfnamefont
  {T.}~\bibnamefont {Chatterji}}, \bibinfo {author} {\bibfnamefont
  {R.}~\bibnamefont {Mittal}}, \bibinfo {author} {\bibfnamefont {L.~J.}\
  \bibnamefont {Chang}}, \bibinfo {author} {\bibfnamefont {S.}~\bibnamefont
  {Nandi}}, \bibinfo {author} {\bibfnamefont {N.}~\bibnamefont {Kumar}},
  \bibinfo {author} {\bibfnamefont {S.~K.}\ \bibnamefont {Dhar}}, \bibinfo
  {author} {\bibfnamefont {A.}~\bibnamefont {Thamizhavel}},\ and\ \bibinfo
  {author} {\bibfnamefont {T.}~\bibnamefont {Brueckel}},\ }\bibfield  {title}
  {\bibinfo {title} {Field-induced spin reorientation and giant spin-lattice
  coupling in {${\text{EuFe}}_{2}{\text{As}}_{2}$}},\ }\href
  {https://doi.org/10.1103/PhysRevB.81.220406} {\bibfield  {journal} {\bibinfo
  {journal} {Phys. Rev. B}\ }\textbf {\bibinfo {volume} {81}},\ \bibinfo
  {pages} {220406(R)} (\bibinfo {year} {2010})}\BibitemShut {NoStop}%
\bibitem [{\citenamefont {Miclea}\ \emph {et~al.}(2009)\citenamefont {Miclea},
  \citenamefont {Nicklas}, \citenamefont {Jeevan}, \citenamefont {Kasinathan},
  \citenamefont {Hossain}, \citenamefont {Rosner}, \citenamefont {Gegenwart},
  \citenamefont {Geibel},\ and\ \citenamefont
  {Steglich}}]{MicleaPhysRevB.79.212509}%
  \BibitemOpen
  \bibfield  {author} {\bibinfo {author} {\bibfnamefont {C.~F.}\ \bibnamefont
  {Miclea}}, \bibinfo {author} {\bibfnamefont {M.}~\bibnamefont {Nicklas}},
  \bibinfo {author} {\bibfnamefont {H.~S.}\ \bibnamefont {Jeevan}}, \bibinfo
  {author} {\bibfnamefont {D.}~\bibnamefont {Kasinathan}}, \bibinfo {author}
  {\bibfnamefont {Z.}~\bibnamefont {Hossain}}, \bibinfo {author} {\bibfnamefont
  {H.}~\bibnamefont {Rosner}}, \bibinfo {author} {\bibfnamefont
  {P.}~\bibnamefont {Gegenwart}}, \bibinfo {author} {\bibfnamefont
  {C.}~\bibnamefont {Geibel}},\ and\ \bibinfo {author} {\bibfnamefont
  {F.}~\bibnamefont {Steglich}},\ }\bibfield  {title} {\bibinfo {title}
  {Evidence for a reentrant superconducting state in
  {${\text{EuFe}}_{2}{\text{As}}_{2}$} under pressure},\ }\href
  {https://doi.org/10.1103/PhysRevB.79.212509} {\bibfield  {journal} {\bibinfo
  {journal} {Phys. Rev. B}\ }\textbf {\bibinfo {volume} {79}},\ \bibinfo
  {pages} {212509} (\bibinfo {year} {2009})}\BibitemShut {NoStop}%
\bibitem [{\citenamefont {Terashima}\ \emph {et~al.}(2009)\citenamefont
  {Terashima}, \citenamefont {Kimata}, \citenamefont {Satsukawa}, \citenamefont
  {Harada}, \citenamefont {Hazama}, \citenamefont {Uji}, \citenamefont
  {S.~Suzuki}, \citenamefont {Matsumoto},\ and\ \citenamefont
  {Murata}}]{TerashimaJPSJ2009}%
  \BibitemOpen
  \bibfield  {author} {\bibinfo {author} {\bibfnamefont {T.}~\bibnamefont
  {Terashima}}, \bibinfo {author} {\bibfnamefont {M.}~\bibnamefont {Kimata}},
  \bibinfo {author} {\bibfnamefont {H.}~\bibnamefont {Satsukawa}}, \bibinfo
  {author} {\bibfnamefont {A.}~\bibnamefont {Harada}}, \bibinfo {author}
  {\bibfnamefont {K.}~\bibnamefont {Hazama}}, \bibinfo {author} {\bibfnamefont
  {S.}~\bibnamefont {Uji}}, \bibinfo {author} {\bibfnamefont {H.}~\bibnamefont
  {S.~Suzuki}}, \bibinfo {author} {\bibfnamefont {T.}~\bibnamefont
  {Matsumoto}},\ and\ \bibinfo {author} {\bibfnamefont {K.}~\bibnamefont
  {Murata}},\ }\bibfield  {title} {\bibinfo {title} {{EuFe$_2$As$_2$} under
  high pressure: An antiferromagnetic bulk superconductor},\ }\href
  {https://doi.org/10.1143/jpsj.78.083701} {\bibfield  {journal} {\bibinfo
  {journal} {J. Phys. Soc. Jpn.}\ }\textbf {\bibinfo {volume} {78}},\ \bibinfo
  {pages} {083701} (\bibinfo {year} {2009})}\BibitemShut {NoStop}%
\bibitem [{\citenamefont {Ren}\ \emph {et~al.}(2009)\citenamefont {Ren},
  \citenamefont {Tao}, \citenamefont {Jiang}, \citenamefont {Feng},
  \citenamefont {Wang}, \citenamefont {Dai}, \citenamefont {Cao},\ and\
  \citenamefont {Xu}}]{RenPhysRevLett.102.137002}%
  \BibitemOpen
  \bibfield  {author} {\bibinfo {author} {\bibfnamefont {Z.}~\bibnamefont
  {Ren}}, \bibinfo {author} {\bibfnamefont {Q.}~\bibnamefont {Tao}}, \bibinfo
  {author} {\bibfnamefont {S.}~\bibnamefont {Jiang}}, \bibinfo {author}
  {\bibfnamefont {C.}~\bibnamefont {Feng}}, \bibinfo {author} {\bibfnamefont
  {C.}~\bibnamefont {Wang}}, \bibinfo {author} {\bibfnamefont {J.}~\bibnamefont
  {Dai}}, \bibinfo {author} {\bibfnamefont {G.}~\bibnamefont {Cao}},\ and\
  \bibinfo {author} {\bibfnamefont {Z.}~\bibnamefont {Xu}},\ }\bibfield
  {title} {\bibinfo {title} {Superconductivity induced by phosphorus doping and
  its coexistence with ferromagnetism in
  {${\mathrm{EuFe}}_{2}({\mathrm{As}}_{0.7}{\mathrm{P}}_{0.3}{)}_{2}$}},\
  }\href {https://doi.org/10.1103/PhysRevLett.102.137002} {\bibfield  {journal}
  {\bibinfo  {journal} {Phys. Rev. Lett.}\ }\textbf {\bibinfo {volume} {102}},\
  \bibinfo {pages} {137002} (\bibinfo {year} {2009})}\BibitemShut {NoStop}%
\bibitem [{\citenamefont {Jeevan}\ \emph {et~al.}(2011)\citenamefont {Jeevan},
  \citenamefont {Kasinathan}, \citenamefont {Rosner},\ and\ \citenamefont
  {Gegenwart}}]{Jeevan2011}%
  \BibitemOpen
  \bibfield  {author} {\bibinfo {author} {\bibfnamefont {H.~S.}\ \bibnamefont
  {Jeevan}}, \bibinfo {author} {\bibfnamefont {D.}~\bibnamefont {Kasinathan}},
  \bibinfo {author} {\bibfnamefont {H.}~\bibnamefont {Rosner}},\ and\ \bibinfo
  {author} {\bibfnamefont {P.}~\bibnamefont {Gegenwart}},\ }\bibfield  {title}
  {\bibinfo {title} {{Interplay of antiferromagnetism, ferromagnetism, and
  superconductivity in
  $\mathrm{Eu}\mathrm{Fe}_{2}(\mathrm{As}_{1-x}\mathrm{P}_{x})_{2}$ single
  crystals}},\ }\href {https://doi.org/10.1103/PhysRevB.83.054511} {\bibfield
  {journal} {\bibinfo  {journal} {Phys. Rev. B}\ }\textbf {\bibinfo {volume}
  {83}},\ \bibinfo {pages} {054511} (\bibinfo {year} {2011})}\BibitemShut
  {NoStop}%
\bibitem [{\citenamefont {Cao}\ \emph {et~al.}(2011)\citenamefont {Cao},
  \citenamefont {Xu}, \citenamefont {Ren}, \citenamefont {Jiang}, \citenamefont
  {Feng},\ and\ \citenamefont {Xu}}]{Cao2011}%
  \BibitemOpen
  \bibfield  {author} {\bibinfo {author} {\bibfnamefont {G.}~\bibnamefont
  {Cao}}, \bibinfo {author} {\bibfnamefont {S.}~\bibnamefont {Xu}}, \bibinfo
  {author} {\bibfnamefont {Z.}~\bibnamefont {Ren}}, \bibinfo {author}
  {\bibfnamefont {S.}~\bibnamefont {Jiang}}, \bibinfo {author} {\bibfnamefont
  {C.}~\bibnamefont {Feng}},\ and\ \bibinfo {author} {\bibfnamefont
  {Z.}~\bibnamefont {Xu}},\ }\bibfield  {title} {\bibinfo {title}
  {{Superconductivity and ferromagnetism in $\mathrm{Eu}\mathrm{Fe}_2
  (\mathrm{As}_{1-x}\mathrm{P}_{x})_2$}},\ }\href@noop {} {\bibfield  {journal}
  {\bibinfo  {journal} {Journal of Physics: Condensed Matter}\ }\textbf
  {\bibinfo {volume} {23}},\ \bibinfo {pages} {464204} (\bibinfo {year}
  {2011})}\BibitemShut {NoStop}%
\bibitem [{\citenamefont {Tokiwa}\ \emph {et~al.}(2012)\citenamefont {Tokiwa},
  \citenamefont {H\"ubner}, \citenamefont {Beck}, \citenamefont {Jeevan},\ and\
  \citenamefont {Gegenwart}}]{TokiwaPhysRevB.86.220505}%
  \BibitemOpen
  \bibfield  {author} {\bibinfo {author} {\bibfnamefont {Y.}~\bibnamefont
  {Tokiwa}}, \bibinfo {author} {\bibfnamefont {S.-H.}\ \bibnamefont
  {H\"ubner}}, \bibinfo {author} {\bibfnamefont {O.}~\bibnamefont {Beck}},
  \bibinfo {author} {\bibfnamefont {H.~S.}\ \bibnamefont {Jeevan}},\ and\
  \bibinfo {author} {\bibfnamefont {P.}~\bibnamefont {Gegenwart}},\ }\bibfield
  {title} {\bibinfo {title} {Unique phase diagram with narrow superconducting
  dome in {EuFe${}_{2}$(As${}_{1\ensuremath{-}x}$P${}_{x}$)${}_{2}$} due to
  {Eu${}^{2+}$} local magnetic moments},\ }\href
  {https://doi.org/10.1103/PhysRevB.86.220505} {\bibfield  {journal} {\bibinfo
  {journal} {Phys. Rev. B}\ }\textbf {\bibinfo {volume} {86}},\ \bibinfo
  {pages} {220505(R)} (\bibinfo {year} {2012})}\BibitemShut {NoStop}%
\bibitem [{\citenamefont {Zapf}\ \emph {et~al.}(2013)\citenamefont {Zapf},
  \citenamefont {Jeevan}, \citenamefont {Ivek}, \citenamefont {Pfister},
  \citenamefont {Klingert}, \citenamefont {Jiang}, \citenamefont {Wu},
  \citenamefont {Gegenwart}, \citenamefont {Kremer},\ and\ \citenamefont
  {Dressel}}]{ZapfPhysRevLett.110.237002}%
  \BibitemOpen
  \bibfield  {author} {\bibinfo {author} {\bibfnamefont {S.}~\bibnamefont
  {Zapf}}, \bibinfo {author} {\bibfnamefont {H.~S.}\ \bibnamefont {Jeevan}},
  \bibinfo {author} {\bibfnamefont {T.}~\bibnamefont {Ivek}}, \bibinfo {author}
  {\bibfnamefont {F.}~\bibnamefont {Pfister}}, \bibinfo {author} {\bibfnamefont
  {F.}~\bibnamefont {Klingert}}, \bibinfo {author} {\bibfnamefont
  {S.}~\bibnamefont {Jiang}}, \bibinfo {author} {\bibfnamefont
  {D.}~\bibnamefont {Wu}}, \bibinfo {author} {\bibfnamefont {P.}~\bibnamefont
  {Gegenwart}}, \bibinfo {author} {\bibfnamefont {R.~K.}\ \bibnamefont
  {Kremer}},\ and\ \bibinfo {author} {\bibfnamefont {M.}~\bibnamefont
  {Dressel}},\ }\bibfield  {title} {\bibinfo {title}
  {{${\mathrm{EuFe}}_{2}({\mathrm{As}}_{1\ensuremath{-}x}{\mathrm{P}}_{x}{)}_{2}$}:
  Reentrant spin glass and superconductivity},\ }\href
  {https://doi.org/10.1103/PhysRevLett.110.237002} {\bibfield  {journal}
  {\bibinfo  {journal} {Phys. Rev. Lett.}\ }\textbf {\bibinfo {volume} {110}},\
  \bibinfo {pages} {237002} (\bibinfo {year} {2013})}\BibitemShut {NoStop}%
\bibitem [{\citenamefont {Nandi}\ \emph {et~al.}(2014)\citenamefont {Nandi},
  \citenamefont {Jin}, \citenamefont {Xiao}, \citenamefont {Su}, \citenamefont
  {Price}, \citenamefont {Shukla}, \citenamefont {Strempfer}, \citenamefont
  {Jeevan}, \citenamefont {Gegenwart},\ and\ \citenamefont
  {Br\"uckel}}]{NandiPhysRevB.89.014512}%
  \BibitemOpen
  \bibfield  {author} {\bibinfo {author} {\bibfnamefont {S.}~\bibnamefont
  {Nandi}}, \bibinfo {author} {\bibfnamefont {W.~T.}\ \bibnamefont {Jin}},
  \bibinfo {author} {\bibfnamefont {Y.}~\bibnamefont {Xiao}}, \bibinfo {author}
  {\bibfnamefont {Y.}~\bibnamefont {Su}}, \bibinfo {author} {\bibfnamefont
  {S.}~\bibnamefont {Price}}, \bibinfo {author} {\bibfnamefont {D.~K.}\
  \bibnamefont {Shukla}}, \bibinfo {author} {\bibfnamefont {J.}~\bibnamefont
  {Strempfer}}, \bibinfo {author} {\bibfnamefont {H.~S.}\ \bibnamefont
  {Jeevan}}, \bibinfo {author} {\bibfnamefont {P.}~\bibnamefont {Gegenwart}},\
  and\ \bibinfo {author} {\bibfnamefont {T.}~\bibnamefont {Br\"uckel}},\
  }\bibfield  {title} {\bibinfo {title} {Coexistence of superconductivity and
  ferromagnetism in {P}-doped {${\text{EuFe}}_{2}{\mathrm{As}}_{2}$}},\ }\href
  {https://doi.org/10.1103/PhysRevB.89.014512} {\bibfield  {journal} {\bibinfo
  {journal} {Phys. Rev. B}\ }\textbf {\bibinfo {volume} {89}},\ \bibinfo
  {pages} {014512} (\bibinfo {year} {2014})}\BibitemShut {NoStop}%
\bibitem [{\citenamefont {Jiao}\ \emph {et~al.}(2011)\citenamefont {Jiao},
  \citenamefont {Tao}, \citenamefont {Bao}, \citenamefont {Sun}, \citenamefont
  {Feng}, \citenamefont {Xu}, \citenamefont {Nowik}, \citenamefont {Felner},\
  and\ \citenamefont {Cao}}]{JiaoEPL2011}%
  \BibitemOpen
  \bibfield  {author} {\bibinfo {author} {\bibfnamefont {W.-H.}\ \bibnamefont
  {Jiao}}, \bibinfo {author} {\bibfnamefont {Q.}~\bibnamefont {Tao}}, \bibinfo
  {author} {\bibfnamefont {J.-K.}\ \bibnamefont {Bao}}, \bibinfo {author}
  {\bibfnamefont {Y.-L.}\ \bibnamefont {Sun}}, \bibinfo {author} {\bibfnamefont
  {C.-M.}\ \bibnamefont {Feng}}, \bibinfo {author} {\bibfnamefont {Z.-A.}\
  \bibnamefont {Xu}}, \bibinfo {author} {\bibfnamefont {I.}~\bibnamefont
  {Nowik}}, \bibinfo {author} {\bibfnamefont {I.}~\bibnamefont {Felner}},\ and\
  \bibinfo {author} {\bibfnamefont {G.-H.}\ \bibnamefont {Cao}},\ }\bibfield
  {title} {\bibinfo {title} {Anisotropic superconductivity in
  {Eu(Fe$_{0.75}$Ru$_{0.25}$)$_2$As$_2$} ferromagnetic superconductor},\ }\href
  {https://doi.org/10.1209/0295-5075/95/67007} {\bibfield  {journal} {\bibinfo
  {journal} {Europhys Lett}\ }\textbf {\bibinfo {volume} {95}},\ \bibinfo
  {pages} {67007} (\bibinfo {year} {2011})}\BibitemShut {NoStop}%
\bibitem [{\citenamefont {Jiang}\ \emph {et~al.}(2009)\citenamefont {Jiang},
  \citenamefont {Xing}, \citenamefont {Xuan}, \citenamefont {Ren},
  \citenamefont {Wang}, \citenamefont {Xu},\ and\ \citenamefont
  {Cao}}]{JiangPhysRevB.80.184514}%
  \BibitemOpen
  \bibfield  {author} {\bibinfo {author} {\bibfnamefont {S.}~\bibnamefont
  {Jiang}}, \bibinfo {author} {\bibfnamefont {H.}~\bibnamefont {Xing}},
  \bibinfo {author} {\bibfnamefont {G.}~\bibnamefont {Xuan}}, \bibinfo {author}
  {\bibfnamefont {Z.}~\bibnamefont {Ren}}, \bibinfo {author} {\bibfnamefont
  {C.}~\bibnamefont {Wang}}, \bibinfo {author} {\bibfnamefont {Z.-a.}\
  \bibnamefont {Xu}},\ and\ \bibinfo {author} {\bibfnamefont {G.}~\bibnamefont
  {Cao}},\ }\bibfield  {title} {\bibinfo {title} {Superconductivity and
  local-moment magnetism in
  {$\text{Eu}{({\text{Fe}}_{0.89}{\text{Co}}_{0.11})}_{2}{\text{As}}_{2}$}},\
  }\href {https://doi.org/10.1103/PhysRevB.80.184514} {\bibfield  {journal}
  {\bibinfo  {journal} {Phys. Rev. B}\ }\textbf {\bibinfo {volume} {80}},\
  \bibinfo {pages} {184514} (\bibinfo {year} {2009})}\BibitemShut {NoStop}%
\bibitem [{\citenamefont {He}\ \emph {et~al.}(2010)\citenamefont {He},
  \citenamefont {Wu}, \citenamefont {Wu}, \citenamefont {Zheng}, \citenamefont
  {Liu}, \citenamefont {Chen}, \citenamefont {Ying}, \citenamefont {Liu},
  \citenamefont {Wang}, \citenamefont {Xie}, \citenamefont {Yan}, \citenamefont
  {Dong}, \citenamefont {Li},\ and\ \citenamefont {Chen}}]{HeJPhys2010}%
  \BibitemOpen
  \bibfield  {author} {\bibinfo {author} {\bibfnamefont {Y.}~\bibnamefont
  {He}}, \bibinfo {author} {\bibfnamefont {T.}~\bibnamefont {Wu}}, \bibinfo
  {author} {\bibfnamefont {G.}~\bibnamefont {Wu}}, \bibinfo {author}
  {\bibfnamefont {Q.~J.}\ \bibnamefont {Zheng}}, \bibinfo {author}
  {\bibfnamefont {Y.~Z.}\ \bibnamefont {Liu}}, \bibinfo {author} {\bibfnamefont
  {H.}~\bibnamefont {Chen}}, \bibinfo {author} {\bibfnamefont {J.~J.}\
  \bibnamefont {Ying}}, \bibinfo {author} {\bibfnamefont {R.~H.}\ \bibnamefont
  {Liu}}, \bibinfo {author} {\bibfnamefont {X.~F.}\ \bibnamefont {Wang}},
  \bibinfo {author} {\bibfnamefont {Y.~L.}\ \bibnamefont {Xie}}, \bibinfo
  {author} {\bibfnamefont {Y.~J.}\ \bibnamefont {Yan}}, \bibinfo {author}
  {\bibfnamefont {J.~K.}\ \bibnamefont {Dong}}, \bibinfo {author}
  {\bibfnamefont {S.~Y.}\ \bibnamefont {Li}},\ and\ \bibinfo {author}
  {\bibfnamefont {X.~H.}\ \bibnamefont {Chen}},\ }\bibfield  {title} {\bibinfo
  {title} {Evidence for competing magnetic and superconducting phases in
  superconducting {Eu$_{1-x}$Sr$_x$Fe$_{2-y}$Co$_y$As$_2$} single crystals},\
  }\href {https://doi.org/10.1088/0953-8984/22/23/235701} {\bibfield  {journal}
  {\bibinfo  {journal} {J. Physics: Condens. Matter}\ }\textbf {\bibinfo
  {volume} {22}},\ \bibinfo {pages} {235701} (\bibinfo {year}
  {2010})}\BibitemShut {NoStop}%
\bibitem [{\citenamefont {Guguchia}\ \emph {et~al.}(2011)\citenamefont
  {Guguchia}, \citenamefont {Bosma}, \citenamefont {Weyeneth}, \citenamefont
  {Shengelaya}, \citenamefont {Puzniak}, \citenamefont {Bukowski},
  \citenamefont {Karpinski},\ and\ \citenamefont
  {Keller}}]{GuguchiaPhysRevB.84.144506}%
  \BibitemOpen
  \bibfield  {author} {\bibinfo {author} {\bibfnamefont {Z.}~\bibnamefont
  {Guguchia}}, \bibinfo {author} {\bibfnamefont {S.}~\bibnamefont {Bosma}},
  \bibinfo {author} {\bibfnamefont {S.}~\bibnamefont {Weyeneth}}, \bibinfo
  {author} {\bibfnamefont {A.}~\bibnamefont {Shengelaya}}, \bibinfo {author}
  {\bibfnamefont {R.}~\bibnamefont {Puzniak}}, \bibinfo {author} {\bibfnamefont
  {Z.}~\bibnamefont {Bukowski}}, \bibinfo {author} {\bibfnamefont
  {J.}~\bibnamefont {Karpinski}},\ and\ \bibinfo {author} {\bibfnamefont
  {H.}~\bibnamefont {Keller}},\ }\bibfield  {title} {\bibinfo {title}
  {Anisotropic magnetic order of the eu sublattice in single crystals of
  {EuFe${}_{2\ensuremath{-}x}$Co${}_{x}$As${}_{2}$} ($x=0,0.2$) studied by
  means of magnetization and magnetic torque},\ }\href
  {https://doi.org/10.1103/PhysRevB.84.144506} {\bibfield  {journal} {\bibinfo
  {journal} {Phys. Rev. B}\ }\textbf {\bibinfo {volume} {84}},\ \bibinfo
  {pages} {144506} (\bibinfo {year} {2011})}\BibitemShut {NoStop}%
\bibitem [{\citenamefont {Jin}\ \emph {et~al.}(2013)\citenamefont {Jin},
  \citenamefont {Nandi}, \citenamefont {Xiao}, \citenamefont {Su},
  \citenamefont {Zaharko}, \citenamefont {Guguchia}, \citenamefont {Bukowski},
  \citenamefont {Price}, \citenamefont {Jiao}, \citenamefont {Cao},\ and\
  \citenamefont {Br\"uckel}}]{JinPhysRevB.88.214516}%
  \BibitemOpen
  \bibfield  {author} {\bibinfo {author} {\bibfnamefont {W.~T.}\ \bibnamefont
  {Jin}}, \bibinfo {author} {\bibfnamefont {S.}~\bibnamefont {Nandi}}, \bibinfo
  {author} {\bibfnamefont {Y.}~\bibnamefont {Xiao}}, \bibinfo {author}
  {\bibfnamefont {Y.}~\bibnamefont {Su}}, \bibinfo {author} {\bibfnamefont
  {O.}~\bibnamefont {Zaharko}}, \bibinfo {author} {\bibfnamefont
  {Z.}~\bibnamefont {Guguchia}}, \bibinfo {author} {\bibfnamefont
  {Z.}~\bibnamefont {Bukowski}}, \bibinfo {author} {\bibfnamefont
  {S.}~\bibnamefont {Price}}, \bibinfo {author} {\bibfnamefont {W.~H.}\
  \bibnamefont {Jiao}}, \bibinfo {author} {\bibfnamefont {G.~H.}\ \bibnamefont
  {Cao}},\ and\ \bibinfo {author} {\bibfnamefont {T.}~\bibnamefont
  {Br\"uckel}},\ }\bibfield  {title} {\bibinfo {title} {Magnetic structure of
  superconducting {Eu(Fe${}_{0.82}$Co${}_{0.18}$)${}_{2}$As${}_{2}$} as
  revealed by single-crystal neutron diffraction},\ }\href
  {https://doi.org/10.1103/PhysRevB.88.214516} {\bibfield  {journal} {\bibinfo
  {journal} {Phys. Rev. B}\ }\textbf {\bibinfo {volume} {88}},\ \bibinfo
  {pages} {214516} (\bibinfo {year} {2013})}\BibitemShut {NoStop}%
\bibitem [{\citenamefont {Paramanik}\ \emph {et~al.}(2014)\citenamefont
  {Paramanik}, \citenamefont {Paulose}, \citenamefont {Ramakrishnan},
  \citenamefont {Nigam}, \citenamefont {Geibel},\ and\ \citenamefont
  {Hossain}}]{Paramanik2014}%
  \BibitemOpen
  \bibfield  {author} {\bibinfo {author} {\bibfnamefont {U.~B.}\ \bibnamefont
  {Paramanik}}, \bibinfo {author} {\bibfnamefont {P.~L.}\ \bibnamefont
  {Paulose}}, \bibinfo {author} {\bibfnamefont {S.}~\bibnamefont
  {Ramakrishnan}}, \bibinfo {author} {\bibfnamefont {A.~K.}\ \bibnamefont
  {Nigam}}, \bibinfo {author} {\bibfnamefont {C.}~\bibnamefont {Geibel}},\ and\
  \bibinfo {author} {\bibfnamefont {Z.}~\bibnamefont {Hossain}},\ }\bibfield
  {title} {\bibinfo {title} {Magnetic and superconducting properties of
  {Ir}-doped {EuFe$_2$As$_2$}},\ }\href
  {http://stacks.iop.org/0953-2048/27/i=7/a=075012} {\bibfield  {journal}
  {\bibinfo  {journal} {Supercond. Sci. Technol.}\ }\textbf {\bibinfo {volume}
  {27}},\ \bibinfo {pages} {075012} (\bibinfo {year} {2014})}\BibitemShut
  {NoStop}%
\bibitem [{\citenamefont {Jeevan}\ \emph
  {et~al.}(2008{\natexlab{b}})\citenamefont {Jeevan}, \citenamefont {Hossain},
  \citenamefont {Kasinathan}, \citenamefont {Rosner}, \citenamefont {Geibel},\
  and\ \citenamefont {Gegenwart}}]{Jeevan2008}%
  \BibitemOpen
  \bibfield  {author} {\bibinfo {author} {\bibfnamefont {H.~S.}\ \bibnamefont
  {Jeevan}}, \bibinfo {author} {\bibfnamefont {Z.}~\bibnamefont {Hossain}},
  \bibinfo {author} {\bibfnamefont {D.}~\bibnamefont {Kasinathan}}, \bibinfo
  {author} {\bibfnamefont {H.}~\bibnamefont {Rosner}}, \bibinfo {author}
  {\bibfnamefont {C.}~\bibnamefont {Geibel}},\ and\ \bibinfo {author}
  {\bibfnamefont {P.}~\bibnamefont {Gegenwart}},\ }\bibfield  {title} {\bibinfo
  {title} {{High-temperature superconductivity in
  $\mathrm{Eu}_{0.5}\mathrm{K}_{0.5}\mathrm{Fe}_{2}\mathrm{As}_{2}$}},\ }\href
  {https://doi.org/10.1103/PhysRevB.78.092406} {\bibfield  {journal} {\bibinfo
  {journal} {Phys. Rev. B}\ }\textbf {\bibinfo {volume} {78}},\ \bibinfo
  {pages} {092406} (\bibinfo {year} {2008}{\natexlab{b}})}\BibitemShut
  {NoStop}%
\bibitem [{\citenamefont {Qi}\ \emph {et~al.}(2008)\citenamefont {Qi},
  \citenamefont {Gao}, \citenamefont {Wang}, \citenamefont {Wang},
  \citenamefont {Zhang},\ and\ \citenamefont {Ma}}]{QiNJP2008}%
  \BibitemOpen
  \bibfield  {author} {\bibinfo {author} {\bibfnamefont {Y.}~\bibnamefont
  {Qi}}, \bibinfo {author} {\bibfnamefont {Z.}~\bibnamefont {Gao}}, \bibinfo
  {author} {\bibfnamefont {L.}~\bibnamefont {Wang}}, \bibinfo {author}
  {\bibfnamefont {D.}~\bibnamefont {Wang}}, \bibinfo {author} {\bibfnamefont
  {X.}~\bibnamefont {Zhang}},\ and\ \bibinfo {author} {\bibfnamefont
  {Y.}~\bibnamefont {Ma}},\ }\bibfield  {title} {\bibinfo {title}
  {{Superconductivity at 34.7$\mathrm{K}$ in the iron arsenide
  $\mathrm{Eu}_{0.7}\mathrm{Na}_{0.3}\mathrm{Fe}_{2}\mathrm{As}_{2}$}},\ }\href
  {http://stacks.iop.org/1367-2630/10/i=12/a=123003} {\bibfield  {journal}
  {\bibinfo  {journal} {New J. Phys.}\ }\textbf {\bibinfo {volume} {10}},\
  \bibinfo {pages} {123003} (\bibinfo {year} {2008})}\BibitemShut {NoStop}%
\bibitem [{\citenamefont {Veshchunov}\ \emph {et~al.}(2017)\citenamefont
  {Veshchunov}, \citenamefont {Vinnikov}, \citenamefont {Stolyarov},
  \citenamefont {Zhou}, \citenamefont {Shi}, \citenamefont {Xu}, \citenamefont
  {Grebenchuk}, \citenamefont {Baranov}, \citenamefont {Golovchanskiy},
  \citenamefont {Pyon}, \citenamefont {Sun}, \citenamefont {Jiao},
  \citenamefont {Cao}, \citenamefont {Tamegai},\ and\ \citenamefont
  {Golubov}}]{VeshchunovJETPLett2017}%
  \BibitemOpen
  \bibfield  {author} {\bibinfo {author} {\bibfnamefont {I.~S.}\ \bibnamefont
  {Veshchunov}}, \bibinfo {author} {\bibfnamefont {L.~Y.}\ \bibnamefont
  {Vinnikov}}, \bibinfo {author} {\bibfnamefont {V.~S.}\ \bibnamefont
  {Stolyarov}}, \bibinfo {author} {\bibfnamefont {N.}~\bibnamefont {Zhou}},
  \bibinfo {author} {\bibfnamefont {Z.~X.}\ \bibnamefont {Shi}}, \bibinfo
  {author} {\bibfnamefont {X.~F.}\ \bibnamefont {Xu}}, \bibinfo {author}
  {\bibfnamefont {S.~Y.}\ \bibnamefont {Grebenchuk}}, \bibinfo {author}
  {\bibfnamefont {D.~S.}\ \bibnamefont {Baranov}}, \bibinfo {author}
  {\bibfnamefont {I.~A.}\ \bibnamefont {Golovchanskiy}}, \bibinfo {author}
  {\bibfnamefont {S.}~\bibnamefont {Pyon}}, \bibinfo {author} {\bibfnamefont
  {Y.}~\bibnamefont {Sun}}, \bibinfo {author} {\bibfnamefont {W.}~\bibnamefont
  {Jiao}}, \bibinfo {author} {\bibfnamefont {G.}~\bibnamefont {Cao}}, \bibinfo
  {author} {\bibfnamefont {T.}~\bibnamefont {Tamegai}},\ and\ \bibinfo {author}
  {\bibfnamefont {A.~A.}\ \bibnamefont {Golubov}},\ }\bibfield  {title}
  {\bibinfo {title} {Visualization of the magnetic flux structure in
  phosphorus-doped {EuFe$_2$As$_2$} single crystals},\ }\href
  {https://doi.org/10.1134/S0021364017020151} {\bibfield  {journal} {\bibinfo
  {journal} {JETP Letters}\ }\textbf {\bibinfo {volume} {105}},\ \bibinfo
  {pages} {98} (\bibinfo {year} {2017})}\BibitemShut {NoStop}%
\bibitem [{\citenamefont {Stolyarov}\ \emph
  {et~al.}(2018{\natexlab{a}})\citenamefont {Stolyarov}, \citenamefont
  {Veshchunov}, \citenamefont {Grebenchuk}, \citenamefont {Baranov},
  \citenamefont {Golovchanskiy}, \citenamefont {Shishkin}, \citenamefont
  {Zhou}, \citenamefont {Shi}, \citenamefont {Xu}, \citenamefont {Pyon},
  \citenamefont {Sun}, \citenamefont {Jiao}, \citenamefont {Cao}, \citenamefont
  {Vinnikov}, \citenamefont {Golubov}, \citenamefont {Tamegai}, \citenamefont
  {Buzdin},\ and\ \citenamefont {Roditchev}}]{StolyarovSciAdv18}%
  \BibitemOpen
  \bibfield  {author} {\bibinfo {author} {\bibfnamefont {V.~S.}\ \bibnamefont
  {Stolyarov}}, \bibinfo {author} {\bibfnamefont {I.~S.}\ \bibnamefont
  {Veshchunov}}, \bibinfo {author} {\bibfnamefont {S.~Y.}\ \bibnamefont
  {Grebenchuk}}, \bibinfo {author} {\bibfnamefont {D.~S.}\ \bibnamefont
  {Baranov}}, \bibinfo {author} {\bibfnamefont {I.~A.}\ \bibnamefont
  {Golovchanskiy}}, \bibinfo {author} {\bibfnamefont {A.~G.}\ \bibnamefont
  {Shishkin}}, \bibinfo {author} {\bibfnamefont {N.}~\bibnamefont {Zhou}},
  \bibinfo {author} {\bibfnamefont {Z.}~\bibnamefont {Shi}}, \bibinfo {author}
  {\bibfnamefont {X.}~\bibnamefont {Xu}}, \bibinfo {author} {\bibfnamefont
  {S.}~\bibnamefont {Pyon}}, \bibinfo {author} {\bibfnamefont {Y.}~\bibnamefont
  {Sun}}, \bibinfo {author} {\bibfnamefont {W.}~\bibnamefont {Jiao}}, \bibinfo
  {author} {\bibfnamefont {G.-H.}\ \bibnamefont {Cao}}, \bibinfo {author}
  {\bibfnamefont {L.~Y.}\ \bibnamefont {Vinnikov}}, \bibinfo {author}
  {\bibfnamefont {A.~A.}\ \bibnamefont {Golubov}}, \bibinfo {author}
  {\bibfnamefont {T.}~\bibnamefont {Tamegai}}, \bibinfo {author} {\bibfnamefont
  {A.~I.}\ \bibnamefont {Buzdin}},\ and\ \bibinfo {author} {\bibfnamefont
  {D.}~\bibnamefont {Roditchev}},\ }\bibfield  {title} {\bibinfo {title}
  {Domain {Meissner} state and spontaneous vortex-antivortex generation in the
  ferromagnetic superconductor {EuFe$_2$(As$_{0.79}$P$_{0.21}$)$_2$}},\ }\href
  {https://doi.org/10.1126/sciadv.aat1061} {\bibfield  {journal} {\bibinfo
  {journal} {Sci. Adv.}\ }\textbf {\bibinfo {volume} {4}},\ \bibinfo {pages}
  {eaat1061} (\bibinfo {year} {2018}{\natexlab{a}})}\BibitemShut {NoStop}%
\bibitem [{\citenamefont {Devizorova}\ \emph {et~al.}(2019)\citenamefont
  {Devizorova}, \citenamefont {Mironov},\ and\ \citenamefont
  {Buzdin}}]{DevizorovaPhysRevLett.122.117002}%
  \BibitemOpen
  \bibfield  {author} {\bibinfo {author} {\bibfnamefont {Z.}~\bibnamefont
  {Devizorova}}, \bibinfo {author} {\bibfnamefont {S.}~\bibnamefont
  {Mironov}},\ and\ \bibinfo {author} {\bibfnamefont {A.}~\bibnamefont
  {Buzdin}},\ }\bibfield  {title} {\bibinfo {title} {Theory of magnetic domain
  phases in ferromagnetic superconductors},\ }\href
  {https://doi.org/10.1103/PhysRevLett.122.117002} {\bibfield  {journal}
  {\bibinfo  {journal} {Phys. Rev. Lett.}\ }\textbf {\bibinfo {volume} {122}},\
  \bibinfo {pages} {117002} (\bibinfo {year} {2019})}\BibitemShut {NoStop}%
\bibitem [{\citenamefont {Liu}\ \emph {et~al.}(2016{\natexlab{a}})\citenamefont
  {Liu}, \citenamefont {Liu}, \citenamefont {Chen}, \citenamefont {Tang},
  \citenamefont {Jiao}, \citenamefont {Tao}, \citenamefont {Xu},\ and\
  \citenamefont {Cao}}]{Liu2016}%
  \BibitemOpen
  \bibfield  {author} {\bibinfo {author} {\bibfnamefont {Y.}~\bibnamefont
  {Liu}}, \bibinfo {author} {\bibfnamefont {Y.-B.}\ \bibnamefont {Liu}},
  \bibinfo {author} {\bibfnamefont {Q.}~\bibnamefont {Chen}}, \bibinfo {author}
  {\bibfnamefont {Z.-T.}\ \bibnamefont {Tang}}, \bibinfo {author}
  {\bibfnamefont {W.-H.}\ \bibnamefont {Jiao}}, \bibinfo {author}
  {\bibfnamefont {Q.}~\bibnamefont {Tao}}, \bibinfo {author} {\bibfnamefont
  {Z.-A.}\ \bibnamefont {Xu}},\ and\ \bibinfo {author} {\bibfnamefont {G.-H.}\
  \bibnamefont {Cao}},\ }\bibfield  {title} {\bibinfo {title} {{A new
  ferromagnetic superconductor:
  $\mathrm{Cs}\mathrm{Eu}\mathrm{Fe}_{4}\mathrm{As}_{4}$}},\ }\href
  {https://doi.org/10.1007/s11434-016-1139-2} {\bibfield  {journal} {\bibinfo
  {journal} {Science Bulletin}\ }\textbf {\bibinfo {volume} {61}},\ \bibinfo
  {pages} {1213} (\bibinfo {year} {2016}{\natexlab{a}})}\BibitemShut {NoStop}%
\bibitem [{\citenamefont {Kawashima}\ \emph {et~al.}(2016)\citenamefont
  {Kawashima}, \citenamefont {Kinjo}, \citenamefont {Nishio}, \citenamefont
  {Ishida}, \citenamefont {Fujihisa}, \citenamefont {Gotoh}, \citenamefont
  {Kihou}, \citenamefont {Eisaki}, \citenamefont {Yoshida},\ and\ \citenamefont
  {Iyo}}]{KawashimaJPSJ2016}%
  \BibitemOpen
  \bibfield  {author} {\bibinfo {author} {\bibfnamefont {K.}~\bibnamefont
  {Kawashima}}, \bibinfo {author} {\bibfnamefont {T.}~\bibnamefont {Kinjo}},
  \bibinfo {author} {\bibfnamefont {T.}~\bibnamefont {Nishio}}, \bibinfo
  {author} {\bibfnamefont {S.}~\bibnamefont {Ishida}}, \bibinfo {author}
  {\bibfnamefont {H.}~\bibnamefont {Fujihisa}}, \bibinfo {author}
  {\bibfnamefont {Y.}~\bibnamefont {Gotoh}}, \bibinfo {author} {\bibfnamefont
  {K.}~\bibnamefont {Kihou}}, \bibinfo {author} {\bibfnamefont
  {H.}~\bibnamefont {Eisaki}}, \bibinfo {author} {\bibfnamefont
  {Y.}~\bibnamefont {Yoshida}},\ and\ \bibinfo {author} {\bibfnamefont
  {A.}~\bibnamefont {Iyo}},\ }\bibfield  {title} {\bibinfo {title}
  {Superconductivity in {Fe}-based compound {EuAFe$_4$As$_4$} ({A = Rb and
  Cs})},\ }\href {https://doi.org/10.7566/jpsj.85.064710} {\bibfield  {journal}
  {\bibinfo  {journal} {J. Phys. Soc. Jpn.}\ }\textbf {\bibinfo {volume}
  {85}},\ \bibinfo {pages} {064710} (\bibinfo {year} {2016})}\BibitemShut
  {NoStop}%
\bibitem [{\citenamefont {Bao}\ \emph {et~al.}(2018)\citenamefont {Bao},
  \citenamefont {Willa}, \citenamefont {Smylie}, \citenamefont {Chen},
  \citenamefont {Welp}, \citenamefont {Chung},\ and\ \citenamefont
  {Kanatzidis}}]{Bao2018}%
  \BibitemOpen
  \bibfield  {author} {\bibinfo {author} {\bibfnamefont {J.-K.}\ \bibnamefont
  {Bao}}, \bibinfo {author} {\bibfnamefont {K.}~\bibnamefont {Willa}}, \bibinfo
  {author} {\bibfnamefont {M.~P.}\ \bibnamefont {Smylie}}, \bibinfo {author}
  {\bibfnamefont {H.}~\bibnamefont {Chen}}, \bibinfo {author} {\bibfnamefont
  {U.}~\bibnamefont {Welp}}, \bibinfo {author} {\bibfnamefont {D.~Y.}\
  \bibnamefont {Chung}},\ and\ \bibinfo {author} {\bibfnamefont {M.~G.}\
  \bibnamefont {Kanatzidis}},\ }\bibfield  {title} {\bibinfo {title} {Single
  crystal growth and study of the ferromagnetic superconductor
  {RbEuFe$_4$As$_4$}},\ }\href {https://doi.org/10.1021/acs.cgd.8b00315}
  {\bibfield  {journal} {\bibinfo  {journal} {Crystal Growth \& Design}\
  }\textbf {\bibinfo {volume} {18}},\ \bibinfo {pages} {3517} (\bibinfo {year}
  {2018})}\BibitemShut {NoStop}%
\bibitem [{\citenamefont {Smylie}\ \emph {et~al.}(2018)\citenamefont {Smylie},
  \citenamefont {Willa}, \citenamefont {Bao}, \citenamefont {Ryan},
  \citenamefont {Islam}, \citenamefont {Claus}, \citenamefont {Simsek},
  \citenamefont {Diao}, \citenamefont {Rydh}, \citenamefont {Koshelev},
  \citenamefont {Kwok}, \citenamefont {Chung}, \citenamefont {Kanatzidis},\
  and\ \citenamefont {Welp}}]{Smylie2018}%
  \BibitemOpen
  \bibfield  {author} {\bibinfo {author} {\bibfnamefont {M.~P.}\ \bibnamefont
  {Smylie}}, \bibinfo {author} {\bibfnamefont {K.}~\bibnamefont {Willa}},
  \bibinfo {author} {\bibfnamefont {J.-K.}\ \bibnamefont {Bao}}, \bibinfo
  {author} {\bibfnamefont {K.}~\bibnamefont {Ryan}}, \bibinfo {author}
  {\bibfnamefont {Z.}~\bibnamefont {Islam}}, \bibinfo {author} {\bibfnamefont
  {H.}~\bibnamefont {Claus}}, \bibinfo {author} {\bibfnamefont
  {Y.}~\bibnamefont {Simsek}}, \bibinfo {author} {\bibfnamefont
  {Z.}~\bibnamefont {Diao}}, \bibinfo {author} {\bibfnamefont {A.}~\bibnamefont
  {Rydh}}, \bibinfo {author} {\bibfnamefont {A.~E.}\ \bibnamefont {Koshelev}},
  \bibinfo {author} {\bibfnamefont {W.-K.}\ \bibnamefont {Kwok}}, \bibinfo
  {author} {\bibfnamefont {D.~Y.}\ \bibnamefont {Chung}}, \bibinfo {author}
  {\bibfnamefont {M.~G.}\ \bibnamefont {Kanatzidis}},\ and\ \bibinfo {author}
  {\bibfnamefont {U.}~\bibnamefont {Welp}},\ }\bibfield  {title} {\bibinfo
  {title} {Anisotropic superconductivity and magnetism in single-crystal
  {${\mathrm{RbEuFe}}_{4}{\mathrm{As}}_{4}$}},\ }\href
  {https://doi.org/10.1103/PhysRevB.98.104503} {\bibfield  {journal} {\bibinfo
  {journal} {Phys. Rev. B}\ }\textbf {\bibinfo {volume} {98}},\ \bibinfo
  {pages} {104503} (\bibinfo {year} {2018})}\BibitemShut {NoStop}%
\bibitem [{\citenamefont {Stolyarov}\ \emph
  {et~al.}(2018{\natexlab{b}})\citenamefont {Stolyarov}, \citenamefont
  {Casano}, \citenamefont {Belyanchikov}, \citenamefont {Astrakhantseva},
  \citenamefont {Grebenchuk}, \citenamefont {Baranov}, \citenamefont
  {Golovchanskiy}, \citenamefont {Voloshenko}, \citenamefont {Zhukova},
  \citenamefont {Gorshunov}, \citenamefont {Muratov}, \citenamefont {Dremov},
  \citenamefont {Vinnikov}, \citenamefont {Roditchev}, \citenamefont {Liu},
  \citenamefont {Cao}, \citenamefont {Dressel},\ and\ \citenamefont
  {Uykur}}]{Stolyarov2018}%
  \BibitemOpen
  \bibfield  {author} {\bibinfo {author} {\bibfnamefont {V.~S.}\ \bibnamefont
  {Stolyarov}}, \bibinfo {author} {\bibfnamefont {A.}~\bibnamefont {Casano}},
  \bibinfo {author} {\bibfnamefont {M.~A.}\ \bibnamefont {Belyanchikov}},
  \bibinfo {author} {\bibfnamefont {A.~S.}\ \bibnamefont {Astrakhantseva}},
  \bibinfo {author} {\bibfnamefont {S.~Y.}\ \bibnamefont {Grebenchuk}},
  \bibinfo {author} {\bibfnamefont {D.~S.}\ \bibnamefont {Baranov}}, \bibinfo
  {author} {\bibfnamefont {I.~A.}\ \bibnamefont {Golovchanskiy}}, \bibinfo
  {author} {\bibfnamefont {I.}~\bibnamefont {Voloshenko}}, \bibinfo {author}
  {\bibfnamefont {E.~S.}\ \bibnamefont {Zhukova}}, \bibinfo {author}
  {\bibfnamefont {B.~P.}\ \bibnamefont {Gorshunov}}, \bibinfo {author}
  {\bibfnamefont {A.~V.}\ \bibnamefont {Muratov}}, \bibinfo {author}
  {\bibfnamefont {V.~V.}\ \bibnamefont {Dremov}}, \bibinfo {author}
  {\bibfnamefont {L.~Y.}\ \bibnamefont {Vinnikov}}, \bibinfo {author}
  {\bibfnamefont {D.}~\bibnamefont {Roditchev}}, \bibinfo {author}
  {\bibfnamefont {Y.}~\bibnamefont {Liu}}, \bibinfo {author} {\bibfnamefont
  {G.-H.}\ \bibnamefont {Cao}}, \bibinfo {author} {\bibfnamefont
  {M.}~\bibnamefont {Dressel}},\ and\ \bibinfo {author} {\bibfnamefont
  {E.}~\bibnamefont {Uykur}},\ }\bibfield  {title} {\bibinfo {title} {Unique
  interplay between superconducting and ferromagnetic orders in
  {${\mathrm{EuRbFe}}_{4}{\mathrm{As}}_{4}$}},\ }\href
  {https://doi.org/10.1103/PhysRevB.98.140506} {\bibfield  {journal} {\bibinfo
  {journal} {Phys. Rev. B}\ }\textbf {\bibinfo {volume} {98}},\ \bibinfo
  {pages} {140506(R)} (\bibinfo {year} {2018}{\natexlab{b}})}\BibitemShut
  {NoStop}%
\bibitem [{\citenamefont {Liu}\ \emph {et~al.}(2016{\natexlab{b}})\citenamefont
  {Liu}, \citenamefont {Liu}, \citenamefont {Tang}, \citenamefont {Jiang},
  \citenamefont {Wang}, \citenamefont {Ablimit}, \citenamefont {Jiao},
  \citenamefont {Tao}, \citenamefont {Feng}, \citenamefont {Xu},\ and\
  \citenamefont {Cao}}]{Liu2016a}%
  \BibitemOpen
  \bibfield  {author} {\bibinfo {author} {\bibfnamefont {Y.}~\bibnamefont
  {Liu}}, \bibinfo {author} {\bibfnamefont {Y.-B.}\ \bibnamefont {Liu}},
  \bibinfo {author} {\bibfnamefont {Z.-T.}\ \bibnamefont {Tang}}, \bibinfo
  {author} {\bibfnamefont {H.}~\bibnamefont {Jiang}}, \bibinfo {author}
  {\bibfnamefont {Z.-C.}\ \bibnamefont {Wang}}, \bibinfo {author}
  {\bibfnamefont {A.}~\bibnamefont {Ablimit}}, \bibinfo {author} {\bibfnamefont
  {W.-H.}\ \bibnamefont {Jiao}}, \bibinfo {author} {\bibfnamefont
  {Q.}~\bibnamefont {Tao}}, \bibinfo {author} {\bibfnamefont {C.-M.}\
  \bibnamefont {Feng}}, \bibinfo {author} {\bibfnamefont {Z.-A.}\ \bibnamefont
  {Xu}},\ and\ \bibinfo {author} {\bibfnamefont {G.-H.}\ \bibnamefont {Cao}},\
  }\bibfield  {title} {\bibinfo {title} {{Superconductivity and ferromagnetism
  in hole-doped ${\mathrm{RbEuFe}}_{4}{\mathrm{As}}_{4}$}},\ }\href
  {https://doi.org/10.1103/PhysRevB.93.214503} {\bibfield  {journal} {\bibinfo
  {journal} {Phys. Rev. B}\ }\textbf {\bibinfo {volume} {93}},\ \bibinfo
  {pages} {214503} (\bibinfo {year} {2016}{\natexlab{b}})}\BibitemShut
  {NoStop}%
\bibitem [{\citenamefont {Willa}\ \emph {et~al.}(2019)\citenamefont {Willa},
  \citenamefont {Willa}, \citenamefont {Bao}, \citenamefont {Koshelev},
  \citenamefont {Chung}, \citenamefont {Kanatzidis}, \citenamefont {Kwok},\
  and\ \citenamefont {Welp}}]{WillaPhysRevB.99.180502}%
  \BibitemOpen
  \bibfield  {author} {\bibinfo {author} {\bibfnamefont {K.}~\bibnamefont
  {Willa}}, \bibinfo {author} {\bibfnamefont {R.}~\bibnamefont {Willa}},
  \bibinfo {author} {\bibfnamefont {J.-K.}\ \bibnamefont {Bao}}, \bibinfo
  {author} {\bibfnamefont {A.~E.}\ \bibnamefont {Koshelev}}, \bibinfo {author}
  {\bibfnamefont {D.~Y.}\ \bibnamefont {Chung}}, \bibinfo {author}
  {\bibfnamefont {M.~G.}\ \bibnamefont {Kanatzidis}}, \bibinfo {author}
  {\bibfnamefont {W.-K.}\ \bibnamefont {Kwok}},\ and\ \bibinfo {author}
  {\bibfnamefont {U.}~\bibnamefont {Welp}},\ }\bibfield  {title} {\bibinfo
  {title} {Strongly fluctuating moments in the high-temperature magnetic
  superconductor {${\mathrm{RbEuFe}}_{4}{\mathrm{As}}_{4}$}},\ }\href
  {https://doi.org/10.1103/PhysRevB.99.180502} {\bibfield  {journal} {\bibinfo
  {journal} {Phys. Rev. B}\ }\textbf {\bibinfo {volume} {99}},\ \bibinfo
  {pages} {180502(R)} (\bibinfo {year} {2019})}\BibitemShut {NoStop}%
\bibitem [{\citenamefont {Hemmida}\ \emph {et~al.}(2020)\citenamefont
  {Hemmida}, \citenamefont {Winterhalter-Stocker}, \citenamefont {Ehlers},
  \citenamefont {von Nidda}, \citenamefont {Yao}, \citenamefont {Bannies},
  \citenamefont {Rienks}, \citenamefont {Kurleto}, \citenamefont {Felser},
  \citenamefont {B{\"u}chner}, \citenamefont {Fink}, \citenamefont {Gorol},
  \citenamefont {F{\"o}rster}, \citenamefont {Arsenijevic}, \citenamefont
  {Fritsch},\ and\ \citenamefont {Gegenwart}}]{hemmida2020topological}%
  \BibitemOpen
  \bibfield  {author} {\bibinfo {author} {\bibfnamefont {M.}~\bibnamefont
  {Hemmida}}, \bibinfo {author} {\bibfnamefont {N.}~\bibnamefont
  {Winterhalter-Stocker}}, \bibinfo {author} {\bibfnamefont {D.}~\bibnamefont
  {Ehlers}}, \bibinfo {author} {\bibfnamefont {H.~A.~K.}\ \bibnamefont {von
  Nidda}}, \bibinfo {author} {\bibfnamefont {M.}~\bibnamefont {Yao}}, \bibinfo
  {author} {\bibfnamefont {J.}~\bibnamefont {Bannies}}, \bibinfo {author}
  {\bibfnamefont {E.~D.~L.}\ \bibnamefont {Rienks}}, \bibinfo {author}
  {\bibfnamefont {R.}~\bibnamefont {Kurleto}}, \bibinfo {author} {\bibfnamefont
  {C.}~\bibnamefont {Felser}}, \bibinfo {author} {\bibfnamefont
  {B.}~\bibnamefont {B{\"u}chner}}, \bibinfo {author} {\bibfnamefont
  {J.}~\bibnamefont {Fink}}, \bibinfo {author} {\bibfnamefont {S.}~\bibnamefont
  {Gorol}}, \bibinfo {author} {\bibfnamefont {T.}~\bibnamefont {F{\"o}rster}},
  \bibinfo {author} {\bibfnamefont {S.}~\bibnamefont {Arsenijevic}}, \bibinfo
  {author} {\bibfnamefont {V.}~\bibnamefont {Fritsch}},\ and\ \bibinfo {author}
  {\bibfnamefont {P.}~\bibnamefont {Gegenwart}},\ }\href@noop {} {\bibinfo
  {title} {Topological magnetic order and superconductivity in
  {EuRbFe$_4$As$_4$}}} (\bibinfo {year} {2020}),\ \Eprint
  {https://arxiv.org/abs/2010.02110} {arXiv:2010.02110 [cond-mat.supr-con]}
  \BibitemShut {NoStop}%
\bibitem [{\citenamefont {Jackson}\ \emph {et~al.}(2018)\citenamefont
  {Jackson}, \citenamefont {VanGennep}, \citenamefont {Bi}, \citenamefont
  {Zhang}, \citenamefont {Materne}, \citenamefont {Liu}, \citenamefont {Cao},
  \citenamefont {Weir}, \citenamefont {Vohra},\ and\ \citenamefont
  {Hamlin}}]{JacksonPhysRevB.98.014518}%
  \BibitemOpen
  \bibfield  {author} {\bibinfo {author} {\bibfnamefont {D.~E.}\ \bibnamefont
  {Jackson}}, \bibinfo {author} {\bibfnamefont {D.}~\bibnamefont {VanGennep}},
  \bibinfo {author} {\bibfnamefont {W.}~\bibnamefont {Bi}}, \bibinfo {author}
  {\bibfnamefont {D.}~\bibnamefont {Zhang}}, \bibinfo {author} {\bibfnamefont
  {P.}~\bibnamefont {Materne}}, \bibinfo {author} {\bibfnamefont
  {Y.}~\bibnamefont {Liu}}, \bibinfo {author} {\bibfnamefont {G.-H.}\
  \bibnamefont {Cao}}, \bibinfo {author} {\bibfnamefont {S.~T.}\ \bibnamefont
  {Weir}}, \bibinfo {author} {\bibfnamefont {Y.~K.}\ \bibnamefont {Vohra}},\
  and\ \bibinfo {author} {\bibfnamefont {J.~J.}\ \bibnamefont {Hamlin}},\
  }\bibfield  {title} {\bibinfo {title} {Superconducting and magnetic phase
  diagram of {${\mathrm{RbEuFe}}_{4}{\mathrm{As}}_{4}$ and
  ${\mathrm{CsEuFe}}_{4}{\mathrm{As}}_{4}$} at high pressure},\ }\href
  {https://doi.org/10.1103/PhysRevB.98.014518} {\bibfield  {journal} {\bibinfo
  {journal} {Phys. Rev. B}\ }\textbf {\bibinfo {volume} {98}},\ \bibinfo
  {pages} {014518} (\bibinfo {year} {2018})}\BibitemShut {NoStop}%
\bibitem [{\citenamefont {Xiang}\ \emph {et~al.}(2019)\citenamefont {Xiang},
  \citenamefont {Bud'ko}, \citenamefont {Bao}, \citenamefont {Chung},
  \citenamefont {Kanatzidis},\ and\ \citenamefont
  {Canfield}}]{XiangPhysRevB.99.144509}%
  \BibitemOpen
  \bibfield  {author} {\bibinfo {author} {\bibfnamefont {L.}~\bibnamefont
  {Xiang}}, \bibinfo {author} {\bibfnamefont {S.~L.}\ \bibnamefont {Bud'ko}},
  \bibinfo {author} {\bibfnamefont {J.-K.}\ \bibnamefont {Bao}}, \bibinfo
  {author} {\bibfnamefont {D.~Y.}\ \bibnamefont {Chung}}, \bibinfo {author}
  {\bibfnamefont {M.~G.}\ \bibnamefont {Kanatzidis}},\ and\ \bibinfo {author}
  {\bibfnamefont {P.~C.}\ \bibnamefont {Canfield}},\ }\bibfield  {title}
  {\bibinfo {title} {Pressure-temperature phase diagram of the
  {${\mathrm{EuRbFe}}_{4}{\mathrm{As}}_{4}$} superconductor},\ }\href
  {https://doi.org/10.1103/PhysRevB.99.144509} {\bibfield  {journal} {\bibinfo
  {journal} {Phys. Rev. B}\ }\textbf {\bibinfo {volume} {99}},\ \bibinfo
  {pages} {144509} (\bibinfo {year} {2019})}\BibitemShut {NoStop}%
\bibitem [{\citenamefont {Iida}\ \emph {et~al.}(2019)\citenamefont {Iida},
  \citenamefont {Nagai}, \citenamefont {Ishida}, \citenamefont {Ishikado},
  \citenamefont {Murai}, \citenamefont {Christianson}, \citenamefont {Yoshida},
  \citenamefont {Inamura}, \citenamefont {Nakamura}, \citenamefont {Nakao},
  \citenamefont {Munakata}, \citenamefont {Kagerbauer}, \citenamefont
  {Eisterer}, \citenamefont {Kawashima}, \citenamefont {Yoshida}, \citenamefont
  {Eisaki},\ and\ \citenamefont {Iyo}}]{IidaPhysRevB.100.014506}%
  \BibitemOpen
  \bibfield  {author} {\bibinfo {author} {\bibfnamefont {K.}~\bibnamefont
  {Iida}}, \bibinfo {author} {\bibfnamefont {Y.}~\bibnamefont {Nagai}},
  \bibinfo {author} {\bibfnamefont {S.}~\bibnamefont {Ishida}}, \bibinfo
  {author} {\bibfnamefont {M.}~\bibnamefont {Ishikado}}, \bibinfo {author}
  {\bibfnamefont {N.}~\bibnamefont {Murai}}, \bibinfo {author} {\bibfnamefont
  {A.~D.}\ \bibnamefont {Christianson}}, \bibinfo {author} {\bibfnamefont
  {H.}~\bibnamefont {Yoshida}}, \bibinfo {author} {\bibfnamefont
  {Y.}~\bibnamefont {Inamura}}, \bibinfo {author} {\bibfnamefont
  {H.}~\bibnamefont {Nakamura}}, \bibinfo {author} {\bibfnamefont
  {A.}~\bibnamefont {Nakao}}, \bibinfo {author} {\bibfnamefont
  {K.}~\bibnamefont {Munakata}}, \bibinfo {author} {\bibfnamefont
  {D.}~\bibnamefont {Kagerbauer}}, \bibinfo {author} {\bibfnamefont
  {M.}~\bibnamefont {Eisterer}}, \bibinfo {author} {\bibfnamefont
  {K.}~\bibnamefont {Kawashima}}, \bibinfo {author} {\bibfnamefont
  {Y.}~\bibnamefont {Yoshida}}, \bibinfo {author} {\bibfnamefont
  {H.}~\bibnamefont {Eisaki}},\ and\ \bibinfo {author} {\bibfnamefont
  {A.}~\bibnamefont {Iyo}},\ }\bibfield  {title} {\bibinfo {title} {Coexisting
  spin resonance and long-range magnetic order of {Eu} in
  {${\mathrm{EuRbFe}}_{4}{\mathrm{As}}_{4}$}},\ }\href
  {https://doi.org/10.1103/PhysRevB.100.014506} {\bibfield  {journal} {\bibinfo
   {journal} {Phys. Rev. B}\ }\textbf {\bibinfo {volume} {100}},\ \bibinfo
  {pages} {014506} (\bibinfo {year} {2019})}\BibitemShut {NoStop}%
\bibitem [{\citenamefont {Islam}\ \emph {et~al.}(2020)\citenamefont {Islam},
  \citenamefont {Chmaissem}, \citenamefont {Koshelev}, \citenamefont {Kim},
  \citenamefont {Cao}, \citenamefont {Rydh}, \citenamefont {Smylie},
  \citenamefont {Willa}, \citenamefont {Bao}, \citenamefont {Chung},
  \citenamefont {Kanatzidis}, \citenamefont {Kwok}, \citenamefont
  {Rosenkranz},\ and\ \citenamefont {Welp}}]{IslamPreprint2019}%
  \BibitemOpen
  \bibfield  {author} {\bibinfo {author} {\bibfnamefont {Z.}~\bibnamefont
  {Islam}}, \bibinfo {author} {\bibfnamefont {O.}~\bibnamefont {Chmaissem}},
  \bibinfo {author} {\bibfnamefont {A.~E.}\ \bibnamefont {Koshelev}}, \bibinfo
  {author} {\bibfnamefont {J.-W.}\ \bibnamefont {Kim}}, \bibinfo {author}
  {\bibfnamefont {H.}~\bibnamefont {Cao}}, \bibinfo {author} {\bibfnamefont
  {A.}~\bibnamefont {Rydh}}, \bibinfo {author} {\bibfnamefont {M.~P.}\
  \bibnamefont {Smylie}}, \bibinfo {author} {\bibfnamefont {K.}~\bibnamefont
  {Willa}}, \bibinfo {author} {\bibfnamefont {J.}~\bibnamefont {Bao}}, \bibinfo
  {author} {\bibfnamefont {D.~Y.}\ \bibnamefont {Chung}}, \bibinfo {author}
  {\bibfnamefont {M.}~\bibnamefont {Kanatzidis}}, \bibinfo {author}
  {\bibfnamefont {W.-K.}\ \bibnamefont {Kwok}}, \bibinfo {author}
  {\bibfnamefont {S.}~\bibnamefont {Rosenkranz}},\ and\ \bibinfo {author}
  {\bibfnamefont {U.}~\bibnamefont {Welp}},\ }\bibfield  {title} {\bibinfo
  {title} {unpublished}} (\bibinfo {year} {2020})\BibitemShut {NoStop}%
\bibitem [{\citenamefont {Van~Kranendonk}\ and\ \citenamefont
  {Van~Vleck}(1958)}]{VanKranendonkRevModPhys.30.1}%
  \BibitemOpen
  \bibfield  {author} {\bibinfo {author} {\bibfnamefont {J.}~\bibnamefont
  {Van~Kranendonk}}\ and\ \bibinfo {author} {\bibfnamefont {J.~H.}\
  \bibnamefont {Van~Vleck}},\ }\bibfield  {title} {\bibinfo {title} {Spin
  waves},\ }\href {https://doi.org/10.1103/RevModPhys.30.1} {\bibfield
  {journal} {\bibinfo  {journal} {Rev. Mod. Phys.}\ }\textbf {\bibinfo {volume}
  {30}},\ \bibinfo {pages} {1} (\bibinfo {year} {1958})}\BibitemShut {NoStop}%
\bibitem [{\citenamefont {Akhiezer}\ \emph {et~al.}(1968)\citenamefont
  {Akhiezer}, \citenamefont {Baryakhtar},\ and\ \citenamefont
  {Peletminskii}}]{AkhiezerBook68}%
  \BibitemOpen
  \bibfield  {author} {\bibinfo {author} {\bibfnamefont {A.~I.}\ \bibnamefont
  {Akhiezer}}, \bibinfo {author} {\bibfnamefont {V.~G.}\ \bibnamefont
  {Baryakhtar}},\ and\ \bibinfo {author} {\bibfnamefont {S.~V.}\ \bibnamefont
  {Peletminskii}},\ }\href@noop {} {\emph {\bibinfo {title} {{Spin waves}}}},\
  North-Holland series in low temperature physics\ (\bibinfo  {publisher}
  {North-Holland},\ \bibinfo {address} {Amsterdam},\ \bibinfo {year}
  {1968})\BibitemShut {NoStop}%
\bibitem [{\citenamefont {Prabhakar}\ and\ \citenamefont
  {Stancil}(2009)}]{prabhakar2009spin}%
  \BibitemOpen
  \bibfield  {author} {\bibinfo {author} {\bibfnamefont {A.}~\bibnamefont
  {Prabhakar}}\ and\ \bibinfo {author} {\bibfnamefont {D.~D.}\ \bibnamefont
  {Stancil}},\ }\href@noop {} {\emph {\bibinfo {title} {Spin waves: Theory and
  applications}}}\ (\bibinfo  {publisher} {Springer},\ \bibinfo {address}
  {Boston},\ \bibinfo {year} {2009})\BibitemShut {NoStop}%
\bibitem [{\citenamefont {Wolf}\ \emph {et~al.}(2001)\citenamefont {Wolf},
  \citenamefont {Awschalom}, \citenamefont {Buhrman}, \citenamefont {Daughton},
  \citenamefont {von Molnár}, \citenamefont {Roukes}, \citenamefont
  {Chtchelkanova},\ and\ \citenamefont {Treger}}]{WolfSci2001}%
  \BibitemOpen
  \bibfield  {author} {\bibinfo {author} {\bibfnamefont {S.~A.}\ \bibnamefont
  {Wolf}}, \bibinfo {author} {\bibfnamefont {D.~D.}\ \bibnamefont {Awschalom}},
  \bibinfo {author} {\bibfnamefont {R.~A.}\ \bibnamefont {Buhrman}}, \bibinfo
  {author} {\bibfnamefont {J.~M.}\ \bibnamefont {Daughton}}, \bibinfo {author}
  {\bibfnamefont {S.}~\bibnamefont {von Molnár}}, \bibinfo {author}
  {\bibfnamefont {M.~L.}\ \bibnamefont {Roukes}}, \bibinfo {author}
  {\bibfnamefont {A.~Y.}\ \bibnamefont {Chtchelkanova}},\ and\ \bibinfo
  {author} {\bibfnamefont {D.~M.}\ \bibnamefont {Treger}},\ }\bibfield  {title}
  {\bibinfo {title} {Spintronics: A spin-based electronics vision for the
  future},\ }\href {https://doi.org/10.1126/science.1065389} {\bibfield
  {journal} {\bibinfo  {journal} {Science}\ }\textbf {\bibinfo {volume}
  {294}},\ \bibinfo {pages} {1488} (\bibinfo {year} {2001})}\BibitemShut
  {NoStop}%
\bibitem [{\citenamefont {\ifmmode \check{Z}\else
  \v{Z}\fi{}uti\ifmmode~\acute{c}\else \'{c}\fi{}}\ \emph
  {et~al.}(2004)\citenamefont {\ifmmode \check{Z}\else
  \v{Z}\fi{}uti\ifmmode~\acute{c}\else \'{c}\fi{}}, \citenamefont {Fabian},\
  and\ \citenamefont {Das~Sarma}}]{ZuticRevModPhys.76.323}%
  \BibitemOpen
  \bibfield  {author} {\bibinfo {author} {\bibfnamefont {I.}~\bibnamefont
  {\ifmmode \check{Z}\else \v{Z}\fi{}uti\ifmmode~\acute{c}\else \'{c}\fi{}}},
  \bibinfo {author} {\bibfnamefont {J.}~\bibnamefont {Fabian}},\ and\ \bibinfo
  {author} {\bibfnamefont {S.}~\bibnamefont {Das~Sarma}},\ }\bibfield  {title}
  {\bibinfo {title} {Spintronics: Fundamentals and applications},\ }\href
  {https://doi.org/10.1103/RevModPhys.76.323} {\bibfield  {journal} {\bibinfo
  {journal} {Rev. Mod. Phys.}\ }\textbf {\bibinfo {volume} {76}},\ \bibinfo
  {pages} {323} (\bibinfo {year} {2004})}\BibitemShut {NoStop}%
\bibitem [{\citenamefont {Neusser}\ and\ \citenamefont
  {Grundler}(2009)}]{NeusserAdvMat09}%
  \BibitemOpen
  \bibfield  {author} {\bibinfo {author} {\bibfnamefont {S.}~\bibnamefont
  {Neusser}}\ and\ \bibinfo {author} {\bibfnamefont {D.}~\bibnamefont
  {Grundler}},\ }\bibfield  {title} {\bibinfo {title} {Magnonics: Spin waves on
  the nanoscale},\ }\href
  {https://doi.org/https://doi.org/10.1002/adma.200900809} {\bibfield
  {journal} {\bibinfo  {journal} {Advanced Materials}\ }\textbf {\bibinfo
  {volume} {21}},\ \bibinfo {pages} {2927} (\bibinfo {year}
  {2009})}\BibitemShut {NoStop}%
\bibitem [{\citenamefont {Kruglyak}\ \emph {et~al.}(2010)\citenamefont
  {Kruglyak}, \citenamefont {Demokritov},\ and\ \citenamefont
  {Grundler}}]{KruglyakJPhys2010}%
  \BibitemOpen
  \bibfield  {author} {\bibinfo {author} {\bibfnamefont {V.~V.}\ \bibnamefont
  {Kruglyak}}, \bibinfo {author} {\bibfnamefont {S.~O.}\ \bibnamefont
  {Demokritov}},\ and\ \bibinfo {author} {\bibfnamefont {D.}~\bibnamefont
  {Grundler}},\ }\bibfield  {title} {\bibinfo {title} {Magnonics},\ }\href
  {https://doi.org/10.1088/0022-3727/43/26/264001} {\bibfield  {journal}
  {\bibinfo  {journal} {Journal of Physics D: Applied Physics}\ }\textbf
  {\bibinfo {volume} {43}},\ \bibinfo {pages} {264001} (\bibinfo {year}
  {2010})}\BibitemShut {NoStop}%
\bibitem [{\citenamefont {Chumak}\ \emph {et~al.}(2015)\citenamefont {Chumak},
  \citenamefont {Vasyuchka}, \citenamefont {Serga},\ and\ \citenamefont
  {Hillebrands}}]{ChumakNatPhys2015}%
  \BibitemOpen
  \bibfield  {author} {\bibinfo {author} {\bibfnamefont {A.~V.}\ \bibnamefont
  {Chumak}}, \bibinfo {author} {\bibfnamefont {V.}~\bibnamefont {Vasyuchka}},
  \bibinfo {author} {\bibfnamefont {A.}~\bibnamefont {Serga}},\ and\ \bibinfo
  {author} {\bibfnamefont {B.}~\bibnamefont {Hillebrands}},\ }\bibfield
  {title} {\bibinfo {title} {Magnon spintronics},\ }\href
  {https://doi.org/10.1038/nphys3347} {\bibfield  {journal} {\bibinfo
  {journal} {Nature Physics}\ }\textbf {\bibinfo {volume} {11}},\ \bibinfo
  {pages} {453} (\bibinfo {year} {2015})}\BibitemShut {NoStop}%
\bibitem [{\citenamefont {Braude}\ and\ \citenamefont
  {Sonin}(2004)}]{BraudePhysRevLett.93.117001}%
  \BibitemOpen
  \bibfield  {author} {\bibinfo {author} {\bibfnamefont {V.}~\bibnamefont
  {Braude}}\ and\ \bibinfo {author} {\bibfnamefont {E.~B.}\ \bibnamefont
  {Sonin}},\ }\bibfield  {title} {\bibinfo {title} {Excitation of spin waves in
  superconducting ferromagnets},\ }\href
  {https://doi.org/10.1103/PhysRevLett.93.117001} {\bibfield  {journal}
  {\bibinfo  {journal} {Phys. Rev. Lett.}\ }\textbf {\bibinfo {volume} {93}},\
  \bibinfo {pages} {117001} (\bibinfo {year} {2004})}\BibitemShut {NoStop}%
\bibitem [{\citenamefont {Braude}(2006)}]{BraudePhysRevB.74.054515}%
  \BibitemOpen
  \bibfield  {author} {\bibinfo {author} {\bibfnamefont {V.}~\bibnamefont
  {Braude}},\ }\bibfield  {title} {\bibinfo {title} {Microwave response and
  spin waves in superconducting ferromagnets},\ }\href
  {https://doi.org/10.1103/PhysRevB.74.054515} {\bibfield  {journal} {\bibinfo
  {journal} {Phys. Rev. B}\ }\textbf {\bibinfo {volume} {74}},\ \bibinfo
  {pages} {054515} (\bibinfo {year} {2006})}\BibitemShut {NoStop}%
\bibitem [{\citenamefont {Buzdin}(1984)}]{BuzdinJETPLett84}%
  \BibitemOpen
  \bibfield  {author} {\bibinfo {author} {\bibfnamefont {A.~I.}\ \bibnamefont
  {Buzdin}},\ }\bibfield  {title} {\bibinfo {title} {Spin-wave spectrum of
  antiferromagnetic superconductors},\ }\href@noop {} {\bibfield  {journal}
  {\bibinfo  {journal} {JETP Lett.}\ }\textbf {\bibinfo {volume} {40}},\
  \bibinfo {pages} {956} (\bibinfo {year} {1984})},\ \bibinfo {note} {[{Pis'ma
  Zh. Eksp. Teor. Fiz., \textbf{40} 193 (1984)}]}\BibitemShut {NoStop}%
\bibitem [{\citenamefont {Volkov}\ and\ \citenamefont
  {Efetov}(2009)}]{VolkovPhysRevLett.103.037003}%
  \BibitemOpen
  \bibfield  {author} {\bibinfo {author} {\bibfnamefont {A.~F.}\ \bibnamefont
  {Volkov}}\ and\ \bibinfo {author} {\bibfnamefont {K.~B.}\ \bibnamefont
  {Efetov}},\ }\bibfield  {title} {\bibinfo {title} {Hybridization of spin and
  plasma waves in josephson tunnel junctions containing a ferromagnetic
  layer},\ }\href {https://doi.org/10.1103/PhysRevLett.103.037003} {\bibfield
  {journal} {\bibinfo  {journal} {Phys. Rev. Lett.}\ }\textbf {\bibinfo
  {volume} {103}},\ \bibinfo {pages} {037003} (\bibinfo {year}
  {2009})}\BibitemShut {NoStop}%
\bibitem [{\citenamefont {Mai}\ \emph {et~al.}(2011)\citenamefont {Mai},
  \citenamefont {Kandelaki}, \citenamefont {Volkov},\ and\ \citenamefont
  {Efetov}}]{MaiPhysRevB.84.144519}%
  \BibitemOpen
  \bibfield  {author} {\bibinfo {author} {\bibfnamefont {S.}~\bibnamefont
  {Mai}}, \bibinfo {author} {\bibfnamefont {E.}~\bibnamefont {Kandelaki}},
  \bibinfo {author} {\bibfnamefont {A.~F.}\ \bibnamefont {Volkov}},\ and\
  \bibinfo {author} {\bibfnamefont {K.~B.}\ \bibnamefont {Efetov}},\ }\bibfield
   {title} {\bibinfo {title} {Interaction of josephson and magnetic
  oscillations in josephson tunnel junctions with a ferromagnetic layer},\
  }\href {https://doi.org/10.1103/PhysRevB.84.144519} {\bibfield  {journal}
  {\bibinfo  {journal} {Phys. Rev. B}\ }\textbf {\bibinfo {volume} {84}},\
  \bibinfo {pages} {144519} (\bibinfo {year} {2011})}\BibitemShut {NoStop}%
\bibitem [{\citenamefont {Koshelev}(2019)}]{KoshelevPhysRevB19}%
  \BibitemOpen
  \bibfield  {author} {\bibinfo {author} {\bibfnamefont {A.~E.}\ \bibnamefont
  {Koshelev}},\ }\bibfield  {title} {\bibinfo {title} {Helical structures in
  layered magnetic superconductors due to indirect exchange interactions
  mediated by interlayer tunneling},\ }\href
  {https://doi.org/10.1103/PhysRevB.100.224503} {\bibfield  {journal} {\bibinfo
   {journal} {Phys. Rev. B}\ }\textbf {\bibinfo {volume} {100}},\ \bibinfo
  {pages} {224503} (\bibinfo {year} {2019})}\BibitemShut {NoStop}%
\bibitem [{\citenamefont {Nagamiya}(1968)}]{Nagamiya1968}%
  \BibitemOpen
  \bibfield  {author} {\bibinfo {author} {\bibfnamefont {T.}~\bibnamefont
  {Nagamiya}},\ }\bibfield  {title} {\bibinfo {title} {Helical spin ordering--1
  {Theory} of helical spin configurations},\ }in\ \href
  {http://www.sciencedirect.com/science/article/pii/S0081194708602209} {\emph
  {\bibinfo {booktitle} {Solid State Physics}}},\ Vol.~\bibinfo {volume} {20},\
  \bibinfo {editor} {edited by\ \bibinfo {editor} {\bibfnamefont
  {F.}~\bibnamefont {Seitz}}, \bibinfo {editor} {\bibfnamefont
  {D.}~\bibnamefont {Turnbull}},\ and\ \bibinfo {editor} {\bibfnamefont
  {H.}~\bibnamefont {Ehrenreich}}}\ (\bibinfo  {publisher} {Academic Press},\
  \bibinfo {year} {1968})\ pp.\ \bibinfo {pages} {305--411}\BibitemShut
  {NoStop}%
\bibitem [{\citenamefont {Johnston}(2015)}]{JohnstonPhysRevB.91.064427}%
  \BibitemOpen
  \bibfield  {author} {\bibinfo {author} {\bibfnamefont {D.~C.}\ \bibnamefont
  {Johnston}},\ }\bibfield  {title} {\bibinfo {title} {Unified molecular field
  theory for collinear and noncollinear {Heisenberg} antiferromagnets},\ }\href
  {https://doi.org/10.1103/PhysRevB.91.064427} {\bibfield  {journal} {\bibinfo
  {journal} {Phys. Rev. B}\ }\textbf {\bibinfo {volume} {91}},\ \bibinfo
  {pages} {064427} (\bibinfo {year} {2015})}\BibitemShut {NoStop}%
\bibitem [{\citenamefont {Kulik}\ and\ \citenamefont
  {Yanson}(1972)}]{kulik1972josephson}%
  \BibitemOpen
  \bibfield  {author} {\bibinfo {author} {\bibfnamefont {I.}~\bibnamefont
  {Kulik}}\ and\ \bibinfo {author} {\bibfnamefont {I.}~\bibnamefont {Yanson}},\
  }\href@noop {} {\emph {\bibinfo {title} {Josephson Effect in Superconducting
  Tunneling Structures}}}\ (\bibinfo  {publisher} {Israel Program for
  Scientific Translations},\ \bibinfo {address} {Jerusalem},\ \bibinfo {year}
  {1972})\BibitemShut {NoStop}%
\bibitem [{\citenamefont {Barone}\ and\ \citenamefont
  {Paterno}(1982)}]{BaroneBook}%
  \BibitemOpen
  \bibfield  {author} {\bibinfo {author} {\bibfnamefont {A.}~\bibnamefont
  {Barone}}\ and\ \bibinfo {author} {\bibfnamefont {G.}~\bibnamefont
  {Paterno}},\ }\href@noop {} {\emph {\bibinfo {title} {Physics and
  Applications of The Josephson Effect}}}\ (\bibinfo  {publisher} {Wiley},\
  \bibinfo {address} {New York},\ \bibinfo {year} {1982})\BibitemShut {NoStop}%
\bibitem [{\citenamefont {Swihart}(1961)}]{SwihartJAP61}%
  \BibitemOpen
  \bibfield  {author} {\bibinfo {author} {\bibfnamefont {J.~C.}\ \bibnamefont
  {Swihart}},\ }\bibfield  {title} {\bibinfo {title} {Field solution for a
  thin-film superconducting strip transmission line},\ }\href
  {https://doi.org/10.1063/1.1736025} {\bibfield  {journal} {\bibinfo
  {journal} {J. Appl. Phys.}\ }\textbf {\bibinfo {volume} {32}},\ \bibinfo
  {pages} {461} (\bibinfo {year} {1961})}\BibitemShut {NoStop}%
\bibitem [{\citenamefont {Cirillo}\ \emph {et~al.}(1998)\citenamefont
  {Cirillo}, \citenamefont {Gr\o{}nbech-Jensen}, \citenamefont {Samuelsen},
  \citenamefont {Salerno},\ and\ \citenamefont
  {Rinati}}]{CirilloPhysRevB.58.12377}%
  \BibitemOpen
  \bibfield  {author} {\bibinfo {author} {\bibfnamefont {M.}~\bibnamefont
  {Cirillo}}, \bibinfo {author} {\bibfnamefont {N.}~\bibnamefont
  {Gr\o{}nbech-Jensen}}, \bibinfo {author} {\bibfnamefont {M.~R.}\ \bibnamefont
  {Samuelsen}}, \bibinfo {author} {\bibfnamefont {M.}~\bibnamefont {Salerno}},\
  and\ \bibinfo {author} {\bibfnamefont {G.~V.}\ \bibnamefont {Rinati}},\
  }\bibfield  {title} {\bibinfo {title} {Fiske modes and {Eck} steps in long
  {Josephson} junctions: {Theory} and experiments},\ }\href
  {https://doi.org/10.1103/PhysRevB.58.12377} {\bibfield  {journal} {\bibinfo
  {journal} {Phys. Rev. B}\ }\textbf {\bibinfo {volume} {58}},\ \bibinfo
  {pages} {12377} (\bibinfo {year} {1998})}\BibitemShut {NoStop}%
\bibitem [{\citenamefont {Kirtley}(2019)}]{KirtleyInBook2019}%
  \BibitemOpen
  \bibfield  {author} {\bibinfo {author} {\bibfnamefont {J.~R.}\ \bibnamefont
  {Kirtley}},\ }\bibfield  {title} {\bibinfo {title} {Magnetic field effects in
  {Josephson} junctions},\ }in\ \href
  {https://doi.org/10.1007/978-3-030-20726-7_6} {\emph {\bibinfo {booktitle}
  {Fundamentals and Frontiers of the Josephson Effect}}},\ \bibinfo {editor}
  {edited by\ \bibinfo {editor} {\bibfnamefont {F.}~\bibnamefont {Tafuri}}}\
  (\bibinfo  {publisher} {Springer International Publishing},\ \bibinfo
  {address} {Cham},\ \bibinfo {year} {2019})\ pp.\ \bibinfo {pages}
  {209--233}\BibitemShut {NoStop}%
\bibitem [{\citenamefont {Carlson}\ and\ \citenamefont
  {Goldman}(1976)}]{CarlsonJLTP1976}%
  \BibitemOpen
  \bibfield  {author} {\bibinfo {author} {\bibfnamefont {R.~V.}\ \bibnamefont
  {Carlson}}\ and\ \bibinfo {author} {\bibfnamefont {A.~M.}\ \bibnamefont
  {Goldman}},\ }\bibfield  {title} {\bibinfo {title} {Dynamics of the order
  parameter of superconducting aluminum films},\ }\href
  {https://doi.org/10.1007/BF00654825} {\bibfield  {journal} {\bibinfo
  {journal} {J. Low Temp. Phys.}\ }\textbf {\bibinfo {volume} {25}},\ \bibinfo
  {pages} {67} (\bibinfo {year} {1976})}\BibitemShut {NoStop}%
\bibitem [{\citenamefont {Kulik}(1965)}]{KulikJETPLett65}%
  \BibitemOpen
  \bibfield  {author} {\bibinfo {author} {\bibfnamefont {I.}~\bibnamefont
  {Kulik}},\ }\bibfield  {title} {\bibinfo {title} {Theory of ``steps'' of
  voltage-current characteristics of the {Josephson} tunnel current},\
  }\href@noop {} {\bibfield  {journal} {\bibinfo  {journal} {JETP Lett.}\
  }\textbf {\bibinfo {volume} {2}},\ \bibinfo {pages} {84} (\bibinfo {year}
  {1965})},\ \bibinfo {note} {{[Zh. Eksper. Teor. Pis. Red. \textbf{2},
  134(1965)]}}\BibitemShut {NoStop}%
\end{thebibliography}%

\end{document}